\documentclass[twocolumn,epjc3]{svjour3}  
\usepackage{amsmath,amssymb,amsfonts}
\usepackage{graphicx,graphics}
\usepackage{graphics}
\usepackage{url}
\usepackage{slashed}
\usepackage{color}
\usepackage{mathtools}

\journalname{Eur. Phys. J. C}

\begin{document}

\title{EPPS16: Nuclear parton distributions with LHC data}

\author{Kari~J.~Eskola$^{1,2,}$\thanksref{e2}, Petja~Paakkinen$^{1,}$\thanksref{e3}, Hannu~Paukkunen$^{1,2,3,}$\thanksref{e1}, Carlos~A.~Salgado$^{3,}$\thanksref{e4}}

\thankstext{e2}{e-mail: kari.eskola@jyu.fi}
\thankstext{e3}{e-mail: petja.paakkinen@jyu.fi‎}
\thankstext{e1}{e-mail: hannu.paukkunen@jyu.fi}
\thankstext{e4}{e-mail: carlos.salgado@usc.es}

\institute{University of Jyvaskyla, Department of Physics, P.O. Box 35, FI-40014 University of Jyvaskyla, Finland \and
Helsinki Institute of Physics, P.O. Box 64, FI-00014 University of Helsinki, Finland \and
Instituto Galego de F\'\i sica de Altas Enerx\'\i as (IGFAE), Universidade de Santiago de Compostela, E-15782 Galicia, Spain
}

\maketitle

\abstract{
We introduce a global analysis of collinearly factorized nuclear parton distribution functions (PDFs) including, for the first time, data constraints from LHC proton-lead collisions. In comparison to our previous analysis, EPS09, where data only from charged-lepton-nucleus deep inelastic scattering (DIS), Drell-Yan (DY) dilepton production in proton-nucleus collisions and inclusive pion production in deuteron-nucleus collisions were the input, we now increase the variety of data constraints to cover also neutrino-nucleus DIS and low-mass DY production in pion-nucleus collisions. The new LHC data significantly extend the kinematic reach of the data constraints. We now allow much more freedom for the flavour dependence of nuclear effects than in other currently available analyses. As a result, especially the uncertainty estimates are more objective flavour by flavour. The neutrino DIS plays a pivotal role in obtaining a mutually consistent behaviour for both up and down valence quarks, and the LHC dijet data clearly constrain gluons at large momentum fraction. Mainly for insufficient statistics, the pion-nucleus DY and heavy gauge boson production in proton-lead collisions impose less visible constraints. The outcome - a new set of next-to-leading order nuclear PDFs called EPPS16 - is made available for applications in high-energy nuclear collisions.
}

\section{Introduction}
\label{Introduction}

Proton-lead (pPb) and lead-lead (PbPb) collisions at the Large Hadron Collider (LHC) have brought heavy-ion physics to the high-energy realm \cite{Salgado:2011wc,ARNP-SW,Armesto:2015ioy,Foka:2016zdb}. A more than ten-fold increase in the center-of-mass energy with respect to the deuteron-gold (DAu) collisions at the Relativistic Heavy-Ion Collider (RHIC) has made it possible to study novel hard-process observables in a heavy-ion environment. For example, production cross sections of heavy gauge bosons (Z and W$^\pm$) and jets have been measured. Because of the new experimental information from the LHC it is now also timely to update the pre-LHC global analyses of collinearly factorized nuclear parton distribution functions (PDFs) ---  
for reviews, see e.g. Refs.~\cite{Eskola:2012rg,Paukkunen:2014nqa}.

The original idea of having nuclear effects in PDFs was data-driven as the early deep inelastic scattering (DIS) experiments unexpectedly revealed significant nuclear effects in the cross sections \cite{Aubert:1983xm,Arneodo:1992wf}. It was then demonstrated \cite{Frankfurt:1990xz,Eskola:1992zb} that such effects in DIS and fixed nuclear-target Drell-Yan (DY) cross sections can be consistently described by modifying the free nucleon PDFs at low $Q^2$
and letting the Dokshitzer-Gribov-Lipatov-Altarelli -Parisi  (DGLAP) evolution \cite{Dokshitzer:1977sg,Gribov:1972ri,Gribov:1972rt,Altarelli:1977zs} take care of the $Q^2$ dependence.
In other words, the data were in line with a concept that the measured nuclear effects are of non-perturbative origin but at sufficiently high $Q^2$ there is no fundamental difference in the scattering off a nucleon or off a nucleus. 
These ideas eventually led to the first global fit and the EKS98 set of leading-order  nuclear PDFs \cite{Eskola:1998iy,Eskola:1998df}. 
Since then, several parametrizations based on the DIS and DY data have been released at leading order (EKPS \cite{Eskola:2007my}, HKM\cite{Hirai:2001np}, HKN04 \cite{Hirai:2004wq}), next-to-leading order (nDS \cite{deFlorian:2003qf}, HKN07 \cite{Hirai:2007sx}, nCTEQ \cite{Schienbein:2009kk}, AT12 \cite{AtashbarTehrani:2012xh}), and next-to-next-to-leading order (KA15 \cite{Khanpour:2016pph}) perturbative QCD.\footnote{For studies addressing origins of the nuclear effects, see e.g. Refs.~\cite{Frankfurt:2011cs,Armesto:2006ph,Kulagin:2004ie}.}
For the rather limited kinematic coverage of the fixed-target data and the fact that only two types of data were used in these fits, significant simplifying assumptions had to be made e.g. with respect to the flavour dependence of the nuclear effects. The constraints on the gluon distribution are also weak in these analyses, and it is only along with the RHIC pion data \cite{Adler:2006wg} that an observable carrying direct information on the nuclear gluons has been added to the global fits --- first in EPS08 \cite{Eskola:2008ca} and EPS09 \cite{Eskola:2009uj}, later in DSSZ \cite{deFlorian:2011fp} and nCTEQ15 \cite{Kovarik:2015cma}. The interpretation of the RHIC pion production data is not, however, entirely unambiguous as the parton-to-pion fragmentation functions (FFs) may as well undergo a nuclear modification \cite{Sassot:2009sh}. This approach was adopted in the DSSZ fit, and consequently their gluons show clearly weaker nuclear effects than in EPS09 (and nCTEQ15) where the FFs were considered to be free from nuclear modifications. To break the tie, more data and new observables were called for. To this end, the recent LHC dijet measurements \cite{Chatrchyan:2014hqa} from pPb collisions have been most essential as a consistent description of these data is obtained with EPS09 and nCTEQ15 but not with DSSZ \cite{Paukkunen:2014pha,Armesto:2015lrg}. 

Another observable that has caused some controversy and debate during the past years is the neutrino-nucleus DIS. It has been claimed \cite{Kovarik:2010uv} (see also Ref.~\cite{Hirai:2016ykc}) that the nuclear PDFs required to correctly describe neutrino data are different than those optimal for the charged-lepton induced DIS measurements. However, it has been demonstrated \cite{Paukkunen:2010hb,Paukkunen:2013grz} that problems appear only in the case of one single data set and, furthermore, that it seems to be largely a normalization issue (which could e.g. be related to the incident neutrino flux which is model-dependent). The neutrino data were also used in the DSSZ fit without visible difficulties.

New data from the LHC 2013 p-Pb run have gradually become available and their impact on the nuclear PDFs has been studied \cite{Armesto:2015lrg,Kusina:2016fxy} in the context of PDF reweighting \cite{Paukkunen:2014zia}. Apart from the aforementioned dijet data \cite{Chatrchyan:2014hqa} which will e.g. require a complete renovation of the DSSZ approach, the available W \cite{Khachatryan:2015hha,Alice:2016wka} and Z \cite{Khachatryan:2015pzs,Aad:2015gta} data were found to have only a rather mild effect mainly for the limited statistical precision of the data. However, the analysis of Ref.~\cite{Armesto:2015lrg} used only nuclear PDFs (EPS09, DSSZ) in which flavour-independent valence and light sea quark distributions were assumed at the parametrization scale. Thus, it could not reveal the possible constraints that these electroweak observables could have for a particular quark flavour. On the other hand, the analysis of Ref.~\cite{Kusina:2016fxy} involves some flavour dependence but the usage of absolute cross sections which are sensitive to the free proton baseline PDFs complicates the interpretation of the results.

In the present paper, we update the EPS09 analysis by adding a wealth of new data from
neutrino DIS \cite{Onengut:2005kv}, pion-nucleus DY process \cite{Badier:1981ci,Bordalo:1987cs,Heinrich:1989cp}, and especially LHC pPb dijet \cite{Chatrchyan:2014hqa}, Z \cite{Khachatryan:2015pzs,Aad:2015gta} and W \cite{Khachatryan:2015hha} production. By this, we take the global nuclear PDF fits onto a completely new level in the variety of data types. In addition, in comparison to EPS09, a large part of the whole framework is upgraded: we switch to a general-mass formalism for the heavy quarks, relax the assumption of the flavour independent nuclear modifications for quarks at the parametrization scale, undo the isospin corrections that some experiments had applied on their data, and also importantly, we now assign no extra weights to any of the data sets. In this updated analysis, we find no significant tension between the data sets considered, which lends support to the assumption of process-independent nuclear PDFs in the studied kinematical region. The result of the analysis presented in this paper is also published as a new set of next-to-leading order (NLO) nuclear PDFs, which we call EPPS16 and which supersedes our earlier set EPS09. The new EPPS16 set will be available at \cite{EPPS16Web}.

\section{Parametrization of nuclear PDFs}
\label{TheNuclearPDFs}

\begin{figure}[htb!]
\centering
\includegraphics[width=\linewidth]{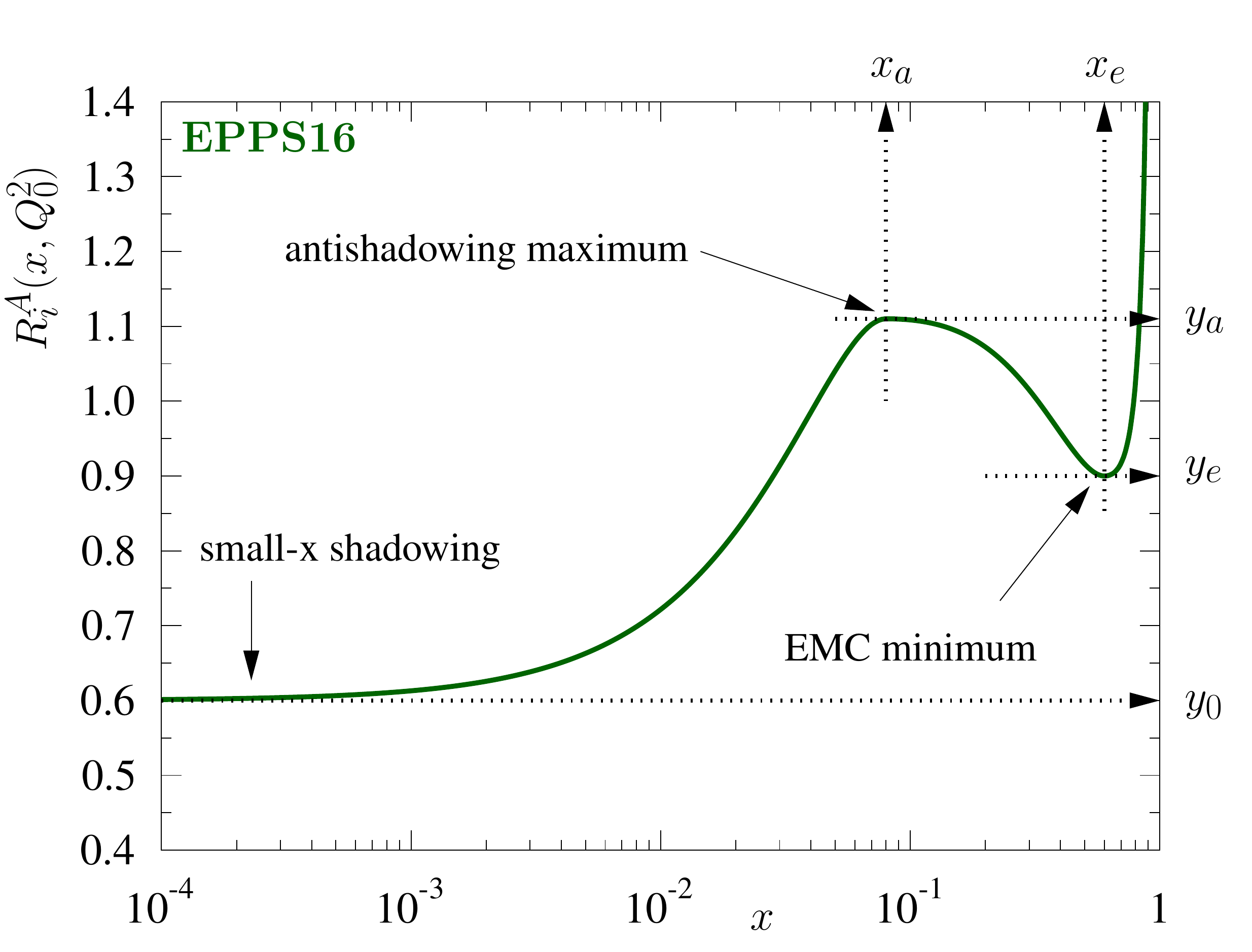}
\caption{Illustration of the EPPS16 fit function $R_i^A(x,Q^2_0)$.}
\label{fig:FitForm}
\end{figure}

Similarly to our earlier works, the bound proton PDF $f_i^{{\rm p}/A}(x,Q^2)$ for mass number $A$ and parton species $i$ is defined relative to the free proton PDF $f_i^{\rm p}(x,Q^2)$ as
\begin{equation}
f_i^{{\rm p}/A}(x,Q^2) = R_i^A(x,Q^2) f_i^{{\rm p}}(x,Q^2),
\end{equation}
where $R_i^A(x,Q^2)$ is the scale-dependent nuclear modification. Our free proton baseline is CT14NLO \cite{Dulat:2015mca}. Consistently with this choice, our analysis here uses the SACOT (simplified Aivazis-Collins-Olness-Tung) gener\-al-mass variable flavour number scheme \cite{Kramer:2000hn,Collins:1998rz,Thorne:2008xf} for the DIS cross sections. The fit function for the nuclear modifications $R_i^A(x,Q^2_0)$ at the parametrization scale $Q^2_0$, illustrated in Fig.~\ref{fig:FitForm}, is also largely inherited from our earlier analyses \cite{Eskola:1998iy,Eskola:2007my,Eskola:2008ca,Eskola:2009uj},
\begin{equation}
R_i^A(x,Q^2_0) = 
\left\{
\begin{array}{ll}
a_0 + a_1(x-x_a)^2  & x \leq x_a \\
b_0 + b_1x^\alpha + b_2x^{2\alpha} + b_3x^{3\alpha} & x_a \leq x \leq x_e \\
c_0 + \left(c_1-c_2x \right) \left(1-x\right)^{-\beta} & x_e \leq x \leq 1,
\end{array}
\right.  
\label{eq:FitForm}
\end{equation}
where $\alpha=10x_a$ and the $i$ and $A$ dependencies of the parameters on the r.h.s. are left implicit.\footnote{See Ref.~\cite{Helenius:2016hcu} for a study experimenting with a more flexible fit function at small $x$.} The purpose of the exponent $\alpha$ is to avoid the ``plateau'' that would otherwise (that is, if $\alpha=1$) develop if $x_a<0.1$. The coefficients $a_i,b_i,c_i$ are fully determined by the
asymptotic small-$x$ limit $y_0=R_i^A(x\rightarrow 0,Q^2_0)$, the antishadowing maximum $y_a=R_i^A(x_a,Q^2_0)$ and the EMC minimum $y_e=R_i^A(x_e,Q^2_0)$, as well as requiring continuity and vanishing first derivatives at the matching points $x_a$ and $x_e$. The $A$ dependencies of $y_0$, $y_a$, $y_e$ are parametrized as
\begin{equation}
 y_i(A) = y_i(A_{\rm ref}) \left(\frac{A}{A_{\rm ref}} \right)^{\gamma_i \left[y_i(A_{\rm ref}) - 1\right]},  \label{eq:Adep}
\end{equation}
where $\gamma_i \geq 0$ and $A_{\rm ref} = 12$. By construction, the nuclear effects (deviations from unity) are now larger for heavier nuclei. Without the factor $y_i(A_{\rm ref})-1$ in the exponent one can more easily fall into a peculiar situation in which e.g. $y_i(A_{\rm ref})<1$, but $y_i(A\gg A_{\rm ref})>1$, which seems physically unlikely. For the valence quarks and gluons the values of $y_0$ are determined by requiring the sum rules
\begin{align}
 &\int_0^1 dx f_{u_{\rm V}}^{{\rm p}/A}(x,Q_0^2) = 2, \label{eq:sum1} \\
 &\int_0^1 dx f_{d_{\rm V}}^{{\rm p}/A}(x,Q_0^2) = 1, \label{eq:sum2} \\
 &\int_0^1 dx x \sum_i f_i^{{\rm p}/A}(x,Q_0^2) = 1,    \label{eq:sum3}
\end{align}
separately for each nucleus and thus the $A$ dependence of these $y_0$ is not parametrized. All other parameters than $y_0$, $y_a$, $y_e$ are $A$-independent. In our present framework we consider the deuteron ($A=2$) to be free from nuclear effects though few-percent effects at high $x$ are found e.g. in Ref.~\cite{Martin:2012da}. The bound neutron PDFs $f_i^{{\rm n}/A}(x,Q^2)$ are obtained from the bound proton PDFs by assuming isospin symmetry,
\begin{align}
f_{u,\overline{u}}^{{\rm n}/A}(x,Q^2) & =  f_{d,\overline{d}}^{{\rm p}/A}(x,Q^2), \\
f_{d,\overline{d}}^{{\rm n}/A}(x,Q^2) & =  f_{u,\overline{u}}^{{\rm p}/A}(x,Q^2), \\
f_{i}^{{\rm n}/A}(x,Q^2) & =  f_{i}^{{\rm p}/A}(x,Q^2) \quad {\rm for} \, {\rm other} \, {\rm flavours}.
\end{align}

Above the parametrization scale $Q^2>Q^2_0$ the nuclear PDFs are obtained by solving the DGLAP evolution equations with 2-loop splitting functions \cite{Furmanski:1980cm,Curci:1980uw}. We use our own DGLAP evolution code which is based on the solution method described in Ref.~\cite{Santorelli:1998yt} and also explained and benchmarked in Ref.~\cite{Paukkunen:2009ks}. Our parametri\-zation scale $Q^2_0$ is fixed to the charm pole mass $Q^2_0 = m_c^2$ where $m_c= 1.3\,{\rm GeV}$. The bottom quark mass is $m_b= 4.75\,{\rm GeV}$ and the value of the strong coupling constant is set by $\alpha_{\rm s}(M_{\rm Z}) = 0.118$, where $M_{\rm Z}$ is the mass of the $Z$ boson. 

As is well known, at NLO and beyond the PDFs do not need to be positive definite and we do not impose such a restriction either. In fact, doing so would be artificial since the parametrization scale is, in principle, arbitrary and positive definite PDFs, say, at $Q^2_0=m_c^2$ may easily correspond to negative small-$x$ PDFs at a scale just slightly below $Q^2_0$. As we could have equally well parametrized the PDFs at such a lower value of $Q^2_0$, we see that restricting the PDFs to be always positive would be an unphysical requirement.

\section{Experimental data}
\label{Experimentaldata}

All the $\ell^- A$ DIS, p$A$ DY and RHIC DAu pion data sets we use in the present analysis are the same as in the EPS09 fit. The only modification on this part is that we now remove the isoscalar corrections of the EMC, NMC and SLAC data (see the next subsection), which is important as we have freed the flavour dependence of the quark nuclear modifications. The $\ell^- A$ DIS data (cross sections or structure functions $F_2$) are always normalized by the $\ell^-$D measurements and, as in EPS09, the only kinematic cut on these data is $Q^2>m_{\rm c}^2$. This is somewhat lower than in typical free-proton fits and the implicit assumption is (also in not setting a cut in the mass of the hadronic final state) that the possible higher-twist effects will cancel in ratios of structure functions/cross sections. While potential signs of $1/Q^2$ effects have been seen in the HERA data \cite{Abt:2016vjh} already around $Q^2=10 \, {\rm GeV}^2$, these effects occur at significantly smaller $x$ than what is the reach of the $\ell^- A$ DIS data. 

From the older measurements, also pion-nucleus DY data from the NA3 \cite{Badier:1981ci}, NA10 \cite{Bordalo:1987cs}, and E615 \cite{Heinrich:1989cp} collaborations are now included. These data have been shown \cite{Dutta:2010pg,Paakkinen:2016wxk} to carry some sensitivity to the flavour-dependent EMC effect. However, more stringent flavour-dependence constraints at large $x$ are provided by the CHORUS (anti)neutrino-Pb DIS data \cite{Onengut:2005kv}, whose treatment in the fit is detailedly explained in Section \ref{Treatmentoftheneutrinodata}.

The present analysis is the first one to directly include LHC data. To this end, we use the currently published pPb data for heavy-gauge boson \cite{Khachatryan:2015hha,Khachatryan:2015pzs,Aad:2015gta} and dijet production \cite{Chatrchyan:2014hqa}. These observables have already been discussed in the literature \cite{Paukkunen:2010qg,Ru:2014yma,Eskola:2013aya,Ru:2016wfx,Armesto:2015lrg,Kusina:2016fxy} in the context of nuclear PDFs. 
Importantly, we include the LHC pPb data always as forward-to-backward ratios in which the cross sections at positive (pseudo)ra\-pidi\-ties $\eta>0$ are divided by the ones at negative rapidities $\eta<0$. This is to reduce the sensitivity to the chosen free-proton baseline PDFs as well as to cancel the experimental luminosity uncertainty. However, upon taking the ratio part of the information is also lost as, for example, the points near $\eta=0$ are, by construction, always close to unity and carry essentially no information. In addition, since the correlations on the systematic errors are not available, all the experimental uncertainties are added in quadrature when forming these ratios
(except for the CMS W measurement \cite{Khachatryan:2015hha} which is taken directly from the publication) which partly undermines the constraining power of these data. The baseline pp measurements performed at the same $\sqrt{s}$ as the pPb runs may, in the future, also facilitate a direct usage of the nuclear modification factors $d\sigma^{\rm pPb}/d\sigma^{\rm pp}$. The technicalities of how the LHC data are included in our analysis are discussed in Section \ref{LookuptablesforLHCobservables}.

\begin{figure}[htb!]
\centering
\includegraphics[width=\linewidth]{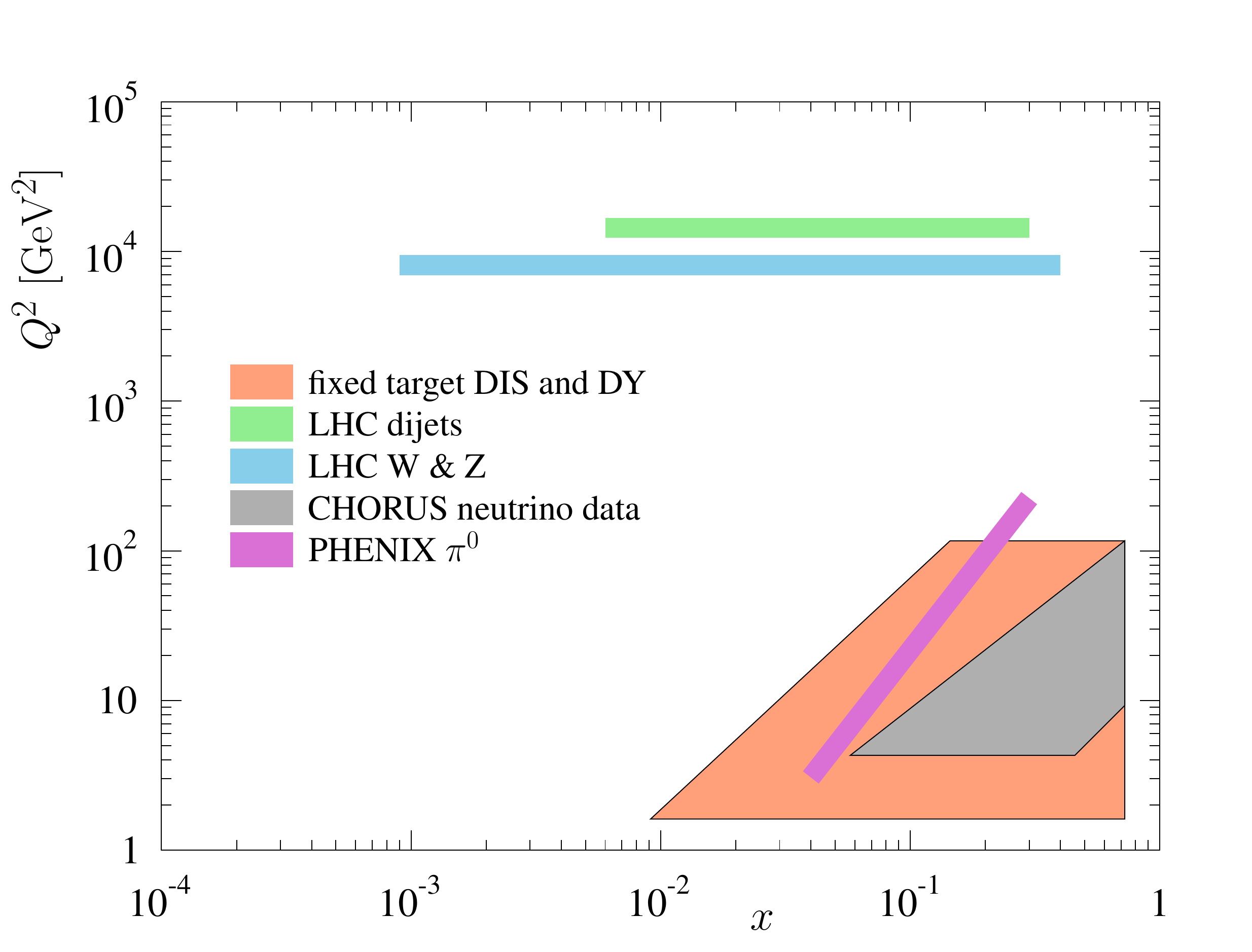}
\caption{The approximate regions in the $(x,Q^2)$ plane at which different data in the EPPS16 fit probe the nuclear PDFs.}
\label{fig:xQ2}
\end{figure}

In Fig.~\ref{fig:xQ2} we illustrate the predominant $x$ and $Q^2$ regions probed by the data. Clearly, the LHC data probe the nuclear PDFs at much higher in $Q^2$ than the earlier DIS and DY data. For the wide rapidity coverage of the LHC detectors the new measurements also reach lower values of $x$ than the old data, but for the limited statistical precision the constraints for the small-$x$ end still remain rather weak. All the exploited data sets including the number of data points, their $\chi^2$ contribution and references are listed in Table~\ref{Table:Data}. We note that, approximately half of the data are now for the $^{208}$Pb nucleus while in the EPS09 analysis only 15 Pb data points (NMC 96) were included. Most of this change is caused by the inclusion of the CHORUS neutrino data.
  
\begin{table*}
\begin{center}
{
\caption{The data sets used in the EPPS16 analysis, listed in the order of growing nuclear mass number. The number of data points and their contribution to $\chi^2$ counts only those data points that fall within the kinematic cuts explained in the text. The new data with respect to the EPS09 analysis are marked with a star.}
\begin{tabular}{lclrrl}
 Experiment & Observable &  Collisions & Data points & $\chi^2$ & Ref. \\
\hline
\hline
 SLAC E139	     & DIS	& $e^-$He(4), $e^-$D        & 21 & 12.2    & \cite{Gomez:1993ri} \\
 CERN NMC 95, re.	     & DIS	& $\mu^-$He(4), $\mu^-$D        & 16 & 18.0    & \cite{Amaudruz:1995tq} \\
 \\
 CERN NMC 95              & DIS	& $\mu^-$Li(6), $\mu^-$D        & 15  & 18.4   & \cite{Arneodo:1995cs} \\
 CERN NMC 95, $Q^2$ dep.  & DIS	& $\mu^-$Li(6), $\mu^-$D        & 153 & 161.2  & \cite{Arneodo:1995cs} \\
  \\
 SLAC E139          & DIS	& $e^-$Be(9), $e^-$D        & 20 & 12.9    & \cite{Gomez:1993ri}  \\
 CERN NMC 96              & DIS	& $\mu^-$Be(9), $\mu^-$C        & 15 & 4.4     & \cite{Arneodo:1996rv} \\
 \\
 SLAC E139	     & DIS	& $e^-$C(12), $e^-$D        &  7  & 6.4    & \cite{Gomez:1993ri}  \\
 CERN NMC 95              & DIS	& $\mu^-$C(12), $\mu^-$D        & 15  & 9.0    & \cite{Arneodo:1995cs} \\
 CERN NMC 95, $Q^2$ dep.  & DIS	& $\mu^-$C(12), $\mu^-$D        & 165 & 133.6  & \cite{Arneodo:1995cs}  \\ 
 CERN NMC 95, re.         & DIS	& $\mu^-$C(12), $\mu^-$D        & 16  & 16.7   & \cite{Amaudruz:1995tq} \\
 CERN NMC 95, re.         & DIS	& $\mu^-$C(12), $\mu^-$Li(6)    & 20  & 27.9   & \cite{Amaudruz:1995tq} \\
 FNAL E772           & DY	& pC(12), pD                &  9  & 11.3   & \cite{Alde:1990im}    \\
\\
 SLAC E139	     & DIS	& $e^-$Al(27), $e^-$D       & 20 & 13.7    & \cite{Gomez:1993ri}   \\
 CERN NMC 96    	     & DIS 	& $\mu^-$Al(27), $\mu^-$C(12)   & 15 & 5.6     & \cite{Arneodo:1996rv} \\
\\
 SLAC E139	     & DIS	& $e^-$Ca(40), $e^-$D       &  7 & 4.8     & \cite{Gomez:1993ri}   \\
 FNAL E772 	     &  DY	& pCa(40), pD               &  9 & 3.33    & \cite{Alde:1990im}  \\
 CERN NMC 95, re. 	     & DIS	& $\mu^-$Ca(40), $\mu^-$D       & 15 & 27.6    & \cite{Amaudruz:1995tq} \\
 CERN NMC 95, re. 	     & DIS	& $\mu^-$Ca(40), $\mu^-$Li(6)   & 20 & 19.5    & \cite{Amaudruz:1995tq} \\
 CERN NMC 96    	     & DIS	& $\mu^-$Ca(40), $\mu^-$C(12)   & 15 & 6.4     & \cite{Arneodo:1996rv} \\
\\
 SLAC E139	     & DIS	& $e^-$Fe(56), $e^-$D       & 26 & 22.6    & \cite{Gomez:1993ri}   \\
 FNAL E772 	     & DY	& $e^-$Fe(56), $e^-$D       &  9 & 3.0     & \cite{Alde:1990im}    \\
 CERN NMC 96    	     & DIS	& $\mu^-$Fe(56), $\mu^-$C(12)   & 15 & 10.8    & \cite{Arneodo:1996rv} \\
 FNAL E866  	     & DY       & pFe(56), pBe(9)           & 28 & 20.1    & \cite{Vasilev:1999fa} \\
\\ 
 CERN EMC           & DIS      & $\mu^-$Cu(64), $\mu^-$D       & 19 & 15.4    & \cite{Ashman:1992kv}   \\
\\
 SLAC E139	     & DIS	& $e^-$Ag(108), $e^-$D      &  7 & 8.0     & \cite{Gomez:1993ri}   \\
\\
 CERN NMC 96  		& DIS  	& $\mu^-$Sn(117), $\mu^-$C(12)  & 15 & 12.5    & \cite{Arneodo:1996rv} \\
 
 CERN NMC 96, $Q^2$ dep.  & DIS	& $\mu^-$Sn(117), $\mu^-$C(12)  & 144 & 87.6   & \cite{Arneodo:1996ru} \\
\\
 FNAL E772 	     & DY	& pW(184), pD     	    &  9  & 7.2    & \cite{Alde:1990im}    \\
 FNAL E866 	     & DY	& pW(184), pBe(9)           & 28  & 26.1   & \cite{Vasilev:1999fa} \\
 CERN NA10${^\bigstar}$   & DY	&  $\pi^-$W(184), $\pi^-$D   &  10  &  11.6   & \cite{Bordalo:1987cs}\\
 FNAL E615${^\bigstar}$   & DY	&  $\pi^+{\rm W}(184)$, $\pi^-{\rm W}$(184)  &  11  &  10.2 & \cite{Heinrich:1989cp} \\
\\
 CERN NA3${^\bigstar}$    & DY	&  $\pi^-$Pt(195), $\pi^-$H  & 7  &  4.6     & \cite{Badier:1981ci} \\
\\
 SLAC E139	     & DIS	& $e^-$Au(197), $e^-$D      & 21 & 8.4     & \cite{Gomez:1993ri}   \\
 RHIC PHENIX              & $\pi^0$ 	& dAu(197), pp              & 20 & 6.9     & \cite{Adler:2006wg}\\

\\
 CERN NMC 96       	     & DIS	        & $\mu^-$Pb(207), $\mu^-$C(12)  & 15 & 4.1     & \cite{Arneodo:1996rv} \\
 CERN CMS${^\bigstar}$    & W$^\pm$	        & pPb(208)                 & 10 & 8.8    & \cite{Khachatryan:2015hha}\\
 CERN CMS${^\bigstar}$    & Z	        & pPb(208)                 & 6  & 5.8    & \cite{Khachatryan:2015pzs}\\
 CERN ATLAS${^\bigstar}$  & Z	        & pPb(208)                 & 7  & 9.6    & \cite{Aad:2015gta}\\
 CERN CMS${^\bigstar}$    & dijet	        & pPb(208)                 & 7  & 5.5    & \cite{Chatrchyan:2014hqa}\\
 CERN CHORUS${^\bigstar}$ & DIS	        & $\nu$Pb(208), $\overline{\nu}$Pb(208)  & 824  &  998.6  & \cite{Onengut:2005kv} \\
\\

 \hline		   
 Total 		     &        &     &  1811 & 1789 &                       \\
\end{tabular}
}
\label{Table:Data}
\end{center}
\end{table*}

\subsection{Isoscalar corrections}
\label{IsoscalarcorectionforDISdata}

Part of the charged-lepton DIS data that have been used in the earlier global nPDF fits had been ``corrected'', in the original publications, for the isospin effects. That is, the experimental collaborations had tried to eliminate the effects emerging from the unequal number of protons and neutrons when making the comparison with the deuteron data. In this way the ratios $F_2^A/F_2^{\rm D}$ could be directly interpreted in terms of nuclear effects in the PDFs. However, this is clearly an unnecessary operation from the viewpoint of global fits, which has previously caused some confusion regarding the nuclear valence quark modifications: the particularly mild effects found in the nDS \cite{deFlorian:2003qf} and DSSZ \cite{deFlorian:2011fp} analyses (see Fig.~\ref{fig:Pb10_with_EPS09} ahead) most likely originate from neglecting such a correction.

The structure function of a nucleus $A$ with $Z$ protons and $N$ neutrons can be written as
\begin{equation}
 F_2^A = \frac{Z}{A} F_2^{{\rm p},A} + \frac{N}{A} F_2^{{\rm n},A}, \label{eq:nstruck}
\end{equation}
where $F_2^{{\rm p},A}$ and $F_2^{{\rm n},A}$ are the structure functions of the bound protons and neutrons. The corresponding isoscalar structure function is defined as the one containing an equal number of protons and neutrons,
\begin{equation}
 \hat{F}_2^A = \frac{1}{2} F_2^{{\rm p},A} + \frac{1}{2} F_2^{{\rm n},A}.
\end{equation}
Using Eq.~(\ref{eq:nstruck}), the isoscalar structure function reads
\begin{equation}
 \hat{F}_2^A  =  \beta F_2^{A},
\end{equation}
where
\begin{equation}
\beta  = \frac{A}{2} \left(1+\frac{F_2^{{\rm n},A}}{F_2^{{\rm p},A}}\right) / \left(Z + N\frac{F_2^{{\rm n},A}}{F_2^{{\rm p},A}}\right) \, .
\end{equation}
Usually, it has been assumed that the ratio ${F_2^{{\rm n},A}}/{F_2^{{\rm p},A}}$ is free from nuclear effects,
\begin{equation}
 \frac{F_2^{{\rm n},A}}{F_2^{{\rm p},A}} = \frac{F_2^{{\rm n}}}{F_2^{{\rm p}}},
\end{equation}
and parametrized according to the DIS data from proton and deuteron targets. Different experiments have used different versions:
\begin{itemize}
\item EMC parametrization \cite{Ashman:1992kv}:
 $$ 
 \frac{F_2^{{\rm n}}}{F_2^{{\rm p}}} = 0.92-0.86x,
 $$
\item SLAC parametrization \cite{Gomez:1993ri}:
 $$ 
 \frac{F_2^{{\rm n}}}{F_2^{{\rm p}}} = 1-0.8x,
 $$
\item NMC parametrization \cite{Amaudruz:1991nw}:
 \begin{eqnarray}
 \frac{F_2^{{\rm n}}}{F_2^{{\rm p}}} & = & A(x) \left( \frac{Q^2}{20} \right)^{B(x)} \left(1 + \frac{x^2}{Q^2} \right) \nonumber \\
A(x) & = & 0.979 - 1.692x + 2.797x^2 - 4.313x^3 + 3.075x^4 \nonumber \\
B(x) & = & -0.171x^2 + 0.244x^3. \nonumber 
\end{eqnarray}
\end{itemize}
Using these functions we calculate the correction factors $\beta$ thereby obtaining the ratios $F_2^A/F_2^{\rm D}$, to be used in the fit, from the isoscalar versions $\hat{F}_2^A/F_2^{\rm D}$ reported by the experiments.

As discussed in Ref.~\cite{Paakkinen:2016wxk}, also the $\pi^- A$ DY data from the NA10 collaboration \cite{Bordalo:1987cs} have been balanced for the neutron excess. The correction was done by utilizing the leading-order DY cross section. Here, we account for this with the isospin correction factor given in Eq.~(8) of Ref.~\cite{Paakkinen:2016wxk}.

\subsection{Treatment of neutrino DIS data}
\label{Treatmentoftheneutrinodata}

In the present work we make use of the CHORUS neutrino and antineutrino DIS data \cite{Onengut:2005kv}. Similar measurements are available also from the CDHSW \cite{Berge:1989hr} and NuTeV \cite{Tzanov:2005kr} collaborations, but only for the CHORUS data the correlations of the systematic uncertainties are directly available in the form we need.\footnote{\url{http://choruswww.cern.ch/Publications/DIS-data/}}
Moreover, the $^{208}$Pb target has a larger neutron excess than the iron targets of CDHSW and NuTeV, thereby carrying more information on the flavour separation. The data are reported as double differential cross sections $d\sigma_{i, {\rm exp}}^{\nu, \overline{\nu}}/dxdy$ 
in the standard DIS variables
and, guided by our free-proton baseline fit CT14NLO~\cite{Dulat:2015mca}, the kinematic cuts we set on these data are $Q^2>4 \, {\rm GeV}^2$ and $W^2>12.25 \, {\rm GeV}^2$.\footnote{The cuts are more stringent here than for other DIS data as only absolute cross sections are available (instead of those relative to a lighter nucleus).} In the computation of these NLO neutrino DIS cross sections, we apply
the dominant electroweak \cite{Arbuzov:2004zr} and target-mass \cite{Accardi:2008ne} corrections as in Refs.~\cite{Paukkunen:2010hb,Paukkunen:2013grz}, together with the SACOT quark-mass scheme.  

In order to suppress the theoretical uncertainties related to the free-proton PDFs,
as well as experimental systematic uncertainties, we treat the data following the normalization prescription laid out in Ref.~\cite{Paukkunen:2013grz}. For each (anti)neutrino beam energy $E$, we compute the total cross section as
\begin{equation}
\sigma_{\rm exp}^{\nu, \overline{\nu}}(E) = \sum_{i} \frac{d\sigma_{i, {\rm exp}}^{\nu, \overline{\nu}}}{dxdy} \Delta_i^{xy} \delta_{E_,E_i},
\label{eq:totxsec}
\end{equation}
where $E_i$ is the beam energy corresponding to the $i$th data point. By $\Delta_i^{xy}$ we mean the size of the $(x,y)$ bin (rectangles) to which the $i$th data point belongs. 
The original data are then normalized by the estimated total cross sections of Eq.~(\ref{eq:totxsec}) as
\begin{equation}
\frac{d\tilde\sigma_{i, {\rm exp}}^{\nu, \overline{\nu}}}{dxdy} \equiv \frac{d\sigma_{i, {\rm exp}}^{\nu, \overline{\nu}}}{dxdy} \bigg/ \sigma_{\rm exp}^{\nu, \overline{\nu}}(E=E_i).
\label{eq:normxsec}
\end{equation}
As discussed e.g. in \cite{Paukkunen:2014zia,Gao:2013xoa}, the $\chi^2$ contribution of data with correlated uncertainties is obtained in terms of the covariance matrix $C$ as
\begin{equation}
 \chi^2 = \sum_{i,j} \left(\frac{d\tilde\sigma_{i, {\rm exp}}^{\nu, \overline{\nu}}}{dxdy} - \frac{d\tilde\sigma_{i, {\rm th}}^{\nu, \overline{\nu}}}{dxdy} \right) C_{ij}^{-1} \left(\frac{d\tilde\sigma_{j, {\rm exp}}^{\nu, \overline{\nu}}}{dxdy} - \frac{d\tilde\sigma_{j, {\rm th}}^{\nu, \overline{\nu}}}{dxdy} \right),
\end{equation}
where now the theory values ${d\tilde\sigma_{j, {\rm th}}^{\nu, \overline{\nu}}}/{dxdy}$ are the computed differential cross sections normalized by the corresponding integrated cross section (similarly to Eq. (\ref{eq:normxsec})). The elements of the covariance matrix  are in our case defined as  
\begin{equation}
C_{ij} \equiv  \left(\tilde\delta_i^{\rm stat}\right)^2 \delta_{ij} + \sum_k \tilde\beta_i^k \tilde\beta_j^k,
\label{Cmatrix}
\end{equation}
where the statistical uncertainty $\tilde \delta_i^{\rm stat}$ on ${d\tilde\sigma_{i, {\rm exp}}^{\nu, \overline{\nu}}}/{dxdy}$ is computed from the original statistical uncertainties $\delta_i^{\rm stat}$ by
\begin{equation}
\tilde \delta_i^{\rm stat} \equiv \delta_i^{\rm stat}/\sigma_{\rm exp}^{\nu, \overline{\nu}}(E_i).
\end{equation}
Here we neglect the statistical uncertainty of $\sigma^{\nu, \overline{\nu}}(E)$ as for this integrated quantity it is always clearly smaller than that of the individual data points. The point-to-point correlated systematic uncertainties $\tilde\beta_i^k$ for the normalized data points we form as
\begin{equation}
\tilde\beta_i^k \equiv \left(\frac{d\sigma_{i, {\rm exp}}^{\nu, \overline{\nu}}}{dxdy}+\beta_i^k\right) \bigg/ \sigma^{\nu, \overline{\nu}}_k(E_i) - \frac{d\tilde\sigma_{i, {\rm exp}}^{\nu, \overline{\nu}}}{dxdy},
\label{betamato}
\end{equation}
where 
\begin{equation}
\sigma_k^{\nu, \overline{\nu}}(E) = \sum_{i} 
\left(\frac{d\sigma_{i, {\rm exp}}^{\nu, \overline{\nu}}}{dxdy} +\beta_i^k \right)\Delta_i^{xy}\delta_{E_,E_i}.
\end{equation}
Above, the index $k$ labels the parameters controlling the experimental systematic uncertainties and 
$\beta_i^k$ are the cross section shifts corresponding to a one standard deviation change in the $k$th parameter. We note that $\tilde\beta_i^k$ in Eq.~\eqref{betamato} for the relative cross sections in Eq.~\eqref{eq:normxsec} are constructed such that if the $\beta_i^k$ correspond only to the same relative normalization shift for all points, then $\tilde\beta_i^k$ are just zero. We also note that in Eq.~\eqref{Cmatrix} we have assumed that the response of ${d\tilde\sigma_{i, {\rm exp}}^{\nu, \overline{\nu}}}/{dxdy}$  to the systematic uncertainty parameters  is linear.

As shown in e.g. Ref.~\cite{Paukkunen:2010hb}, the $Q^2$ dependence of nuclear effects in neutrino DIS data is weak. Hence, 
for a concise graphical presentation of the data as a function of $x$, we integrate over the $y$ variable by
\begin{equation}
\frac{d\tilde\sigma_{{\rm exp}}^{\nu, \overline{\nu}}}{dx}(E) = \sum_{j} \frac{d\tilde\sigma_{j, {\rm exp}}^{\nu, \overline{\nu}}}{dxdy}  \Delta_j^{y} \delta_{x,x_j}\delta_{E_,E_j}, \label{eq:nuplot1}
\end{equation}
where $\Delta_j^{y}$ is the size of the $y$ bin to which the $j$th data point belongs, and $x_j$ the corresponding value of the $x$ variable. The overall statistical uncertainty 
to the relative cross section in Eq.~\eqref{eq:nuplot1} is computed as
\begin{equation}
\delta^{\rm stat} 
(E,x)
= \sqrt{\sum_{j} \left(\tilde\delta^{\rm stat}_j \Delta_j^{y} \right)^2 \delta_{x,x_j}\delta_{E,E_j}} \, , \label{eq:nuplot2}
\end{equation}
and the total systematic uncertainty is given by
\begin{equation}
 \delta^{\rm sys} (E,x)
= \sqrt{
 \sum_k \left[ \delta_k^{\rm sys} 
(E,x)
\right]^2
 }, 
\end{equation}
where
\begin{equation}
 \delta_k^{\rm sys} 
(E,x) =  
\sum_{j} \tilde\beta_j^k \Delta_j^{y} \delta_{x,x_j}\delta_{E_,E_j}.
\label{eq:nuplot3}
\end{equation}
In the plots for ${d\tilde\sigma_{{\rm exp}}^{\nu, \overline{\nu}}} / {dx}$ presented in Section \ref{Results} (Figs.~\ref{fig:nudatat} and \ref{fig:anudatat} ahead), the statistical and total systematic uncertainties have been added in quadrature. We also divide by the theory values obtained by using the CT14\-NLO free proton PDFs (but still with the correct amount of protons and neutrons). We stress that Eqs.~(\ref{eq:nuplot1})--(\ref{eq:nuplot3}) are used only for a simple graphical presentation of the data but not for the actual fit.

\subsection{Look-up tables for LHC observables and others}
\label{LookuptablesforLHCobservables}

\begin{figure}[htb!]
\centering
\includegraphics[width=\linewidth]{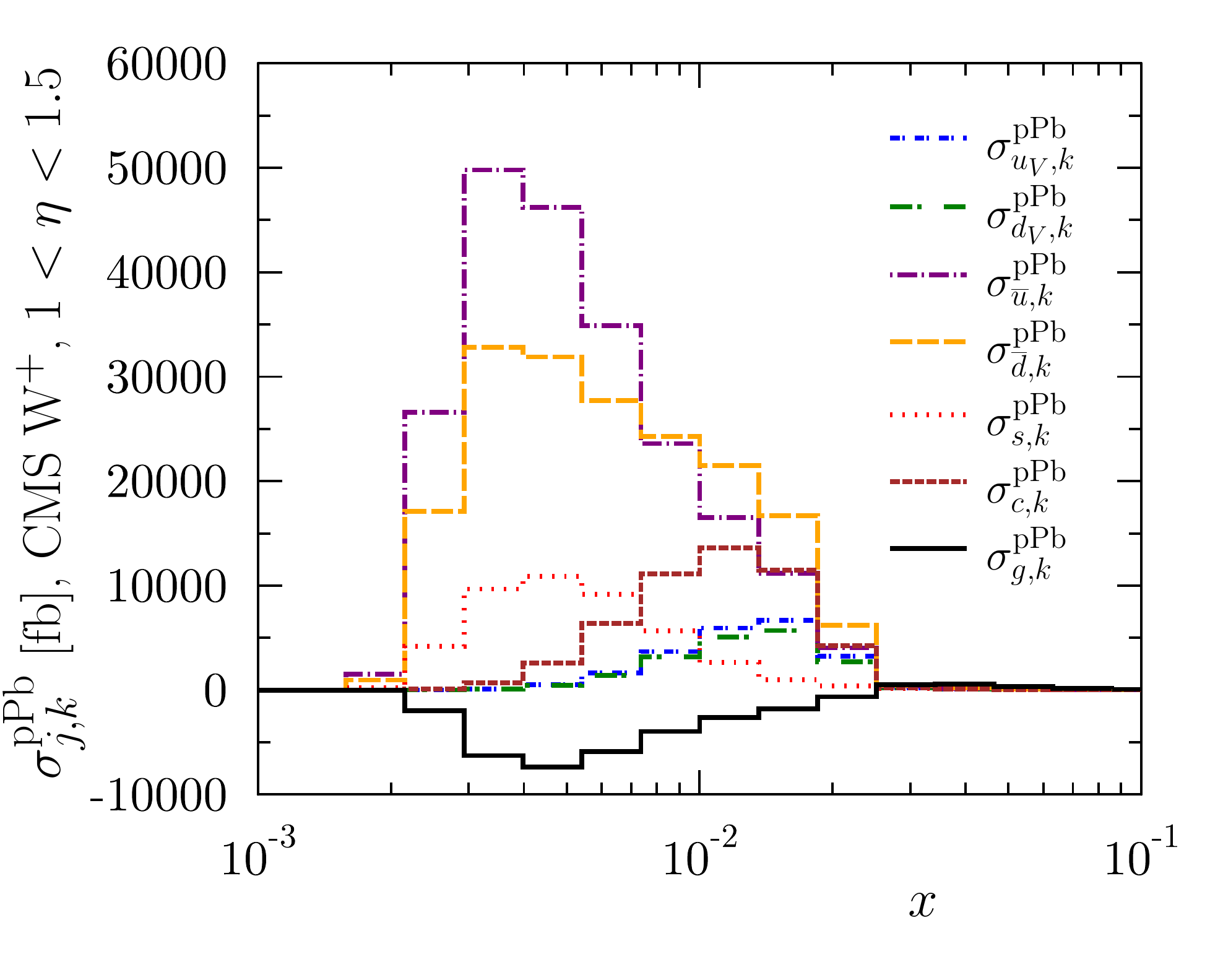}
\caption{An example of the $\sigma_{j,k}^{\rm pPb}$ histograms used in evaluating the LHC pPb cross sections in Eq.~\eqref{eq:LHCxsec}. The cross section $\sigma^{\rm pPb}$ is computed as a sum of all the bins weighted by the appropriate nuclear modification factors. The sum of all the bins gives the cross section with no nuclear modifications ($R_i^{\rm Pb}=1$).}
\label{fig:illustrate}
\end{figure}

In order to efficiently include the LHC observables in our fit at the NLO level, a fast method to evaluate the cross sections is essential. We have adopted the following pragmatic approach: For a given observable, a hard-process cross section $\sigma^{\rm pPb}$ in pPb collisions,
we set up a grid in the $x$ variable of the Pb nucleus, $x_0,\ldots,x_N=1$, and evaluate, for each $x$ bin $k$ and parton flavor $j$ 
\begin{equation}
 \sigma_{j,k}^{\rm pPb} = \sum_{i} f_i^{\rm p} \otimes \hat \sigma_{ij} \otimes f_{j,k}^{\rm Pb}, 
\end{equation}
where $\hat \sigma_{ij}$ are the coefficient functions appropriate for a given process and $f_{j,k}^{\rm Pb}$ involve only proton PDFs with no nuclear modifications,
\begin{align}
f_{j,k}^{\rm Pb}(x) & \equiv \sum_\ell \left[Zf_{\ell}^{\rm p, Pb}(x) + Nf_{\ell}^{\rm n, Pb}(x)\right]\bigg|_{R_j^{\rm Pb}=1, R_{i\neq j }^{\rm Pb}= 0 } \nonumber \\ 
& \times  \theta\left(x-x_{k-1} \right) \theta\left(x_{k}-x \right). 
\end{align}
Thus, the functions $f_{j,k}^{\rm Pb}$ pick up the partonic weight of the nuclear modification $R_j^{\rm Pb}$ in a given interval $x_{k-1} < x < x_{k}$. Since the nuclear modification factors $R_i^A$ are relatively slowly varying functions in $x$ (e.g. in comparison to the absolute PDFs), the observable $\sigma^{\rm pPb}$ can be computed as a sum of $\sigma_{j,k}^{\rm pPb}$ weighted by the appropriate nuclear modification,
\begin{equation}
\sigma^{\rm pPb} = \sum_{j,k} \sigma_{j,k}^{\rm pPb} R_j^{\rm Pb}(x_{k-1} < x < x_{k}). \label{eq:LHCxsec}
\end{equation}
As an illustration, in Fig.~\ref{fig:illustrate}, we show the histograms of $\sigma_{j,k}^{\rm pPb}$ corresponding to W$^+$ production measured by CMS in the bin $1 < \eta_{\rm lab} < 1.5$. For the electroweak LHC observables  we have used the MCFM code \cite{Campbell:2015qma} to compute the grids, and for dijet production the modified EKS code \cite{Gao:2012he,Kunszt:1992tn,Ellis:1992en}. 

We set up similar grids also for inclusive pion production in DAu collisions at RHIC using the INCNLO \cite{Aversa:1988vb} code
with KKP FFs \cite{Kniehl:2000fe}, and for the DY process in $\pi A$ collisions using MCFM with the GRV pion PDFs \cite{Gluck:1991ey}. In all cases, we have checked that the grids reproduce a direct evaluation of the observables within 1\% accuracy in the case of EPS09 nuclear PDFs.

\section{Analysis procedure}
\label{Analysisprocedure}

The standard statistical procedure for comparing experimental data to theory
is to inspect the behaviour of the overall $\chi^2$ function, defined as
\begin{equation}
\chi^2 \left( \vec a \right)  \equiv  \sum_k  \chi^2_k \left( \vec a \right), \label{eq:chi2global}
\end{equation}
where $\vec a$ is a set of theory parameters and 
$\chi^2_k \left( \vec a \right)$ denotes the contribution of each independent data set $k$,
\begin{equation}
\chi^2_k \left( \vec a \right)  \equiv  \sum_{i,j} \left[ T_i\left( \vec a \right) - D_i \right] C_{ij}^{-1}
                                          \left[ T_j\left( \vec a \right) - D_j \right]. \label{eq:chi2individual}
\end{equation}
Here, $T_i\left( \vec a \right)$ denote the theoretical values of the observables in the data set $k$, $D_i$ are the corresponding experimental values, and $C_{ij}$ is the covariance matrix. In most cases, only the total uncertainty is known, and in this case $C_{ij} = (\delta^{\rm uncorr.}_i)^2\delta_{ij}$, where $\delta^{\rm uncorr.}_i$ is the point-to-point uncorrelated data uncertainty. In the case that the only correlated uncertainty is the overall normalization $\delta^{\rm norm.}$, we can also write
\begin{equation}
\chi^2_k \left( \vec a \right)  = \left( \frac{1-f_N}{\delta^{\rm norm.}} \right)^2 + \sum_{i}  \left[ \frac{T_i\left( \vec a \right) - f_N D_i}{\delta^{\rm uncorr.}_i} \right]^2, \label{eq:chi2onlynorm}
\end{equation}
which is to be minimized with respect to $f_N$. All the uncertainties are considered  additive (e.g. the possible D'Agostini bias \cite{D'Agostini:1993uj} or equivalent is neglected). The central fit is then defined to correspond to the minimum value of the global $\chi^2$ obtainable with a given set of free parameters,
\begin{equation}
 \chi^2 \left( \vec a^0 \right) \equiv \min \left[ \chi^2 \left( \vec a \right) \right]. \label{eq:defmin}
\end{equation}
In practice, we minimize the $ \chi^2$ function using the Leven\-berg-Marquardt method \cite{Levenberg,Marquardt,Press:1992zz}. 

In our previous EPS09 analysis, additional weight factors were included in Eq.~(\ref{eq:chi2global}) to increase the importance of some hand-picked data sets. 
We emphasize that in the present EPPS16 study
we have abandoned this practice due to the subjectiveness it entails. In the EPS09 analysis the use of such data weights was also partially related to technical difficulties in finding a stable minimum of $\chi^2 \left( \vec a \right)$ when using the \texttt{MINUIT} \cite{James:1994vla} library. In the EPS09 analysis an additional penalty term was also introduced to the $\chi^2 \left( \vec a \right)$ function to avoid unphysical $A$ dependence at small $x$ (i.e. to have larger nuclear effects for larger nuclei). Here, such a term is not 
required because of the improved functional form discussed in Section~\ref{TheNuclearPDFs}. 

As the nuclear PDFs are here allowed to go negative it is also possible to drift to a situation in which the longitudinal structure function $F_{\rm L}^A$ becomes negative. To avoid this, we include penalty terms in $\chi^2 \left( \vec a \right)$ at small $x$ that grow quickly if $F_{\rm L}^A < 0$. We observe, however, that the final results in EPPS16 are not sensitive to such a positiveness requirement.

\subsection{Uncertainty analysis}
\label{Uncertaintyanalysis}

As in our earlier analysis EPS09, we use the Hessian-matrix based approach to estimate the PDF uncertainties \cite{Pumplin:2001ct}. The dominant behaviour of the global $\chi^2$ about the fitted minimum can be written as
\begin{equation}
 \chi^2(\vec a) \approx \chi^2_0 + \sum_{ij} \delta a_i H_{ij} \delta a_j,   
\label{eq:chi2orig}
\end{equation}
where $\delta a_j \equiv a_j-a_j^0$ are differences from the best-fit values and $\chi^2_0 \equiv \chi^2(\vec a^0)$ is the lowest attainable $\chi^2$ of Eq.~\eqref{eq:defmin}. The Hessian matrix $H_{ij}$ can be diagonalized by defining a new set of parameters by
\begin{equation}
 z_k \equiv \sum_j D_{kj}  \delta a_j, \label{eq:diag} 
\end{equation}
with
\begin{equation}
 D_{kj}  \equiv  \sqrt{\epsilon_k} v_j^{(k)}, \label{eq:directions} 
\end{equation}
where $\epsilon_k$ are the eigenvalues and $v_j^{(k)}$ are the components of the corresponding orthonormal eigenvectors of the Hessian matrix,
\begin{align}
H_{ij} v_j^{(k)} & =  \epsilon_k v_i^{(k)} \, , \\
\sum_i v_i^{(k)} v_i^{(\ell)} & =  \sum_i v_k^{(i)} v_\ell^{(i)} = \delta_{k\ell}.
\end{align}
In these new coordinates,
\begin{equation}
 \chi^2(\vec z) \approx \chi^2_0  + \sum_i  z_i^2 \, . \label{eq:chi2diagonalized}
\end{equation}

In comparison to Eq.~(\ref{eq:chi2orig}), here in Eq.~(\ref{eq:chi2diagonalized}) all the correlations among the original parameters $a_i$ are hidden in the definition Eq.~(\ref{eq:diag}), which facilitates a very simple error propagation \cite{Pumplin:2001ct}. Indeed, since the directions $z_i$ are uncorrelated, the 
upward/downward-symm\-etric 
uncertainty for any PDF-dependent quantity $\mathcal{O}$ can be written as
\begin{equation}
\Delta \mathcal{O} = \sqrt{\sum_i \left(\Delta z_i\right)^2 \left( \frac{\partial \mathcal{O}}{\partial z_i} \, \right)^2 }\, , \label{eq:err}
\end{equation}
with an uncertainty interval $\Delta z_i = (t_i^+ + t_i^-)/2$ where 
$t_i^\pm$ are $z_i$-interval limits  which depend on the chosen tolerance criterion. The partial derivatives in Eq.~\eqref{eq:err} are evaluated with the aid of PDF error sets $S^\pm_i$ defined in the space of $z_i$ coordinates in terms of $t_i^\pm$ as
\begin{align}
\vec z({S^\pm_1}) & = \pm {t_1^\pm} \left(1,0,...,0 \right), \nonumber\\
&\vdots \label{eq:errset}\\
\vec z({S^\pm_N}) & = \pm {t_N^\pm} \left(0,0,...,1 \right),\nonumber 
\end{align}
where $N$ is the number of the original parameters $a_i$. It then follows that
\begin{equation}
\Delta \mathcal{O} = \frac{1}{2} \sqrt{\sum_i \left[\mathcal{O}\left(S^+_i \right) - \mathcal{O}\left(S^-_i \right) \right]^2}\, . \label{eq:err2}
\end{equation}

Although simple on paper, in practice it is a non-trivial task to obtain a sufficiently accurate Hessian matrix in a multivariate fit such that Eq.~(\ref{eq:chi2diagonalized}) would be accurate. One possibility, used e.g. in Ref.~\cite{Martin:2009iq}, is to use the linearized Hessian matrix obtained from Eq.~\eqref{eq:chi2individual}
\begin{equation}
 H_{ij}^{\rm linearized} \equiv \sum_{k,\ell} \frac{\partial T_k}{\partial a_i} C^{-1}_{k\ell} \frac{\partial T_\ell}{\partial a_j},
\end{equation}
where the partial derivatives are evaluated by finite differences. The advantage is that by this definition, the Hessian matrix is always positive definite and thereby has automatically positive eigenvalues and e.g. Eq.~(\ref{eq:diag}) is always well-defined. 

Another possibility, which is the option chosen in the present study, is to scan the neighborhood of the minimum $\chi^2$ and fit it with an ansatz 
\begin{equation}
 \chi^2(\vec a) = \chi^2_0 + \sum_{i,j} \delta a_i h_{ij} \delta a_j,
\end{equation}
whose parameters $h_{ij}$ then correspond to the components of the Hessian matrix. While this gives more accurate results than the linearized method (where some information is thrown away), the eigenvalues of the Hessian become easily negative for the presence of third- and higher-order components in the true $\chi^2$ profile. Hence, to arrive at positive-definite eigenvalues,
some manual labour is typically required e.g. in tuning the parameter intervals used when scanning the global $\chi^2$. Yet, the resulting uncertainties always depend somewhat on the chosen parameter intervals, especially when the uncertainties are large. To improve the precision, we have adopted an iterative procedure similar to the one in Ref.~\cite{Pumplin:2000vx}: After having obtained the first estimate for the Hessian matrix and the $z$ coordinates, we recompute the Hessian matrix in the $z$ space by re-scanning the vicinity of $\vec z = 0$ and fitting it with a polynomial
\begin{equation}
 \chi^2(\vec z) = \chi^2_0 + \sum_{i,j} z_i \hat h_{ij} z_j,
\label{eq:ziteration}
\end{equation}
where $\hat h_{ij}$ is an estimate for the Hessian matrix in the $z$ space. We then re-define the $z$ coordinates by
\begin{equation}
    z_k \rightarrow \sum_{\ell} \hat D_{k\ell} \delta a_\ell , \label{eq:diag2} 
\end{equation}
where
\begin{equation}
  \hat D_{k\ell}  \equiv  \sum_{j} \sqrt{\hat \epsilon_k} \hat v_j^{(k)} D_{j\ell},
\label{eq:D} 
\end{equation}
and $\hat \epsilon_k$ and $\hat {\vec v}^{(k)}$ are now the eigenvalues and eigenvectors of the matrix $\hat h_{ij}$. Then we repeat the iteration a few times,
using $\hat D_{ij}$ of the previous round as $D_{ij}$ in Eq.~\eqref{eq:D}. Ideally, one should find that the eigenvalues $\hat \epsilon_k$ converge to unity during the iteration but in practice, some deviations will always persist for the presence of non-quadratic components in the true $\chi^2$ profile. We have also noticed that, despite the iteration, the resulting uncertainty bands still depend somewhat on the finite step sizes and grids used in the $\chi^2$-profile scanning especially in the regions where the uncertainties are large. In such regions the Hessian method starts to be unreliable and the found uncertainties represent only the lower limits for the true uncertainties.

\begin{figure*}[htb!]
\centering
\includegraphics[width=1.0\linewidth]{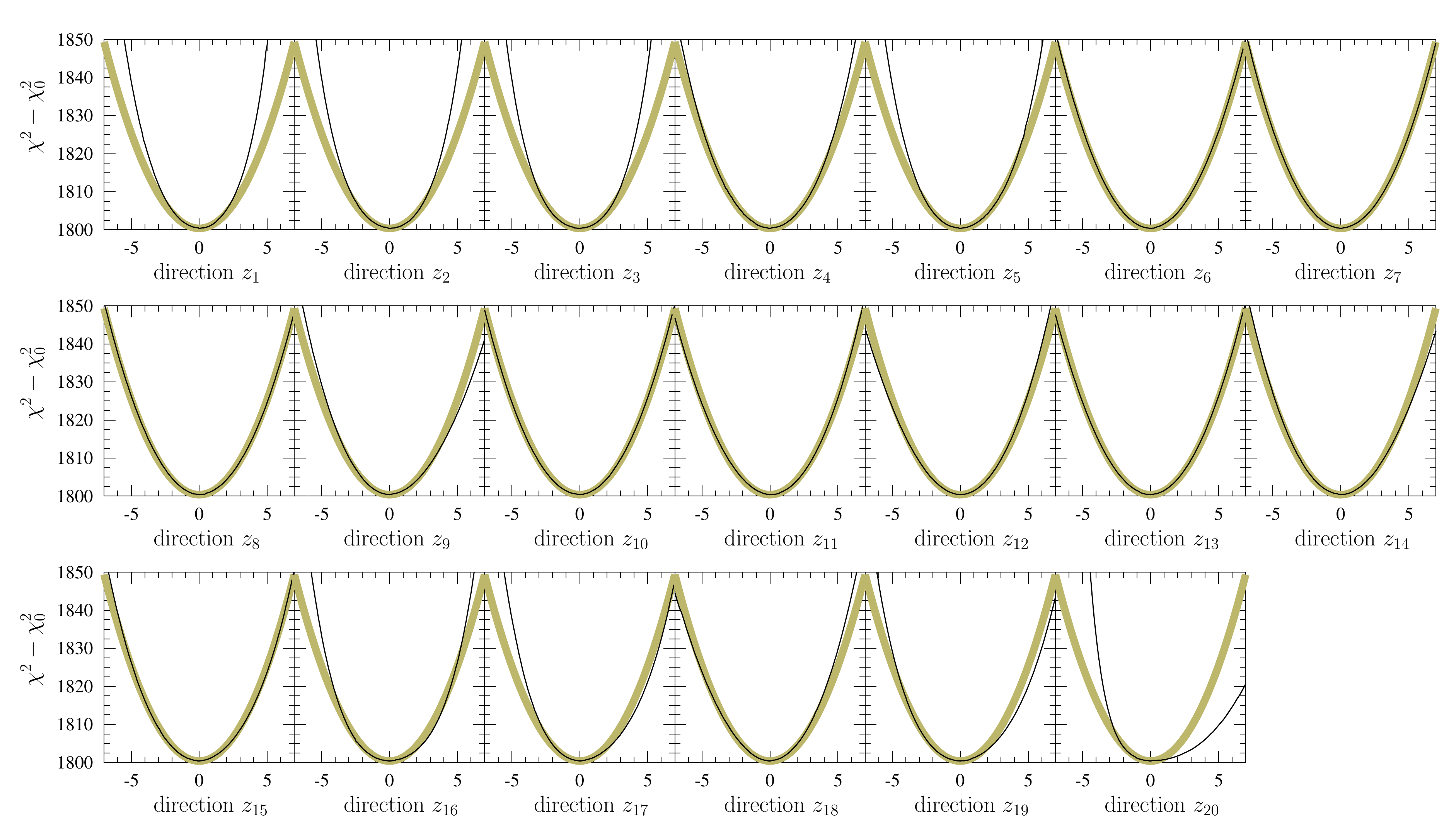}
\caption{The $\chi^2$ profiles (black curves) as a function of final eigenvector directions $z_i$ compared to ideal behaviour $\chi^2_0 + z_i^2$ (thicker colored curves).}
\label{fig:QuadraticTest}
\end{figure*} 

The global $\chi^2$ profiles as a function of the final eigenvector directions, which we arrive at in the present EPPS16 analysis, are shown in Fig.~\ref{fig:QuadraticTest}. In obtaining these, during the iteration, the finite step sizes ($z_i$ in Eq.~\eqref{eq:ziteration}) along each provisional eigenvector direction were adjusted such that the total $\chi^2$ increased by 5 units from the minimum. 
As seen in the figure, in most cases, the quadratic approximation gives a very good description of the true behaviour of $\chi^2$, but in some cases higher-order (e.g. cubic and quartic) components are evidently present. The effects of higher-order components can be partly compensated by using larger step sizes during the iteration such that the quadratic polynomial  approximates the true $\chi^2$ better up to larger deviations from the minimum (but is less accurate near the minimum). However, we have noticed that with increasing step sizes 
the resulting PDF uncertainties get eventually smaller, which indicates that some corners of the parameter space are not covered as completely as with the now considered 5-unit increase in $\chi^2$.

\begin{figure}[htb!]
\centering
\includegraphics[width=\linewidth]{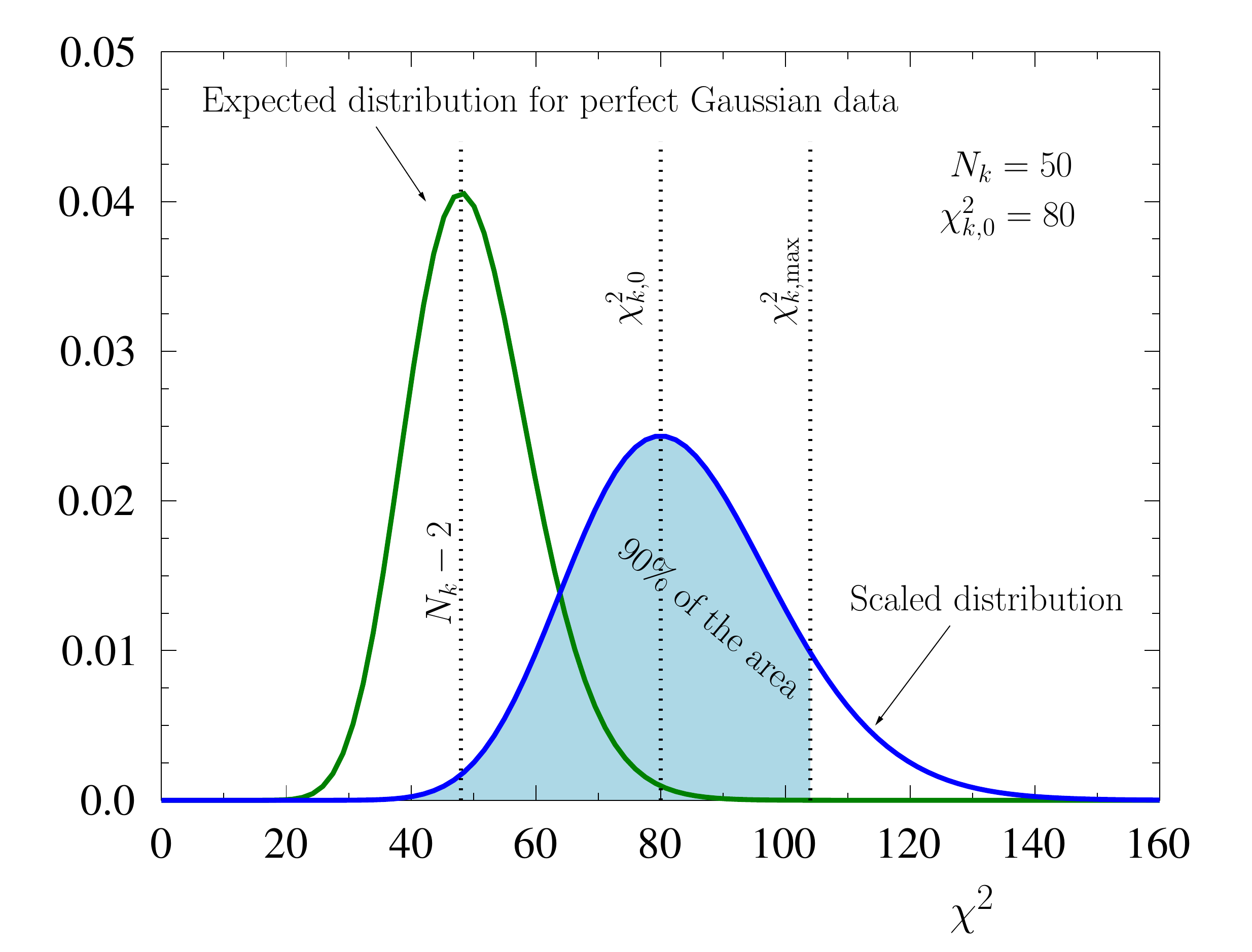}
\caption{Determination of 90\% confidence limit for an individual data set with $N_k=50$ data points and for which the global minimum corresponds to $\chi^2_{k,0}=80$.}
\label{fig:conflimdemo}
\end{figure}

The basic idea in the determination of the PDF uncertainty sets in the present work is similar to that in the EPS09 analysis. As in EPS09, for each data set $k$ with $N_k$ data points we determine a 90\% confidence limit $\chi^2_{k,{\rm max}}$ by solving
\begin{equation}
\int_0^{M_k} 
\frac{d\chi^2}{2\Gamma(N_{k}/2)} \left( \frac{\chi^2}{2} \right)^{N_{k}/2-1} \exp \left(-\chi^2/2 \right) = 0.90, \label{eq:chi2dist}
\end{equation}
where 
\begin{equation}
M_k = \chi^2_{k,{\rm max}} \times \frac{N_{k}-2}{\chi^2_{k,0}}, \label{eq:scalethevariable}
\end{equation}
and in which $\chi^2_{k,0}$ is the value of $\chi^2$ for $k$th data set at the global minimum. The integrand in Eq.~(\ref{eq:chi2dist}) is the usual $\chi^2$ distribution --- the probability density to obtain a given value of $\chi^2$ when the data are Gaussianly distributed around the known truth. The effect of Eq.~(\ref{eq:scalethevariable}) is, as sketched in Fig.~\ref{fig:conflimdemo}, to scale the $\chi^2$ distribution such that its maximum occurs at the central value of the fit $\chi^2_{0,k}$, against which the confidence limit is defined. In other words, we assume that if the experiment would be repeated several times the outcome would follow the scaled distribution (the blue curve in Fig.~\ref{fig:conflimdemo}) and not the ideal one (the green curve in Fig.~\ref{fig:conflimdemo}). This procedure allows to define confidence limits also for data sets which happen to give very large $\chi^2_k/N_k$ for e.g. underestimated uncertainties or particularly large fluctuations \cite{Pumplin:2002vw}. 

\begin{figure*}[htb!]
\centering
\includegraphics[width=0.245\linewidth]{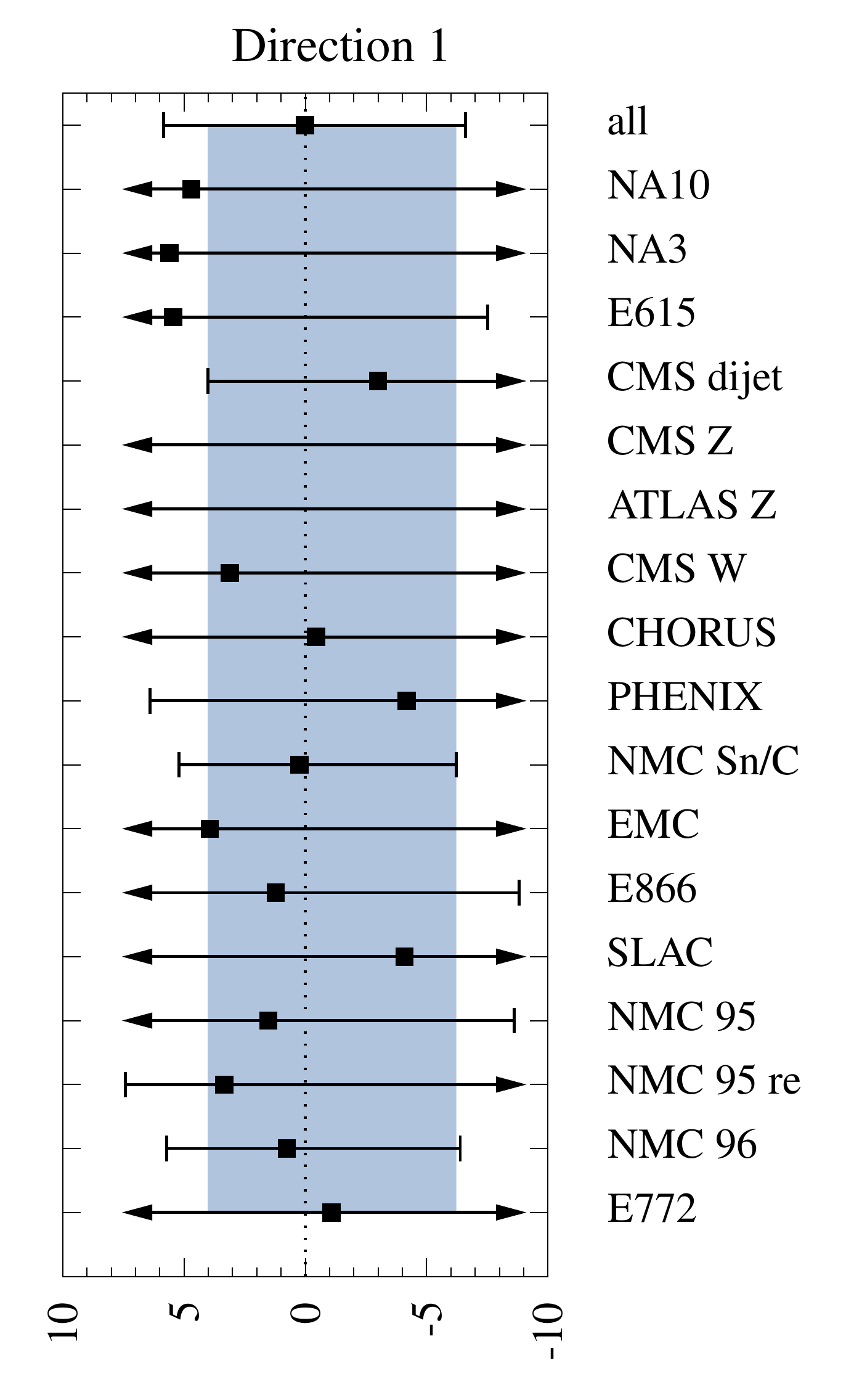}
\includegraphics[width=0.245\linewidth]{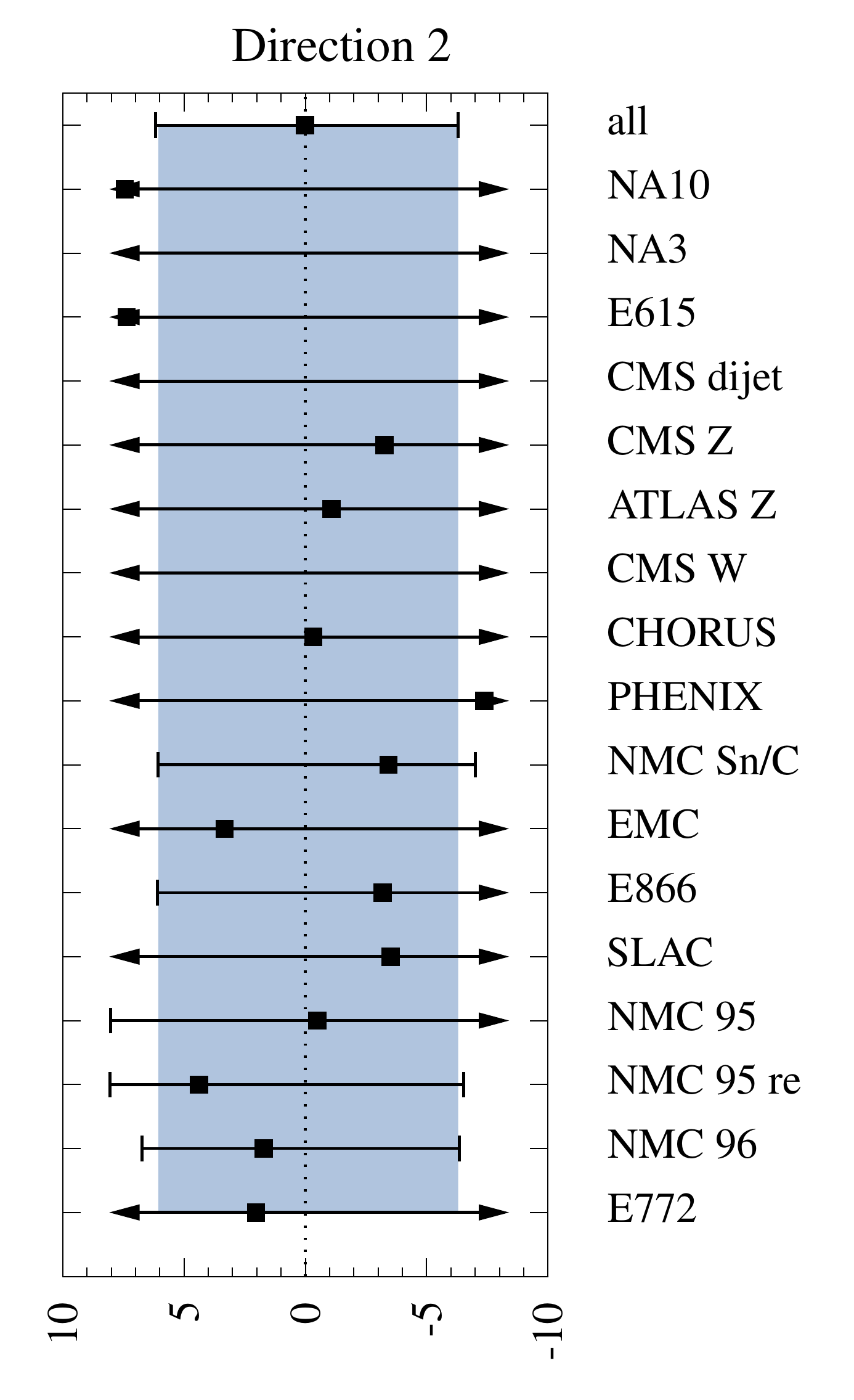}
\includegraphics[width=0.245\linewidth]{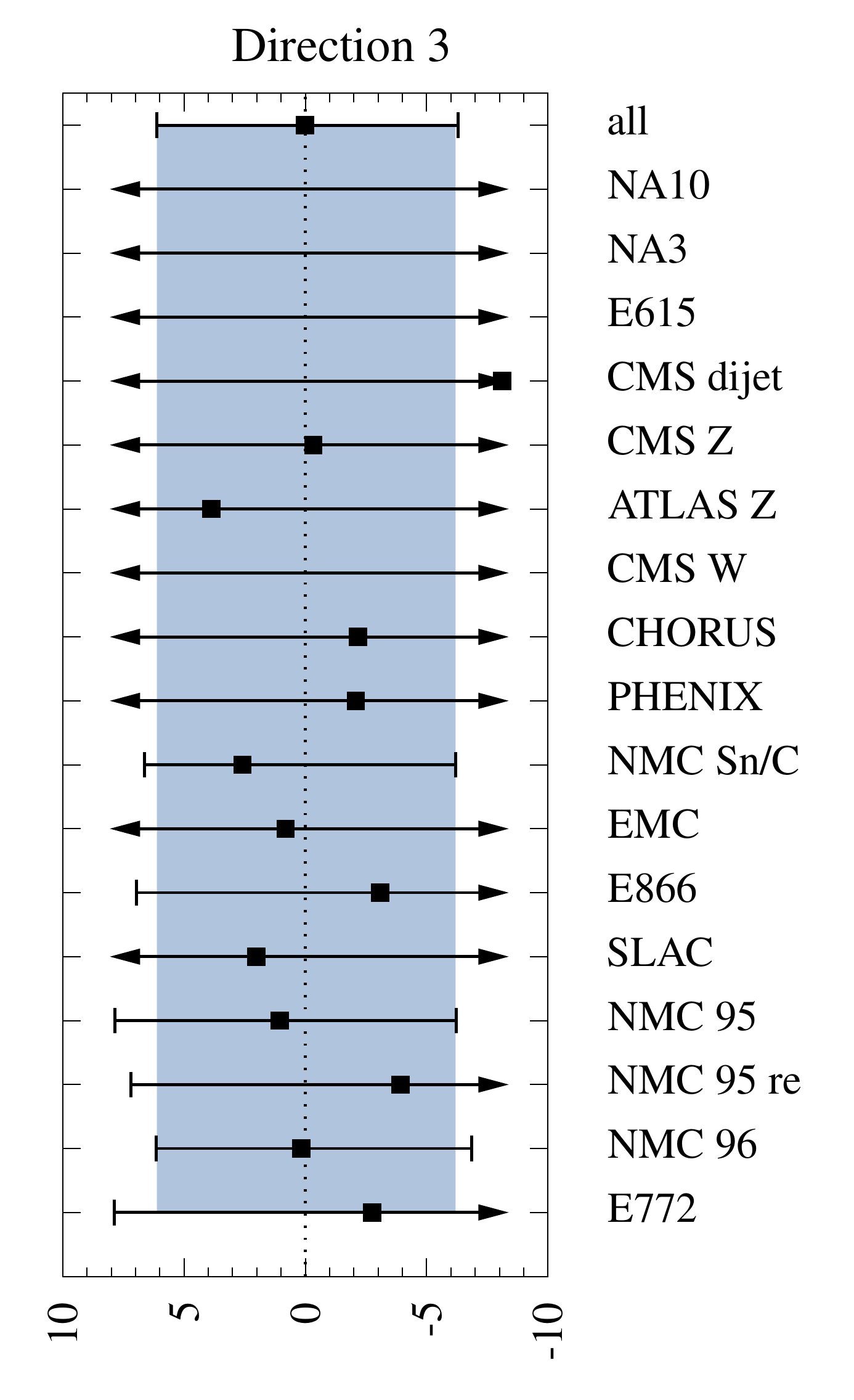}
\includegraphics[width=0.245\linewidth]{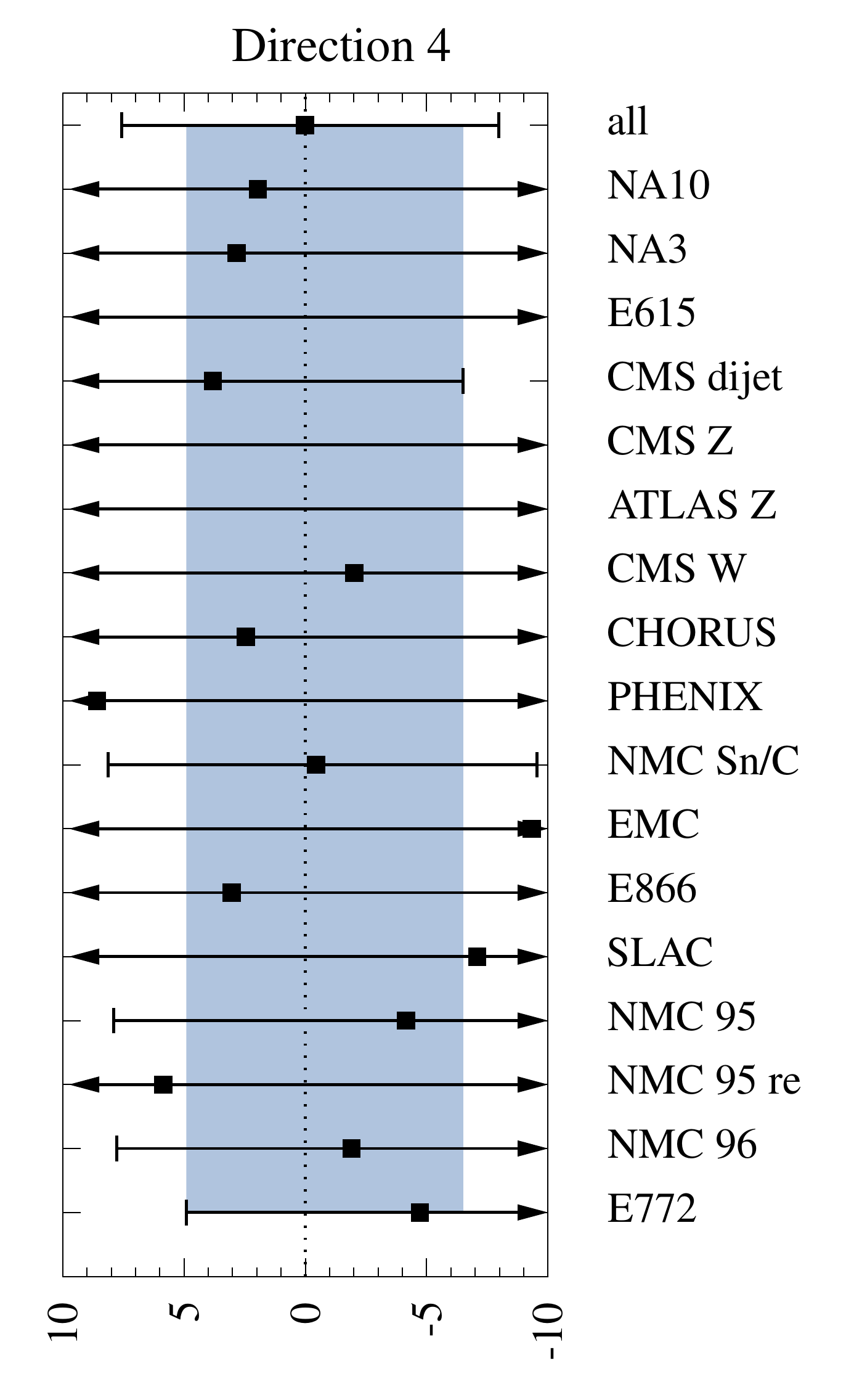}
\includegraphics[width=0.245\linewidth]{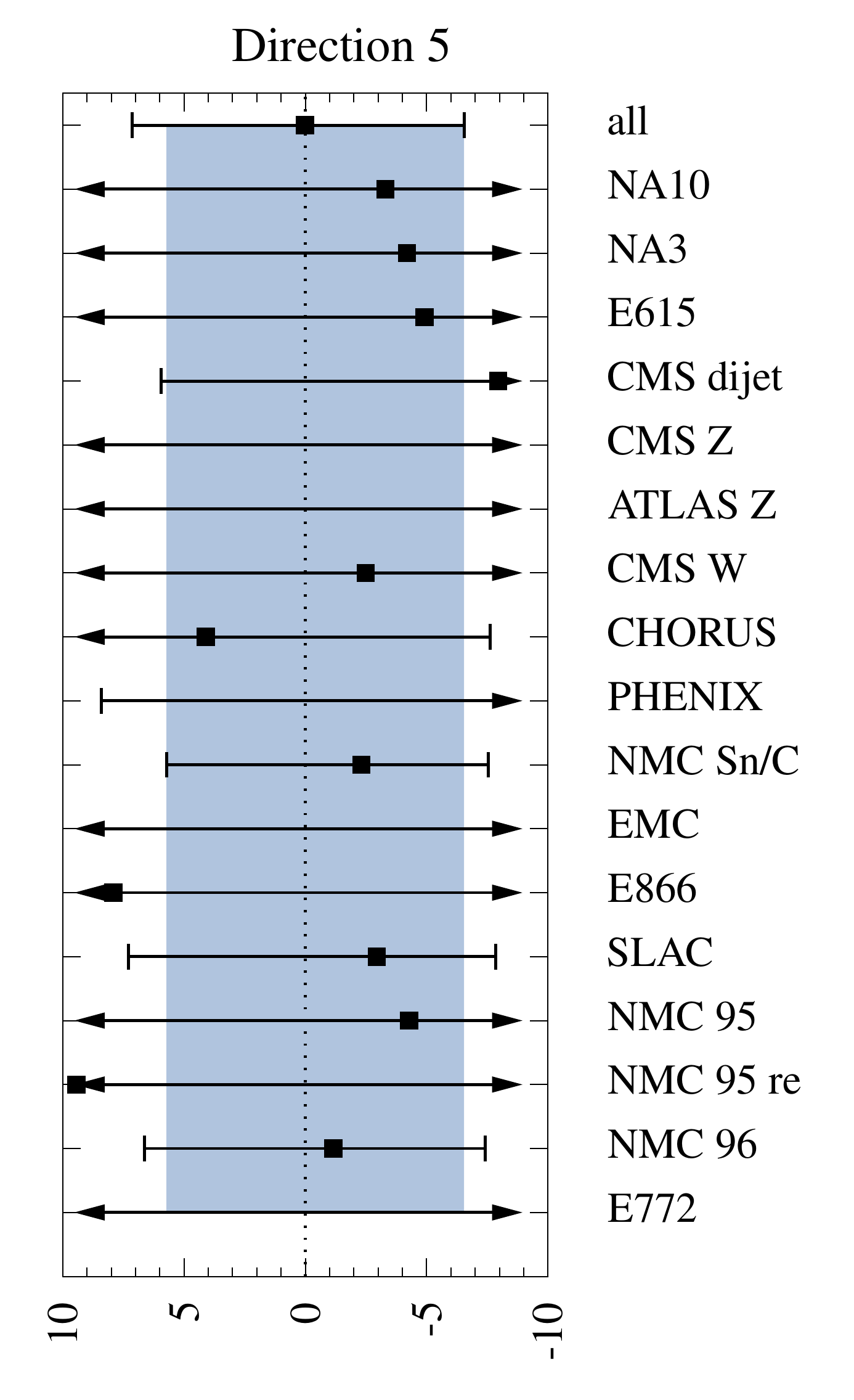}
\includegraphics[width=0.245\linewidth]{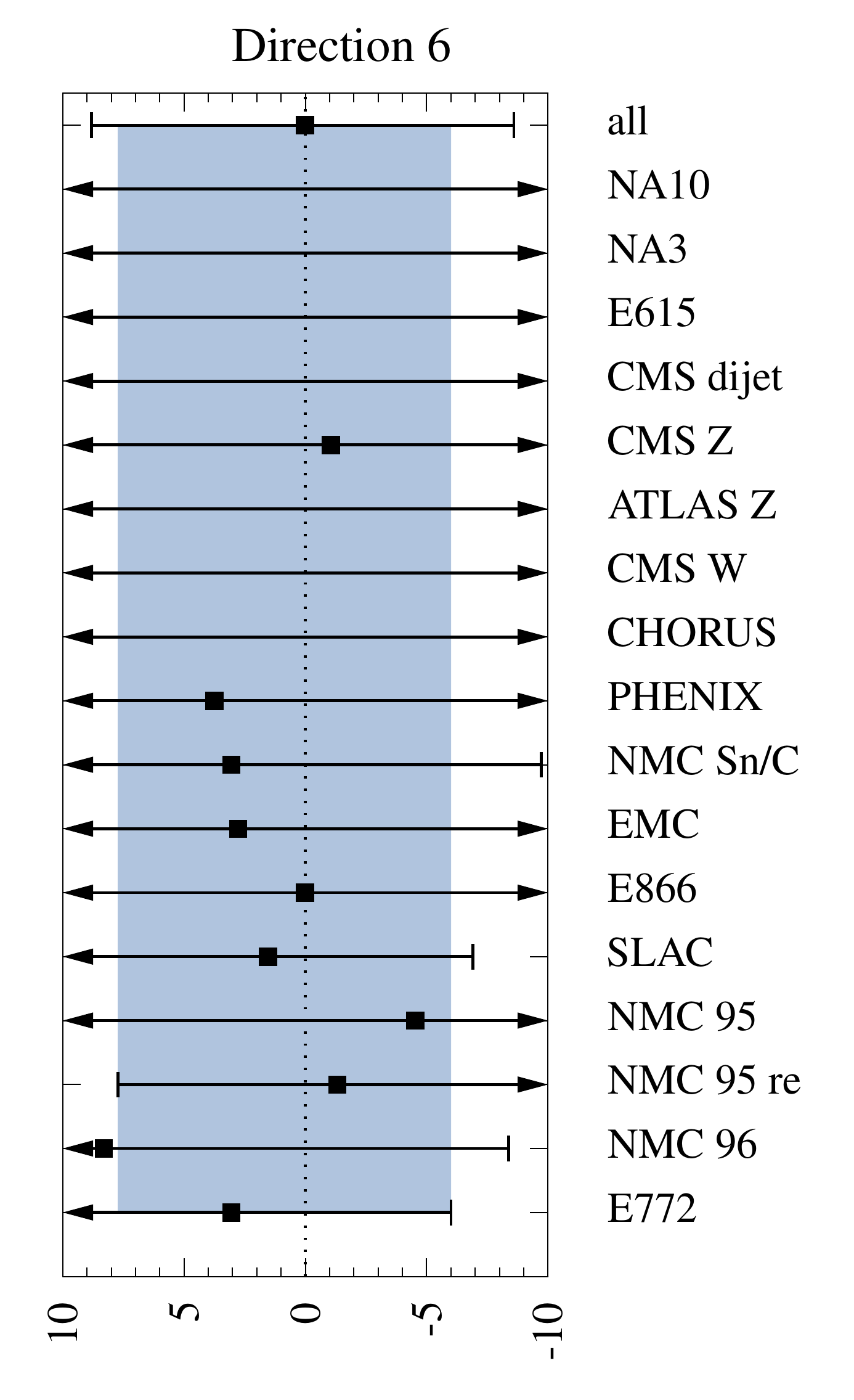}
\includegraphics[width=0.245\linewidth]{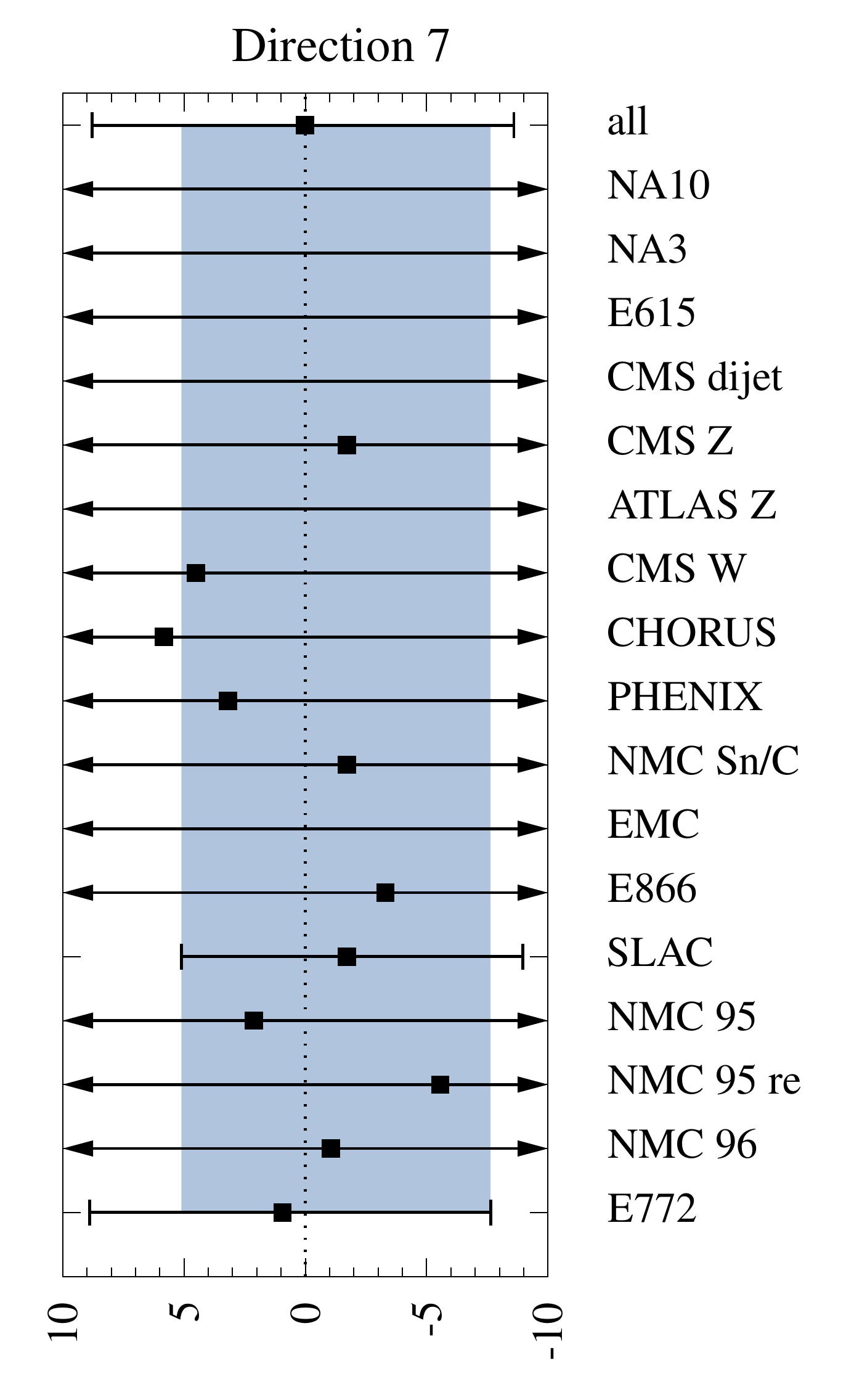}
\includegraphics[width=0.245\linewidth]{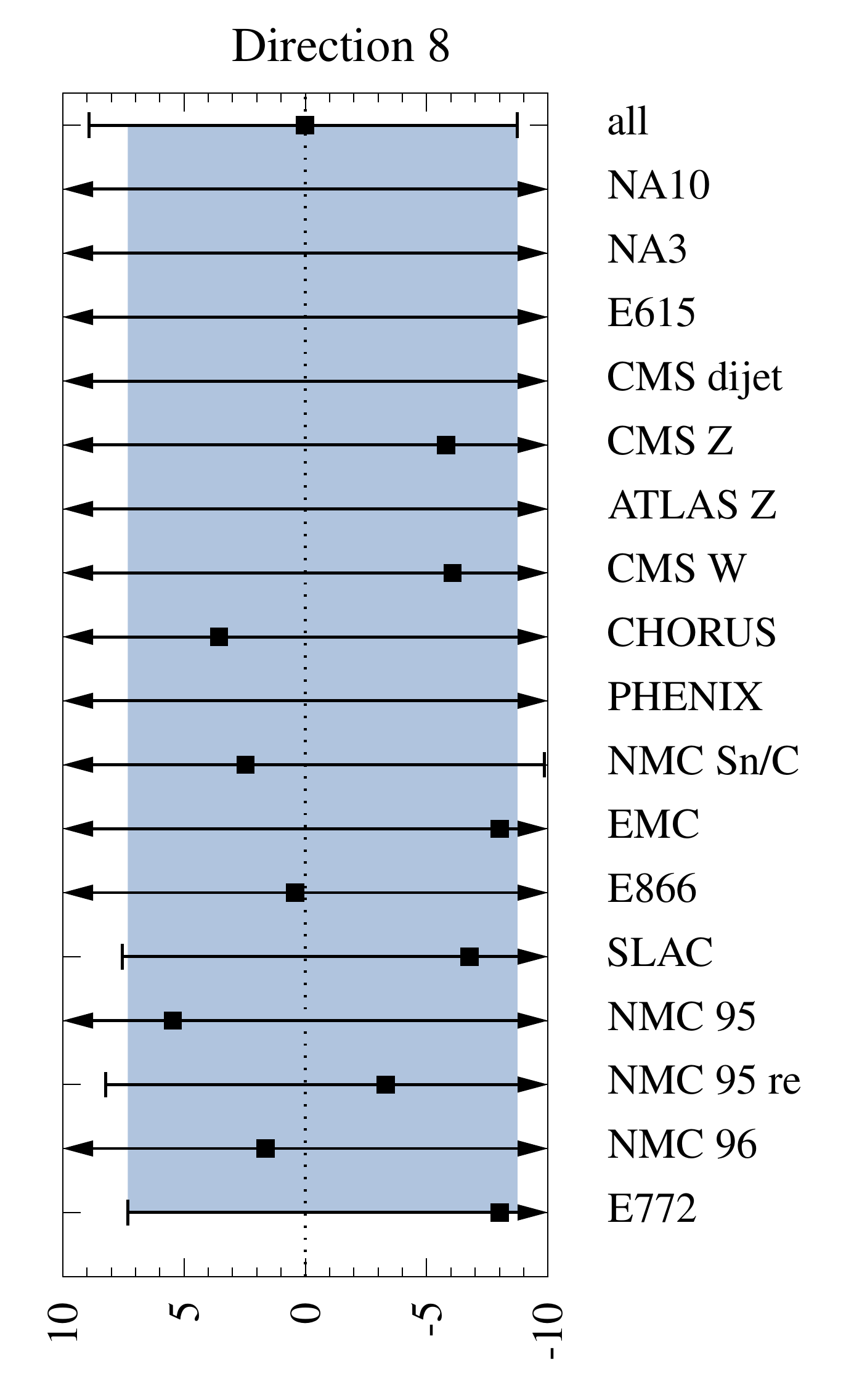}
\includegraphics[width=0.245\linewidth]{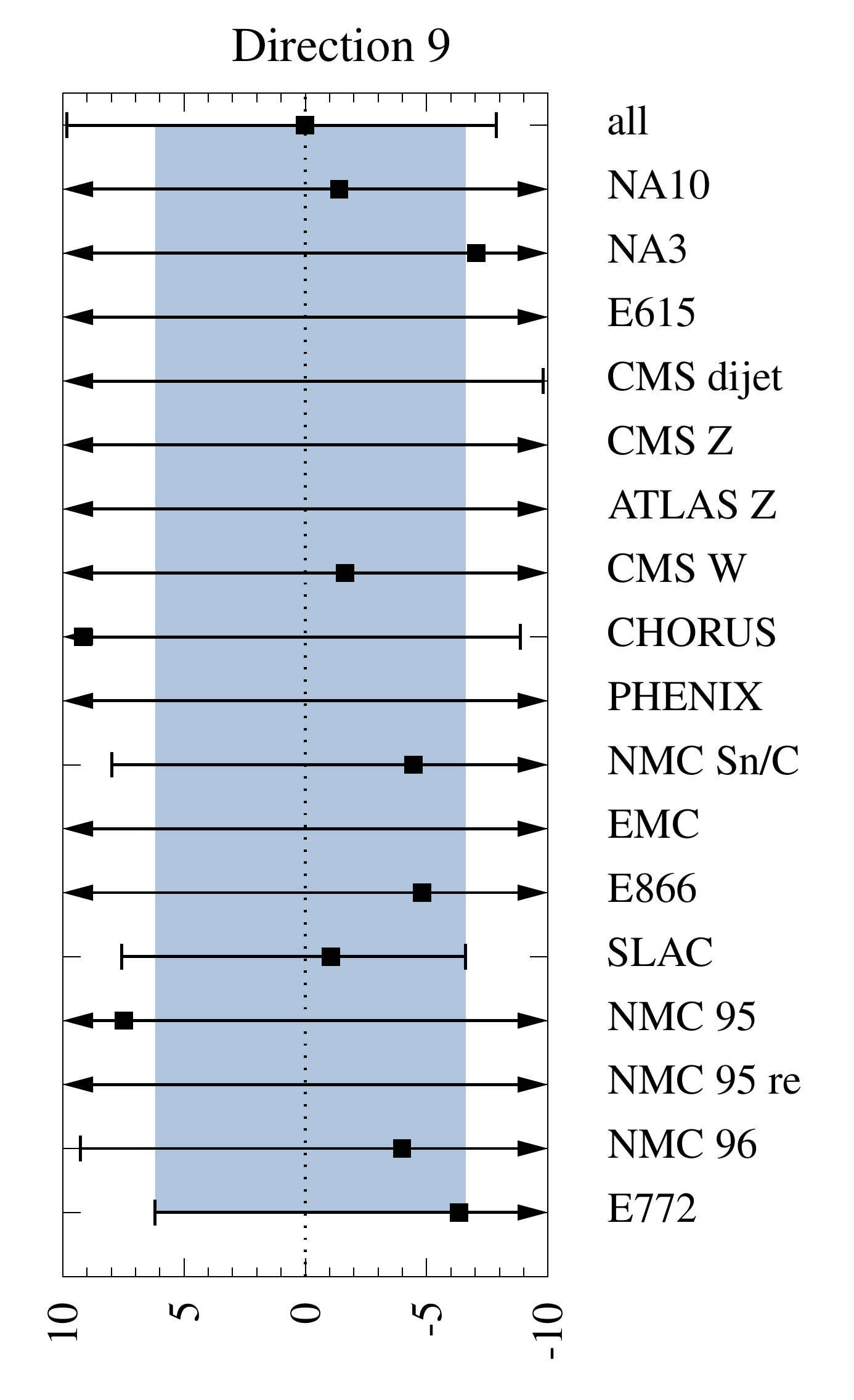}
\includegraphics[width=0.245\linewidth]{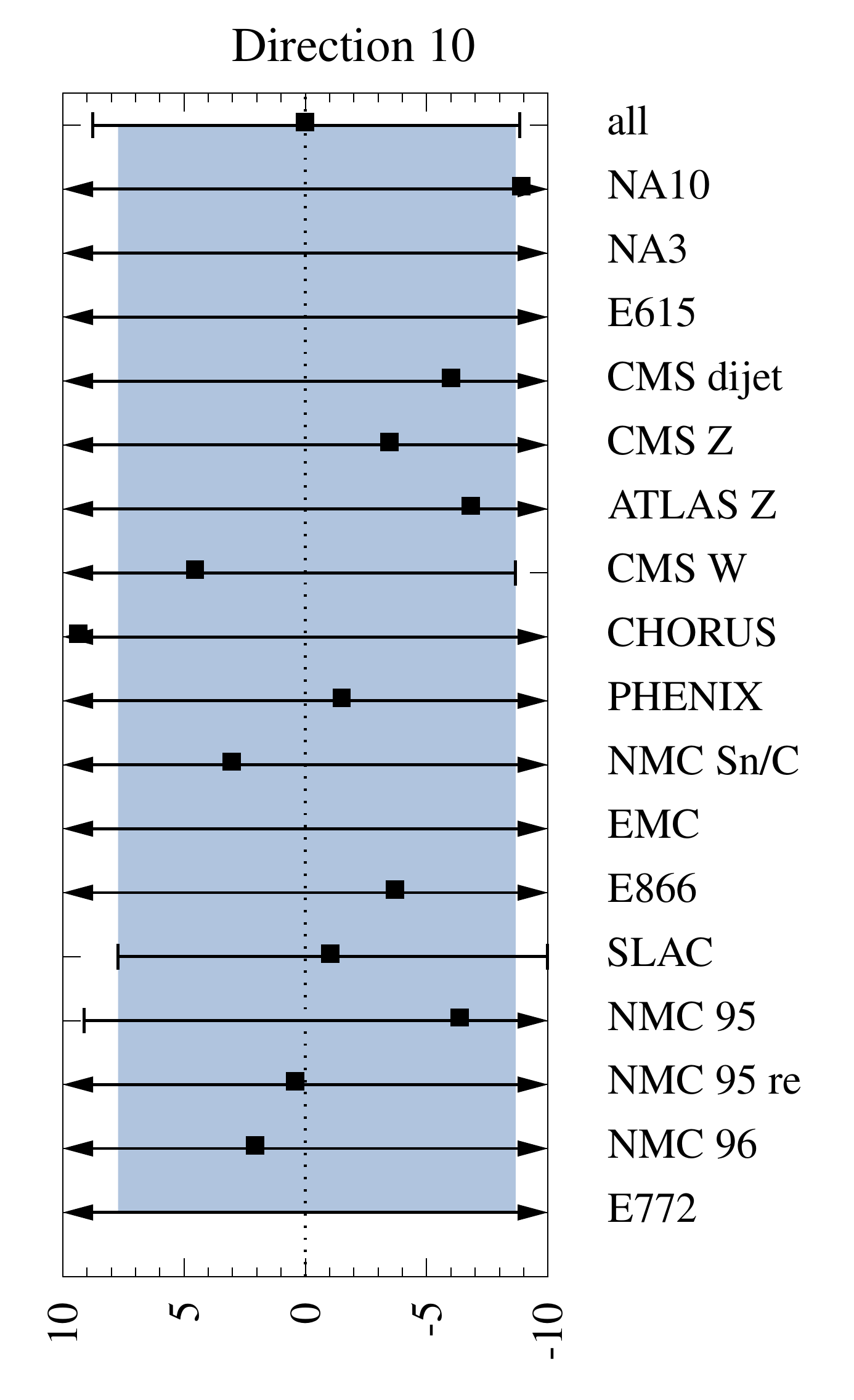}
\includegraphics[width=0.245\linewidth]{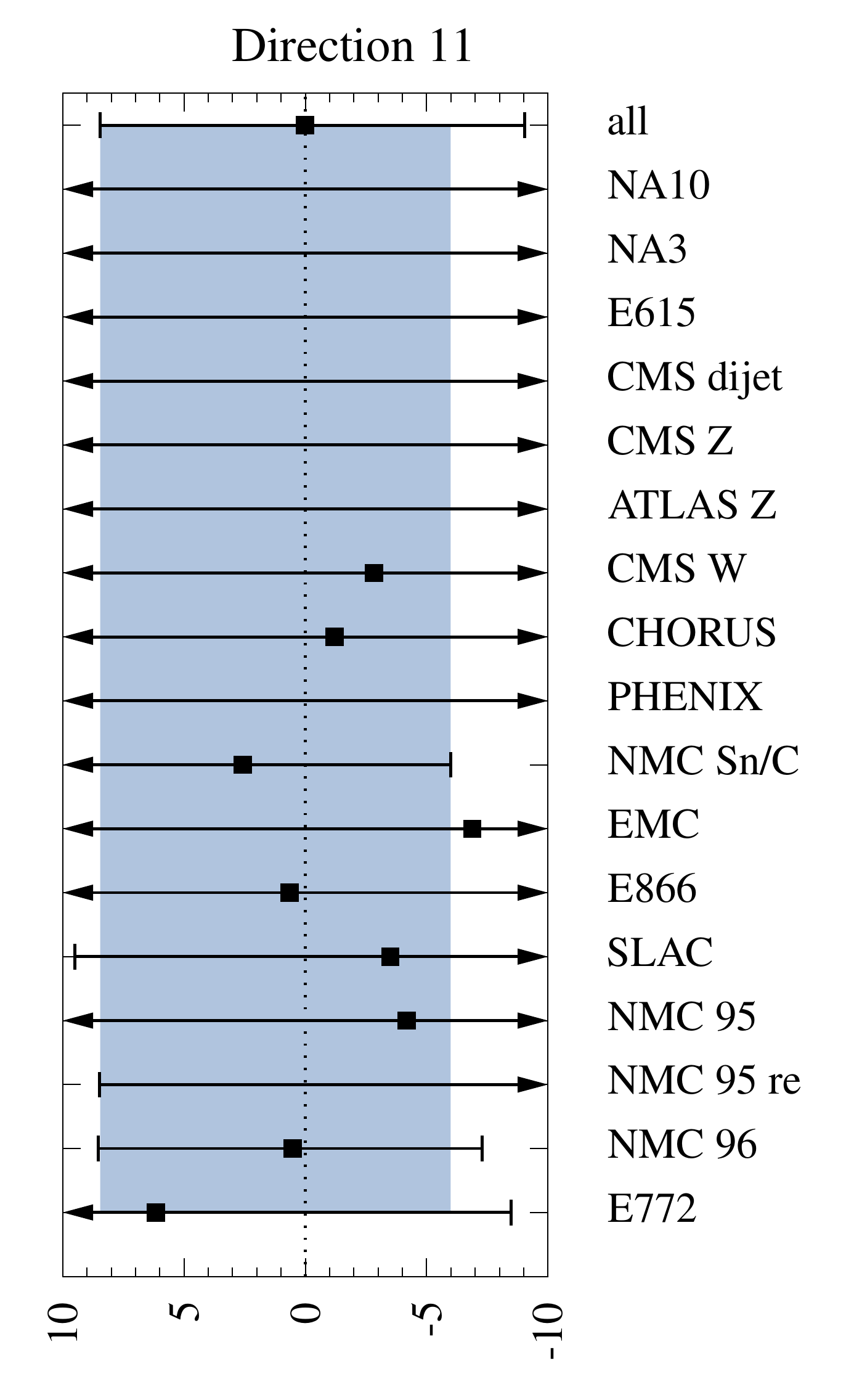}
\includegraphics[width=0.245\linewidth]{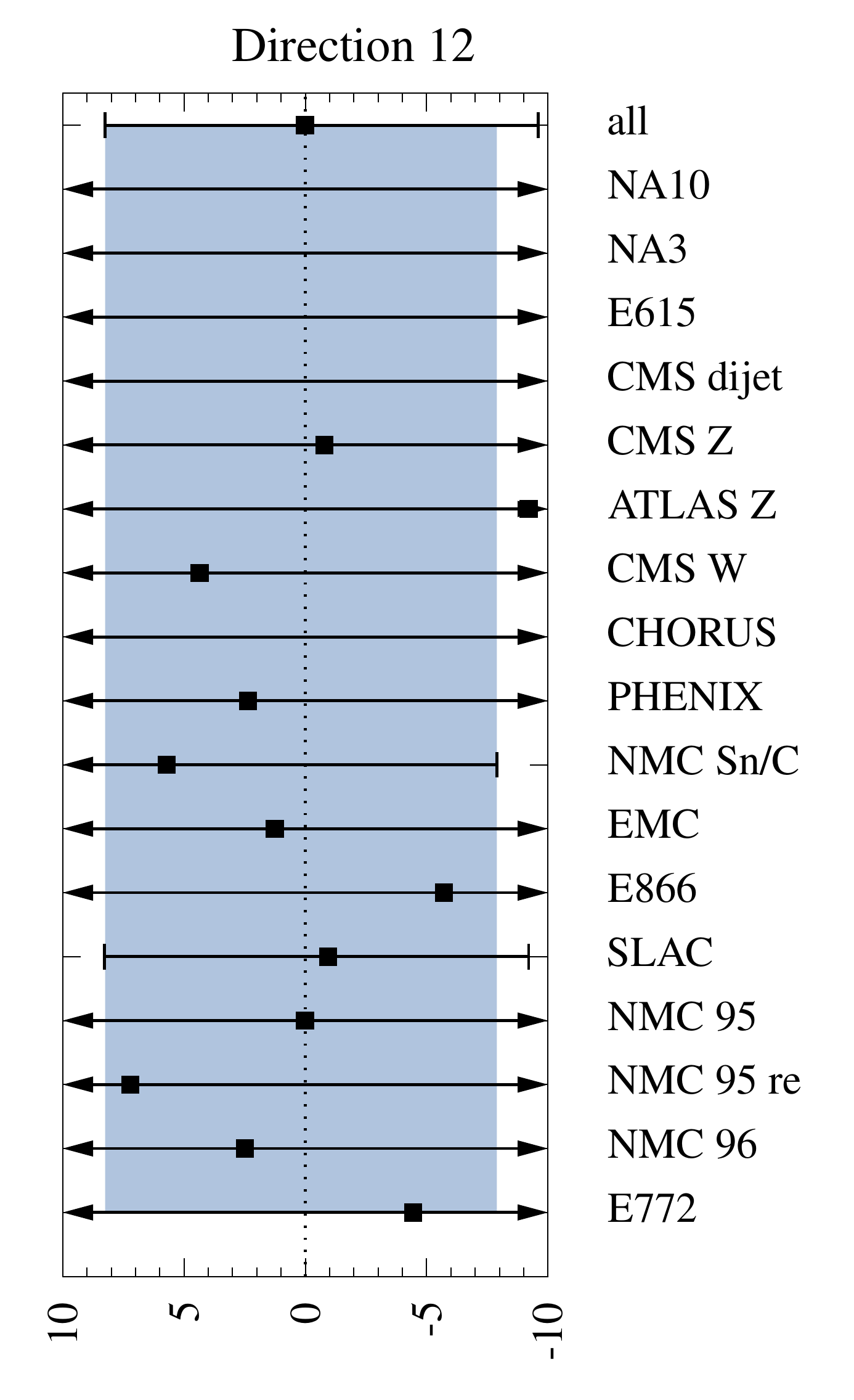}
\caption{Determination of the confidence limits for the eigendirections 1 to 12. The bars show the limits $z_{i,{\rm min}}^k$, $z_{i,{\rm max}}^k$ for each individual (or grouped) data set $k$ and the marker in between indicates where the minimum $\chi^2_{k,0}$  of that data set is reached.  The set "all" refers to all data combined.
An arrow signifies that the confidence limit has not yet been reached in the scanned interval. The gray bands are the intersection intervals $\left[z_{i,{\rm min}},z_{i,{\rm max}}\right]$ explained in the text.
}
\label{fig:Conflimsall}
\end{figure*} 

\begin{figure*}[htb!]
\centering
\includegraphics[width=0.245\linewidth]{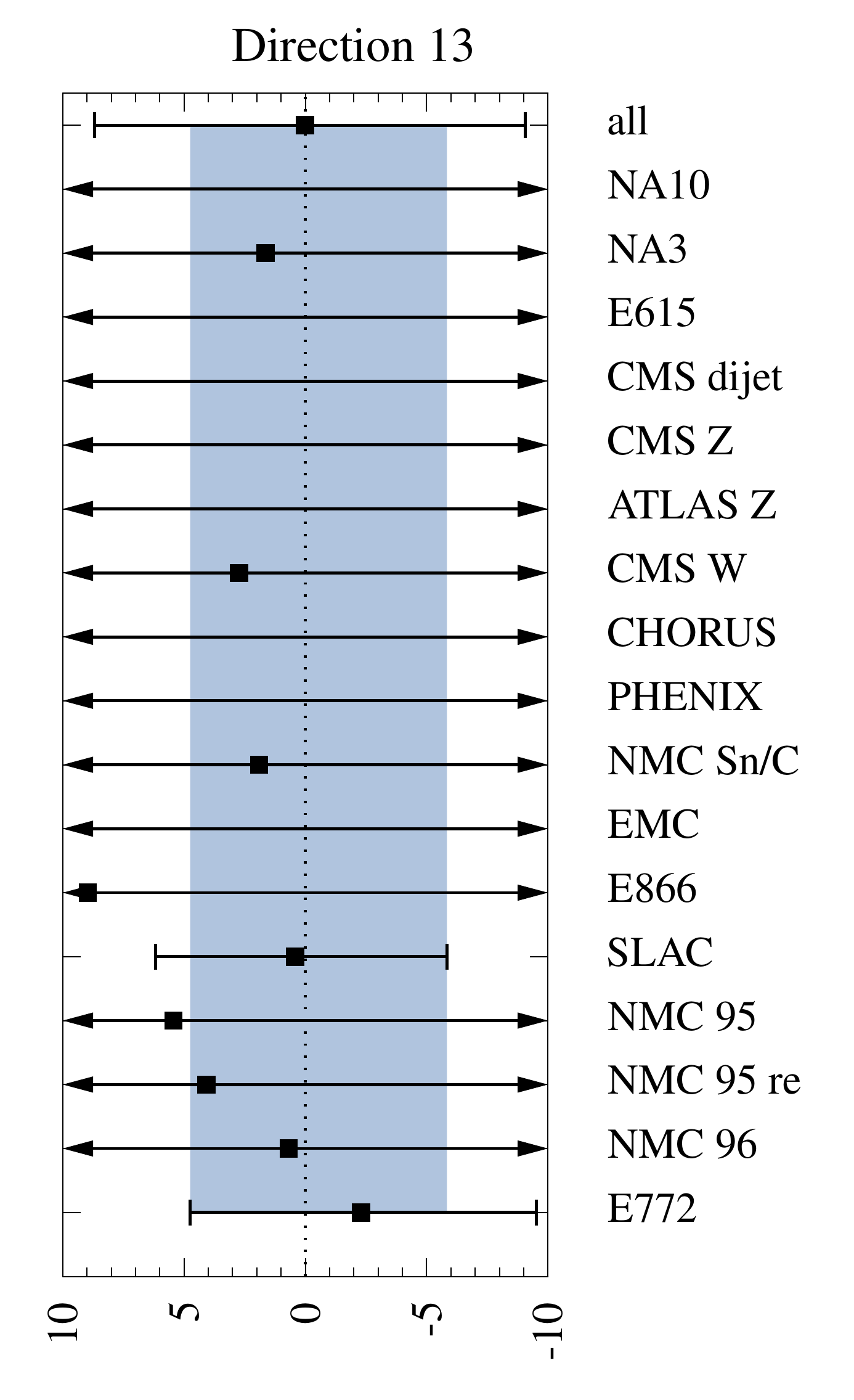}
\includegraphics[width=0.245\linewidth]{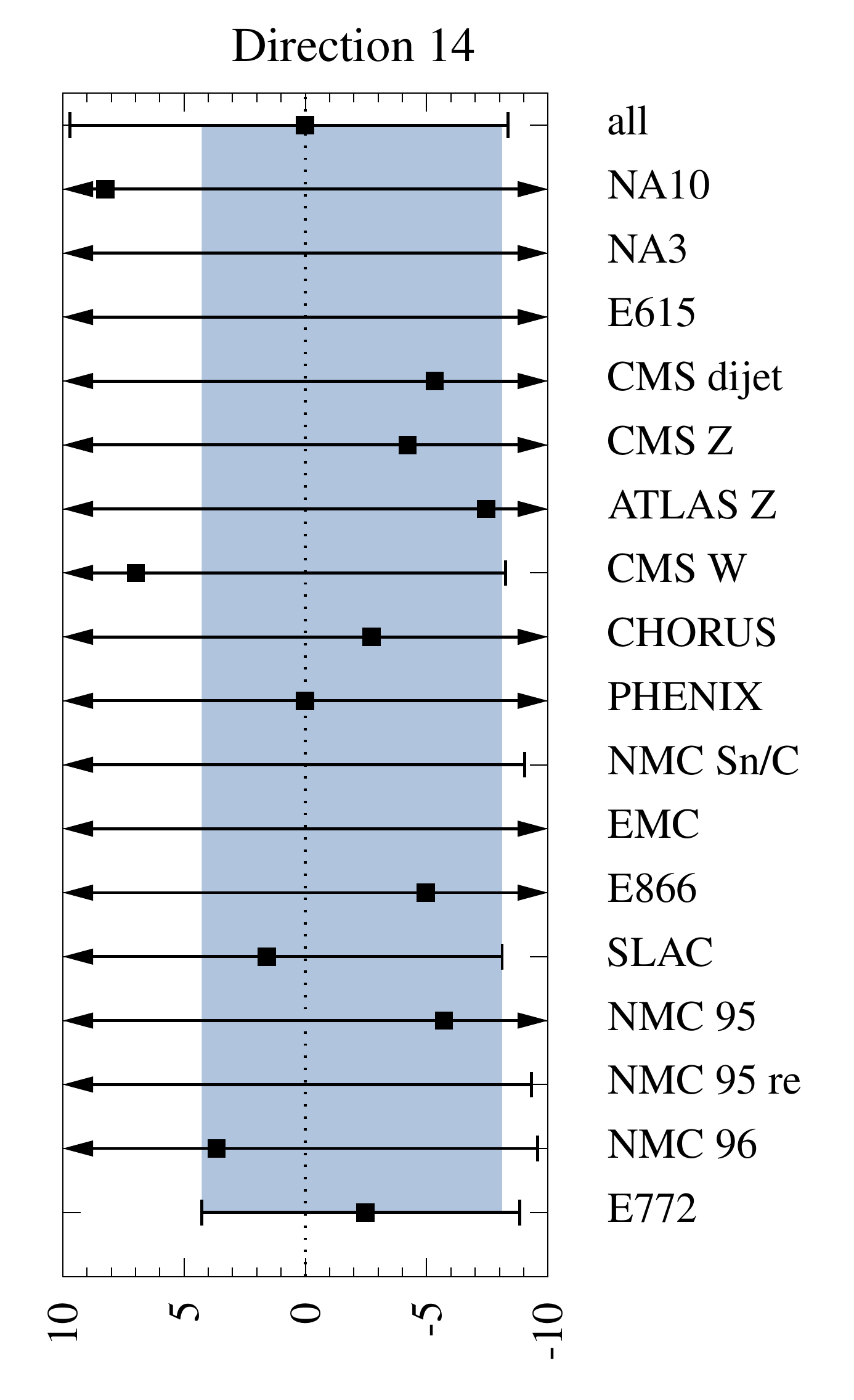}
\includegraphics[width=0.245\linewidth]{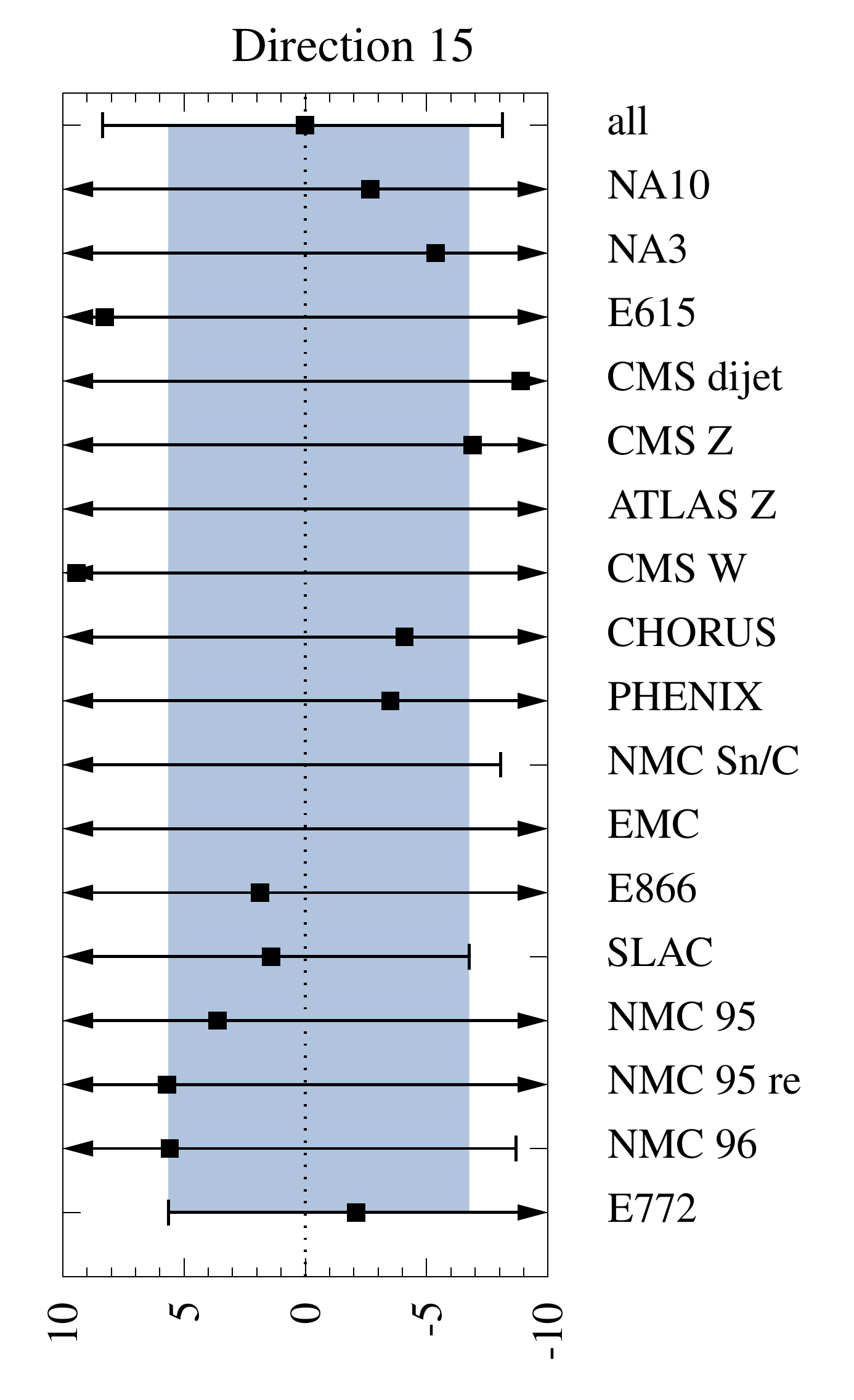}
\includegraphics[width=0.245\linewidth]{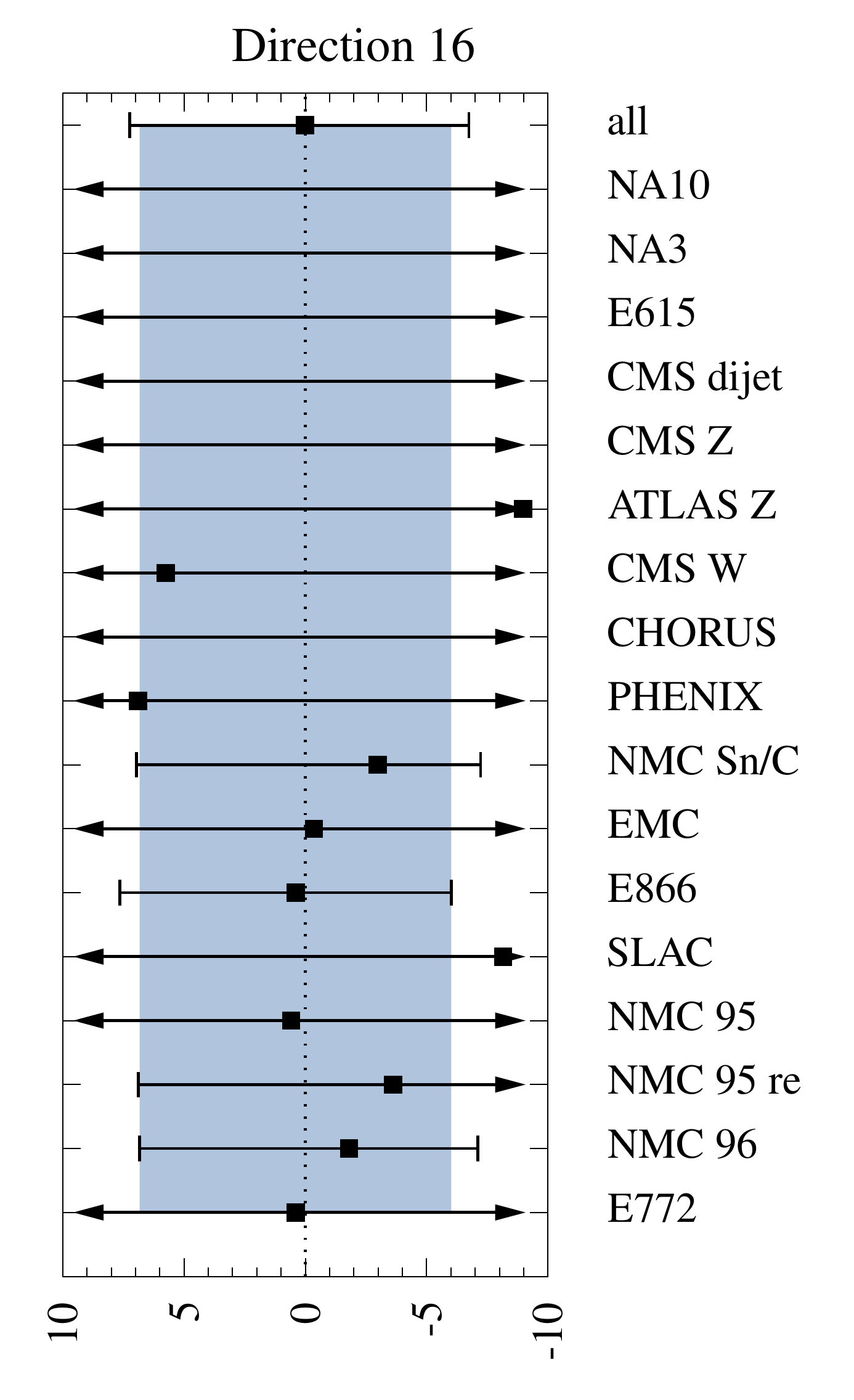}
\includegraphics[width=0.245\linewidth]{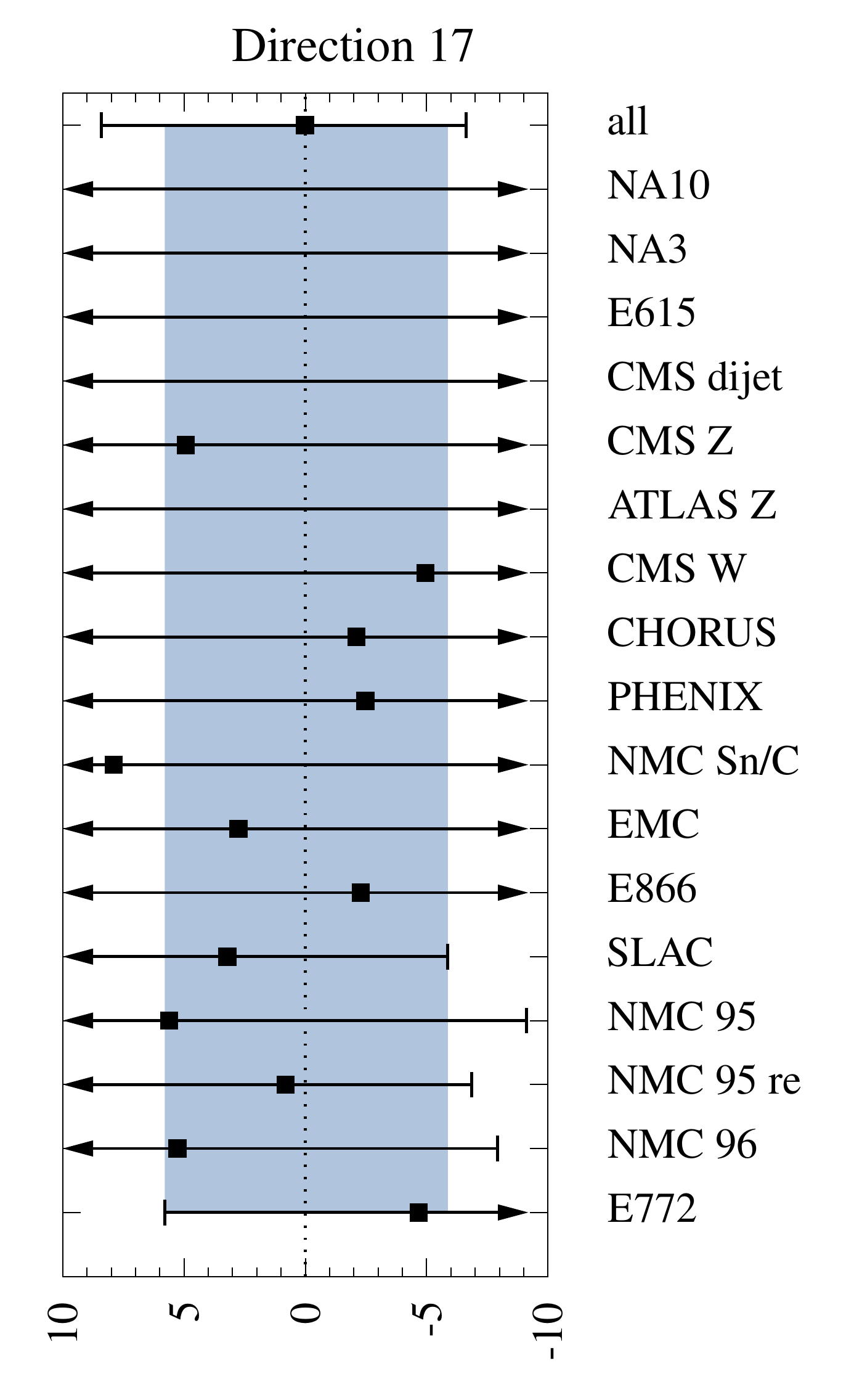}
\includegraphics[width=0.245\linewidth]{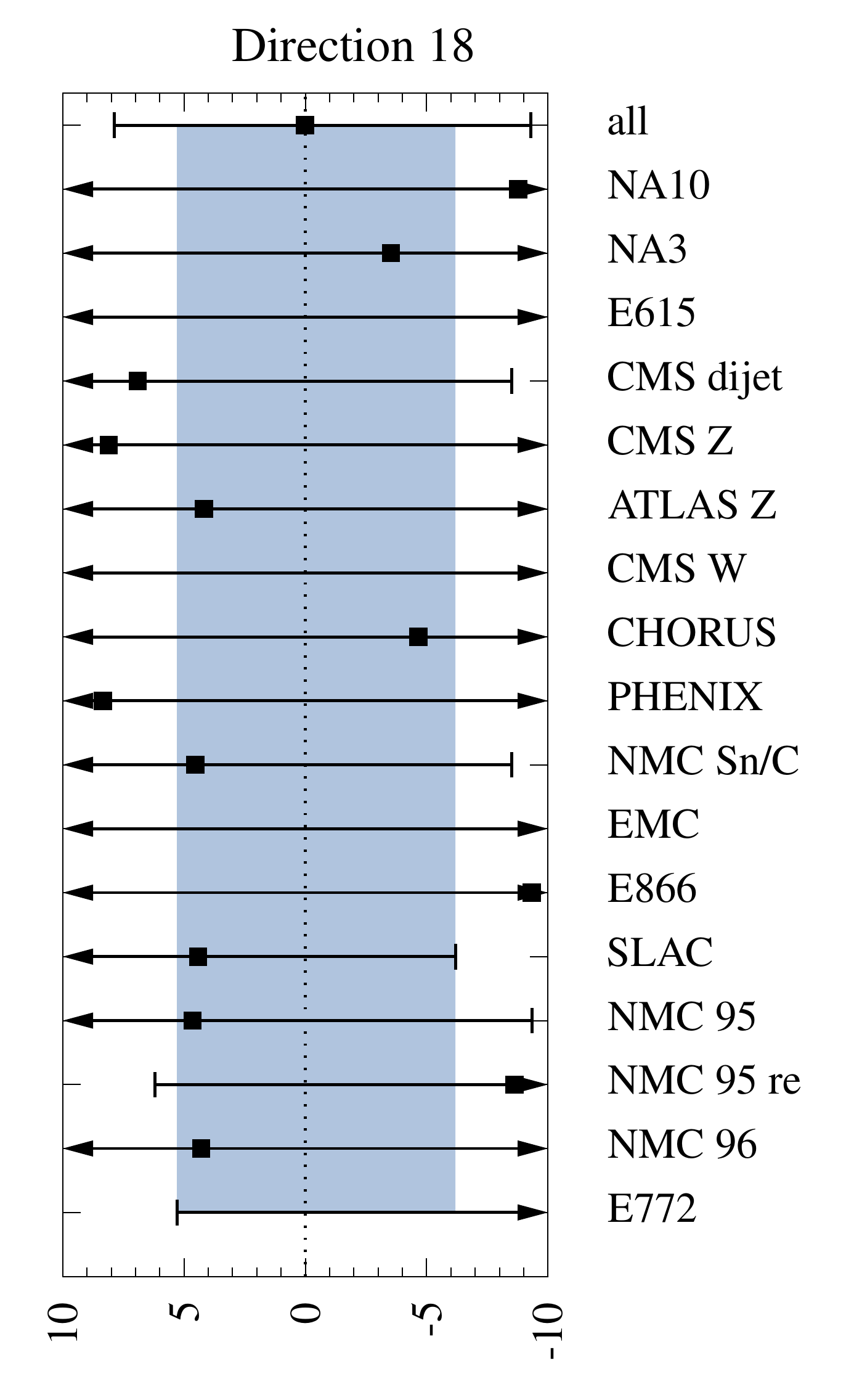}
\includegraphics[width=0.245\linewidth]{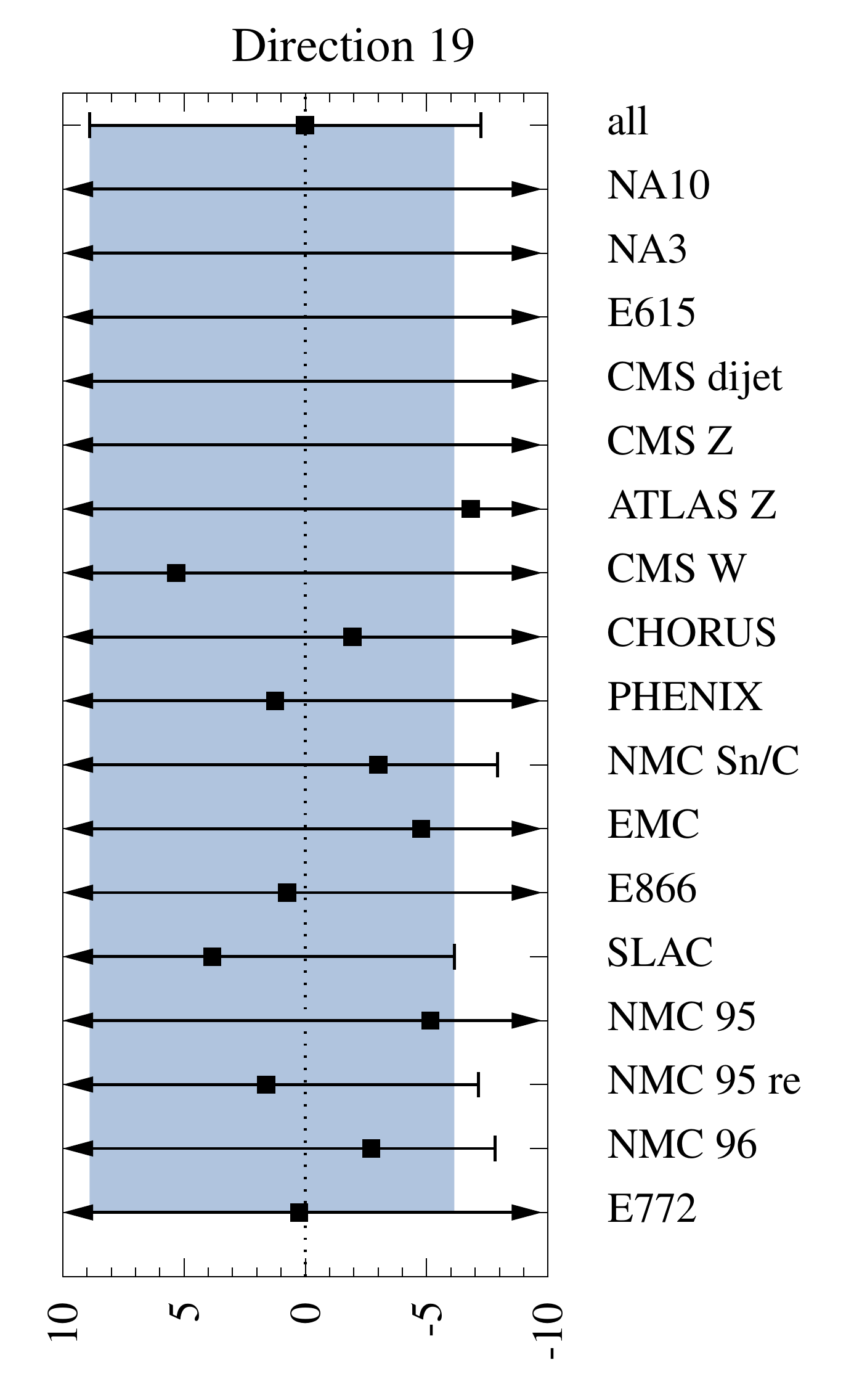}
\includegraphics[width=0.245\linewidth]{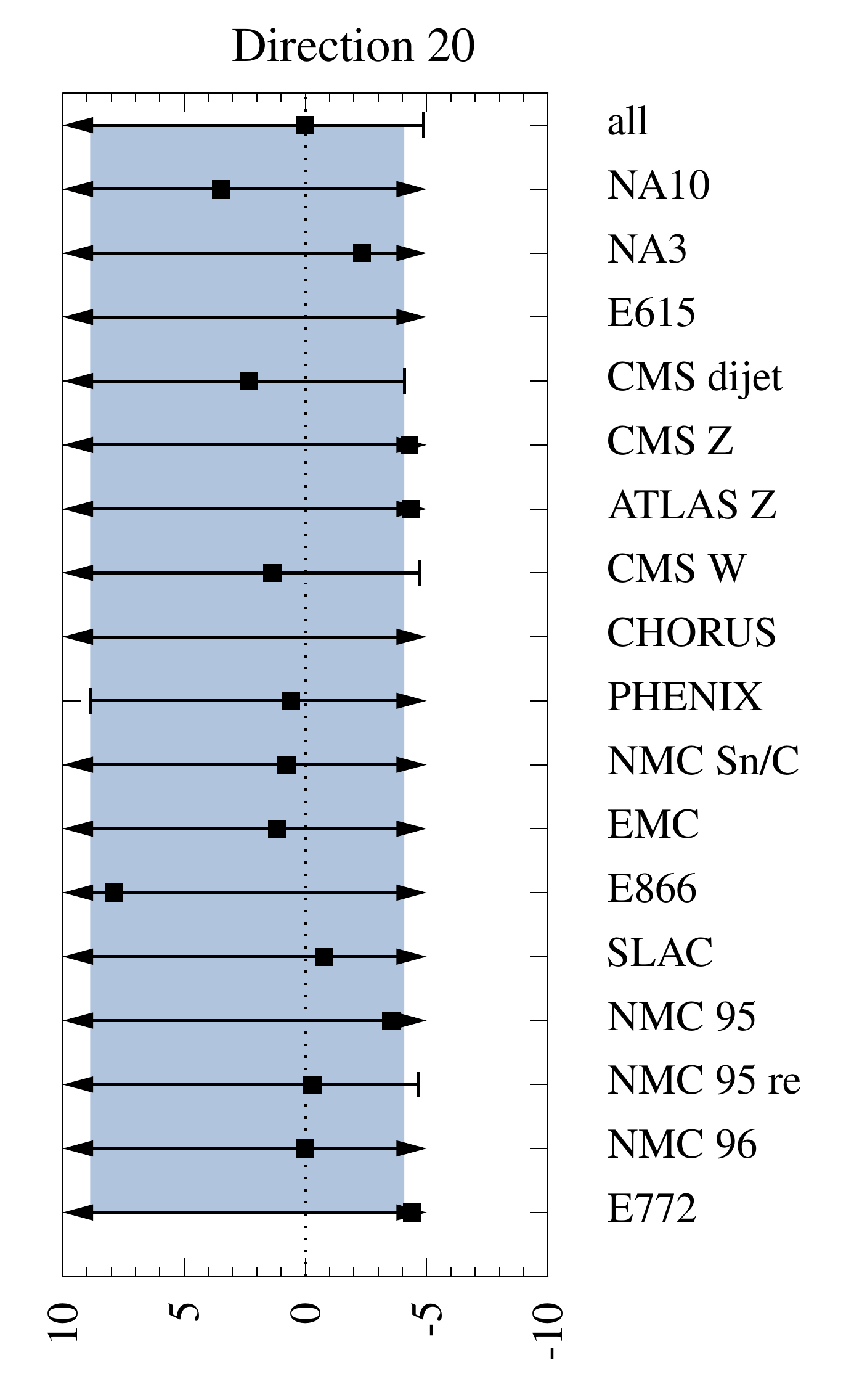}
\caption{As Fig.~\ref{fig:Conflimsall} but for eigendirections 13 to 20.}
\label{fig:Conflimsall2}
\end{figure*}

For each eigenvector direction $z_i$ and data set $k$ we find the interval $\left[z_{i,{\rm min}}^k,z_{i,{\rm max}}^k\right]$ for which $\chi^2_k < \chi^2_{k,{\rm max}}$. Looping over all the data sets $k$ we then find the intersection 
of the intervals $\left[z_{i,{\rm min}}^k,z_{i,{\rm max}}^k\right]$ for each $i$. In other words, we require all the individual data sets to remain within the defined 90\% limit,
\begin{align}
z_{i,{\rm min}} & \equiv  \max \left\{ z_{i,{\rm min}}^k \right\}, \nonumber\\
z_{i,{\rm max}} & \equiv  \min \left\{ z_{i,{\rm max}}^k \right\}.
\end{align}
The outcome of this process is shown in Figs.~\ref{fig:Conflimsall} and \ref{fig:Conflimsall2} for all eigendirections.  The individual limits $\left[z_{i,{\rm min}}^k,z_{i,{\rm max}}^k\right]$ are shown as solid lines (with bars or arrows) and the intersection $\left[z_{i,{\rm min}},z_{i,{\rm max}}\right]$ as a gray band. This procedure is repeated for all eigendirections $i$. 
We note that we have here grouped together all the data (summing the $\chi^2$ contributions) from a given experiment and thus, in Figs.~\ref{fig:Conflimsall} and \ref{fig:Conflimsall2} there are less labels than individual contributions in Table~\ref{Table:Data}. 
Motivation for such a grouping is that even if an experiment gives data for various nuclei (e.g. SLAC E139) these are not unrelated e.g. for the baseline measurement and detector systematics. Furthermore, it may also happen (e.g. direction 8, lower limit, in Fig.~\ref{fig:Conflimsall}) that none of the individual experiments
(with grouped data) places stringent uncertainty limits, i.e. the intervals $\left[z_{i,{\rm min}},z_{i,{\rm max}}\right]$ become rather wide and the total $\chi^2$grows substantially above $\chi^2_0$. In such a case, the data from various experiments together may provide a better constraint than an individual experiment.
To take this into account, we treat the aggregate of all the data as a one additional ``experiment'' (the first rows in the panels of Figs.~\ref{fig:Conflimsall} and \ref{fig:Conflimsall2}.

\begin{figure*}[bht!]
\centering
\includegraphics[width=\linewidth]{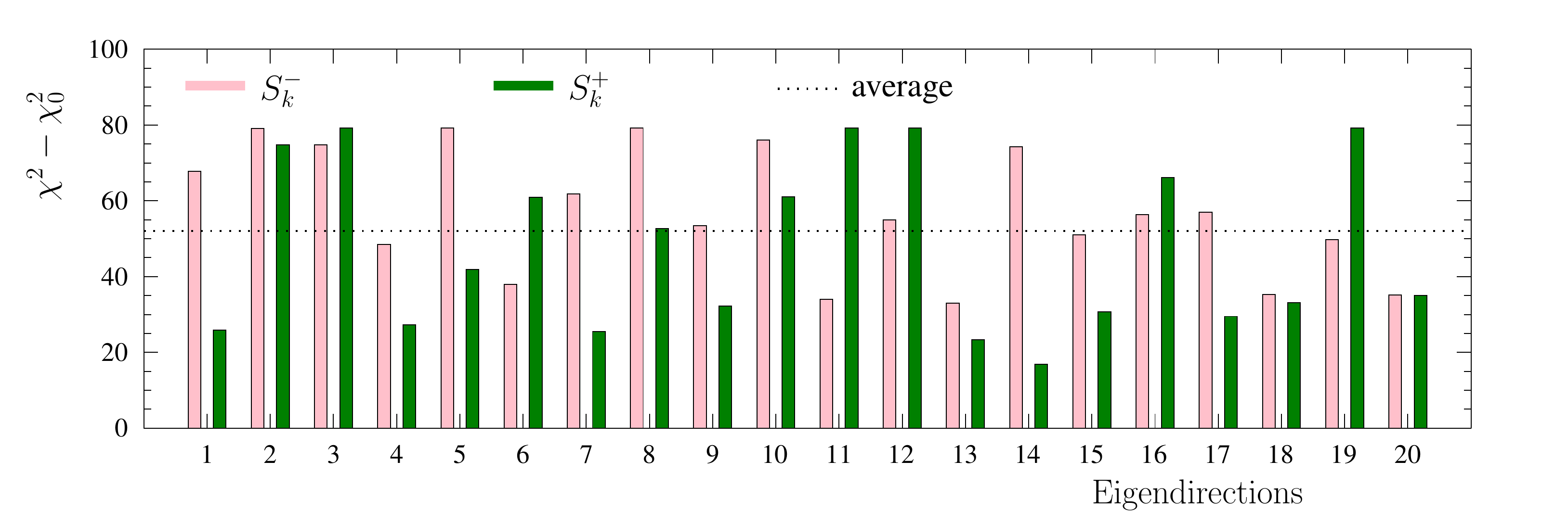}
\caption{The individual values of $\chi^2(S_k^\pm)-\chi^2_0$ compared with the average $\Delta \chi^2 = 52$.}
\label{fig:DeltaChi2s}
\end{figure*} 

We study two options to define the PDF uncertainty sets $S_i^\pm$.
In the first one, we set $t_i^+ = z_{i,{\rm max}}$ and $t_i^- = -z_{i,{\rm min}}$ in 
Eq.~\eqref{eq:errset}, i.e.,
\begin{align}
\vec z({S^+_1}\left[{\rm dyn}\right]) & = z_{1,\max} \left(1,0,...,0 \right) \nonumber\\
\vec z({S^-_1}\left[{\rm dyn}\right]) & = z_{1,\min} \left(1,0,...,0 \right) \nonumber\\
& \vdots \label{eq:EPPS16errset} \\
\vec z({S^+_N}\left[{\rm dyn}\right]) & = z_{N,\max}   \left(0,0,...,1 \right) \nonumber\\
\vec z({S^-_N}\left[{\rm dyn}\right]) & = z_{N,\min}   \left(0,0,...,1 \right), \nonumber
\end{align}
where the numbers $z_{i,{\rm min/max}}$ are obtained as described above. This is sometimes referred to as dynamic tolerance determination \cite{Martin:2009iq}. For the second option, we specify an average tolerance $\Delta\chi^2$ as
\begin{equation}
\Delta\chi^2 \equiv \frac{1}{N} \sum_i \frac{\chi^2\left(S_{i}^-\left[{\rm dyn}\right]\right) + \chi^2\left(S_{i}^+\left[{\rm dyn}\right]\right) - 2\chi^2_0}{2},
\end{equation}
where $\chi^2\left(S_{i}^\pm\left[{\rm dyn}\right]\right)$ are the values of $\chi^2$ that correspond to the error sets $S_i^\pm\left[{\rm dyn}\right]$ defined above. For the present fit with all the data, we find $\Delta\chi^2 \approx 52$. This averaging process is illustrated in Fig.~\ref{fig:DeltaChi2s} which shows the individual differences $\chi^2\left(S_{i}^-\left[{\rm dyn}\right]\right)-\chi^2_0$ and $\chi^2\left(S_{i}^+\left[{\rm dyn}\right]\right)-\chi^2_0$ as bars together with the found average. 
In this case the PDF uncertainty sets $S_i^\pm\left[ \Delta\chi^2\right]$ are defined by
imposing a fixed global tolerance $\Delta \chi^2=52$,
\begin{align}
\vec z\left({S^\pm_1}\left[ \Delta\chi^2\right]\right) & = \delta z_1^\pm \left(1,0,\ldots,0 \right)
\nonumber\\
&\vdots  \label{eq:2nderrset} \\
\vec z\left({S^\pm_N}\left[ \Delta\chi^2\right]\right) & = \delta z_N^\pm  \left(0,0,...,1 \right)
\nonumber
\end{align}
where the numbers $\delta z_i^\pm$ are the deviations in positive and negative direction chosen such that the $\chi^2$ grows by 52. The obtained values for $\delta z_i^\pm$ are listed in Table \ref{Table:kaikkideltat}.

\begin{table}[!htb]
\caption[]{\small The parameter deviations  $\delta z_i^\pm$ defining the EPPS16 error sets in Eq.~\eqref{eq:2nderrset}.}
\label{Table:kaikkideltat}
\begin{center}
\begin{tabular}{ll|ll}
$\delta z_i^-$       &  Value & $\delta z_i^+$ &  Value \\
\hline
$\delta z_1^-$      & -5.620   & $\delta z_1^+$      &  5.121 \\
$\delta z_2^-$      & -5.489   & $\delta z_2^+$      &  5.395  \\
$\delta z_3^-$      & -5.496   & $\delta z_3^+$      &  5.344 \\
$\delta z_4^-$      & -6.705   & $\delta z_4^+$      &  6.412 \\
$\delta z_5^-$      & -5.631   & $\delta z_5^+$      &  6.194 \\
$\delta z_6^-$      & -7.013   & $\delta z_6^+$      &  7.148  \\
$\delta z_7^-$      & -7.021   & $\delta z_7^+$      &  7.219  \\
$\delta z_8^-$      & -7.092   & $\delta z_8^+$      &  7.268 \\
$\delta z_9^-$      & -6.532   & $\delta z_9^+$      &  7.935 \\
$\delta z_{10}^-$ & -7.231   & $\delta z_{10}^+$   &  7.133  \\
$\delta z_{11}^-$ & -7.396   & $\delta z_{11}^+$   &  6.968  \\
$\delta z_{12}^-$ & -7.674   & $\delta z_{12}^+$   &  6.814   \\
$\delta z_{13}^-$ & -7.343   & $\delta z_{13}^+$   &  7.065  \\
$\delta z_{14}^-$ & -6.863   & $\delta z_{14}^+$   &  7.749  \\
$\delta z_{15}^-$ & -6.810   & $\delta z_{15}^+$   &  7.080  \\
$\delta z_{16}^-$ & -5.847   & $\delta z_{16}^+$   &  6.327  \\
$\delta z_{17}^-$ & -5.669   & $\delta z_{17}^+$   &  7.238  \\
$\delta z_{18}^-$ & -7.531   & $\delta z_{18}^+$   &  6.510  \\
$\delta z_{19}^-$ & -6.240   & $\delta z_{19}^+$   &  7.576  \\
$\delta z_{20}^-$ & -4.485   & $\delta z_{20}^+$   &  10.53  
\end{tabular}
\end{center}
\end{table}

As expected, Fig.~\ref{fig:DeltaChi2s} shows rather significant variations in 
$\chi^2\left(S_{i}^\pm\left[{\rm dyn}\right]\right)-\chi^2_0$
depending on which eigendirection one looks at. However, the corresponding variations in 
$z_{i,{\rm min/max}}\sim \sqrt{\chi^2\left(S_{i}^\pm\left[{\rm dyn}\right]\right)-\chi^2_0}$ which determine the error sets are much milder. Hence, it can be expected that 
the two error-set options,  $S_{i}^\pm\left[{\rm dyn}\right]$ and $S_i^\pm\left[ \Delta\chi^2\right]$, will eventually lead to rather similar uncertainty estimates.
In what follows (see Fig.~\ref{fig:global_vs_dynamic} ahead), we will verify that this indeed is the case.   
Hence, and also to enable PDF reweighting \cite{Paukkunen:2014zia}, 
we choose the $S_i^\pm\left[ \Delta\chi^2\right]$ 
with the single global tolerance $\Delta\chi^2$ as the final EPPS16 error sets.

As in EPS09, the propagation of PDF uncertainties into an observable $\mathcal{O}$ will be here computed separately for the upward and downward directions,
\begin{align}
& \left( \delta \mathcal{O}^\pm \right)^2 = \label{eq:asymerr} \\ 
& \sum_i \left[ \substack{ \max \\ \min} \left\{ \mathcal{O}\left(S^+_i\right)-\mathcal{O}\left(S_0\right),\mathcal{O}\left(S^-_i\right)-\mathcal{O}\left(S_0\right),0 \right\} \right]^2 \nonumber ,
\end{align}
where $\mathcal{O}\left(S_0\right)$ denotes the prediction with the central set and $\mathcal{O}\left(S_i^\pm\right)$ are the values computed with the error sets \cite{Nadolsky:2001yg}. 

\begin{figure*}[htb!]
\centering
\includegraphics[width=0.220\linewidth]{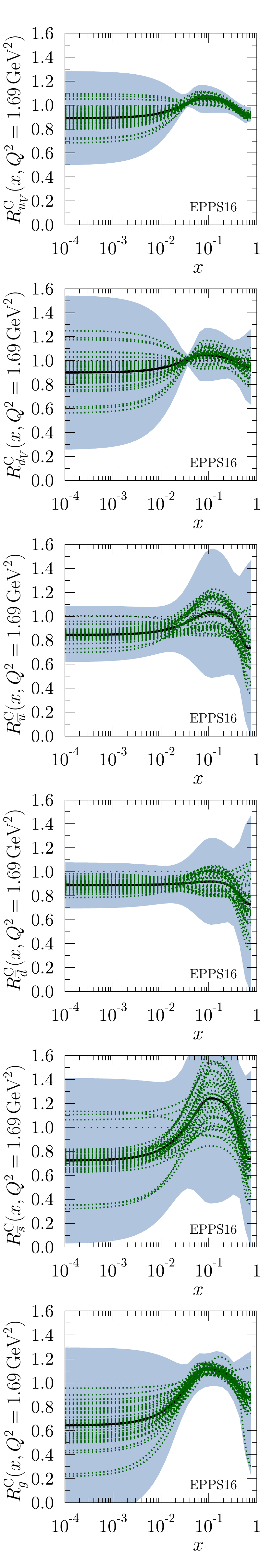}
\includegraphics[width=0.220\linewidth]{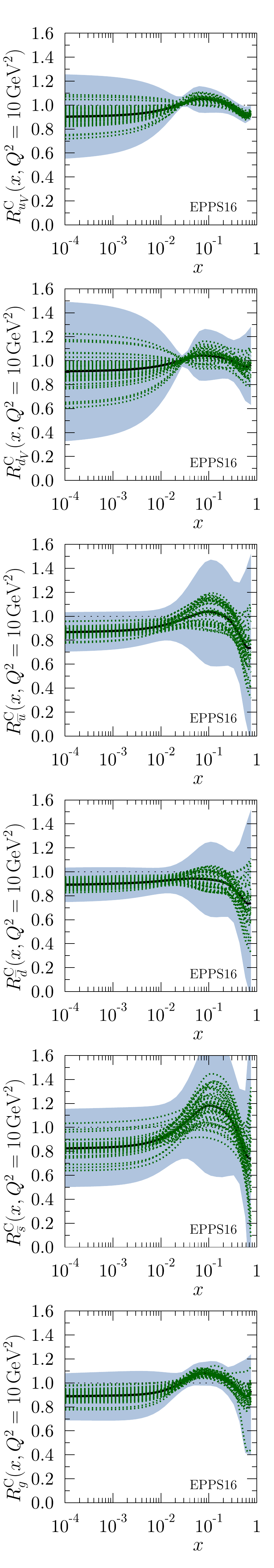}
\includegraphics[width=0.220\linewidth]{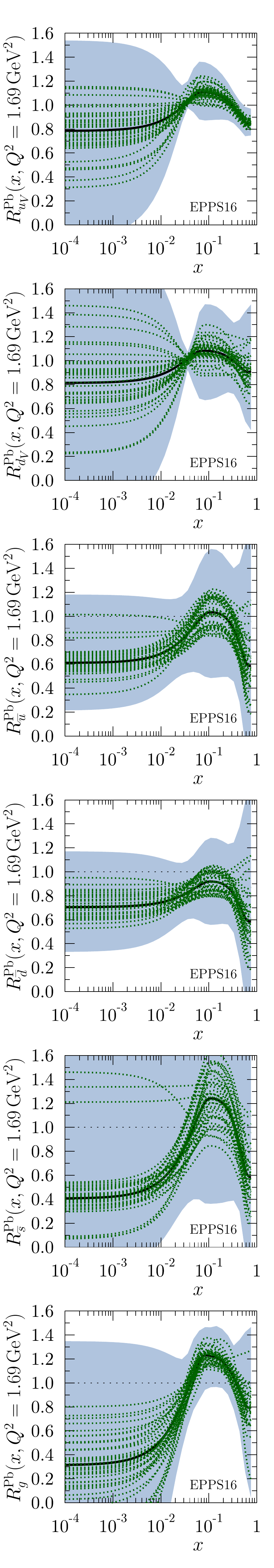}
\includegraphics[width=0.220\linewidth]{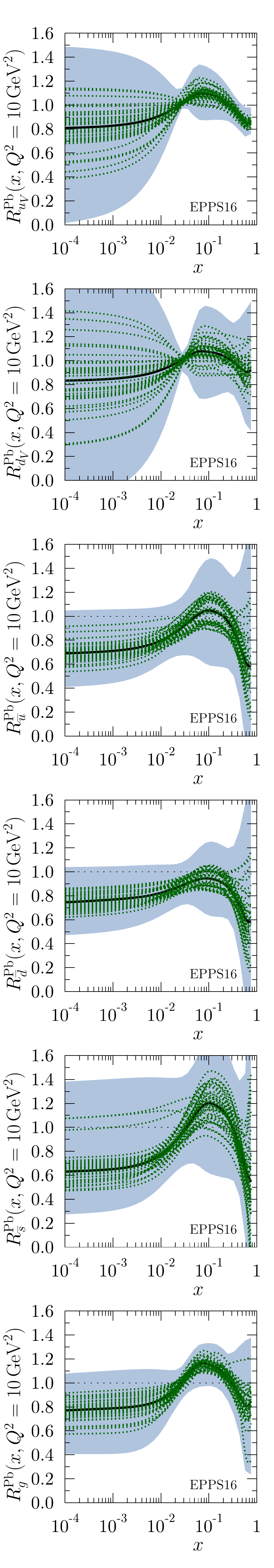}
\caption{The EPPS16 nuclear modifications for Carbon (leftmost columns) and Lead (rightmost columns) at the parametrization scale $Q^2=1.69\,{\rm GeV}^2$ and at $Q^2=10\,{\rm GeV}^2$. The thick black curves correspond to the central fit $S_0$ and the dotted curves to the individual error sets $S_i^\pm\left[ \Delta\chi^2\right]$ of Eq.~\eqref{eq:2nderrset}. The total uncertainties are shown as blue bands.}
\label{fig:allsets_lowQ2}
\end{figure*} 

\section{Results}
\label{Results}

\subsection{Parametrization and its uncertainties}
\label{Parametrization_and_its_uncertainties}

The parameter values that define the fit functions, the nuclear modifications $R_i^A$ in Eq.~\eqref{eq:FitForm}  at the initial scale $Q^2_0$ are listed in Table~\ref{Table:Params} where we also indicate the parameters that were fixed to those of other parton species or assumed to have some particular value. The fixed value of $\beta=1.3$ for all flavours as well as setting $\gamma_{y_a}=0$ for sea quarks are motivated by the EPS09 analysis. Freeing the latter easily leads to an unphysical case ($\gamma_{y_a}<0$) and thus we have decided to keep it fixed at this stage. 

\begin{table}[!htb]
\caption[]{\small List of parameters defining the central set of EPPS16 at the initial scale $Q_0^2=1.69\,{\rm GeV}^2$. The numbers in bold indicate the 20 parameters that were free in the fit.}
\label{Table:Params}
\begin{tabular}{c|llllll}
 Parameter   	&  ${u_{\rm V}}$ & ${d_{\rm V}}$ &  ${\overline{u}}$ \\
\hline
   $y_0(A_{\rm ref})$	& {sum rule}	   & {sum rule}      &  \textbf{0.844} \\
   $\gamma_{y_0}$    & {sum rule}	   & {sum rule}      &  \textbf{0.731} \\ 
   $x_a$    	&  \textbf{0.0717} & as $u_{\rm V}$  &  \textbf{0.104} \\
   $x_e$    	&  \textbf{0.693}  & as $u_{\rm V}$  & as $u_{\rm V}$  \\
   $y_a(A_{\rm ref})$    	&  \textbf{1.06}   &  \textbf{1.05}  &  \textbf{1.03}  \\
   $\gamma_{y_a}$    &  \textbf{0.278}  & as $u_{\rm V}$  & 0, fixed        \\
   $y_e(A_{\rm ref})$        &  \textbf{0.908}  &  \textbf{0.943} &  \textbf{0.725} \\
   $\gamma_{y_e}$    &  \textbf{0.288}  & as $u_{\rm V}$  & as $u_{\rm V}$  \\
   $\beta$      & 1.3, fixed       & 1.3, {fixed}    & 1.3, {fixed}    \\
\hline \\
 Parameter   	&  ${\overline{d}}$ &  ${s}$ & $g$\\
\hline
   $y_0(A_{\rm ref})$	& \textbf{0.889}    & \textbf{0.723}    & {sum rule}      \\
   $\gamma_{y_0}$    & as $\overline{u}$ & as $\overline{u}$ & {sum rule}      \\ 
   $x_a$    	& as $\overline{u}$ & as $\overline{u}$ & \textbf{0.0820} \\
   $x_e$    	& as $u_{\rm V}$    & as $u_{\rm V}$    & as $u_{\rm V}$  \\
   $y_a(A_{\rm ref})$    	& \textbf{0.919}    &  \textbf{1.24}    &  \textbf{1.12}  \\
   $\gamma_{y_a}$    & 0, fixed          & 0, fixed          & as $u_{\rm V}$  \\
   $y_e(A_{\rm ref})$        & as $\overline{u}$ & as $\overline{u}$ & \textbf{0.874}  \\
   $\gamma_{y_e}$    & as $u_{\rm V}$    & as $u_{\rm V}$    & as $u_{\rm V}$  \\
   $\beta$      & 1.3, {fixed}      &  1.3, {fixed}     &  1.3, {fixed}   \\
\end{tabular}
\end{table}

\begin{figure}[htb!]
\centering
\includegraphics[width=1.0\linewidth]{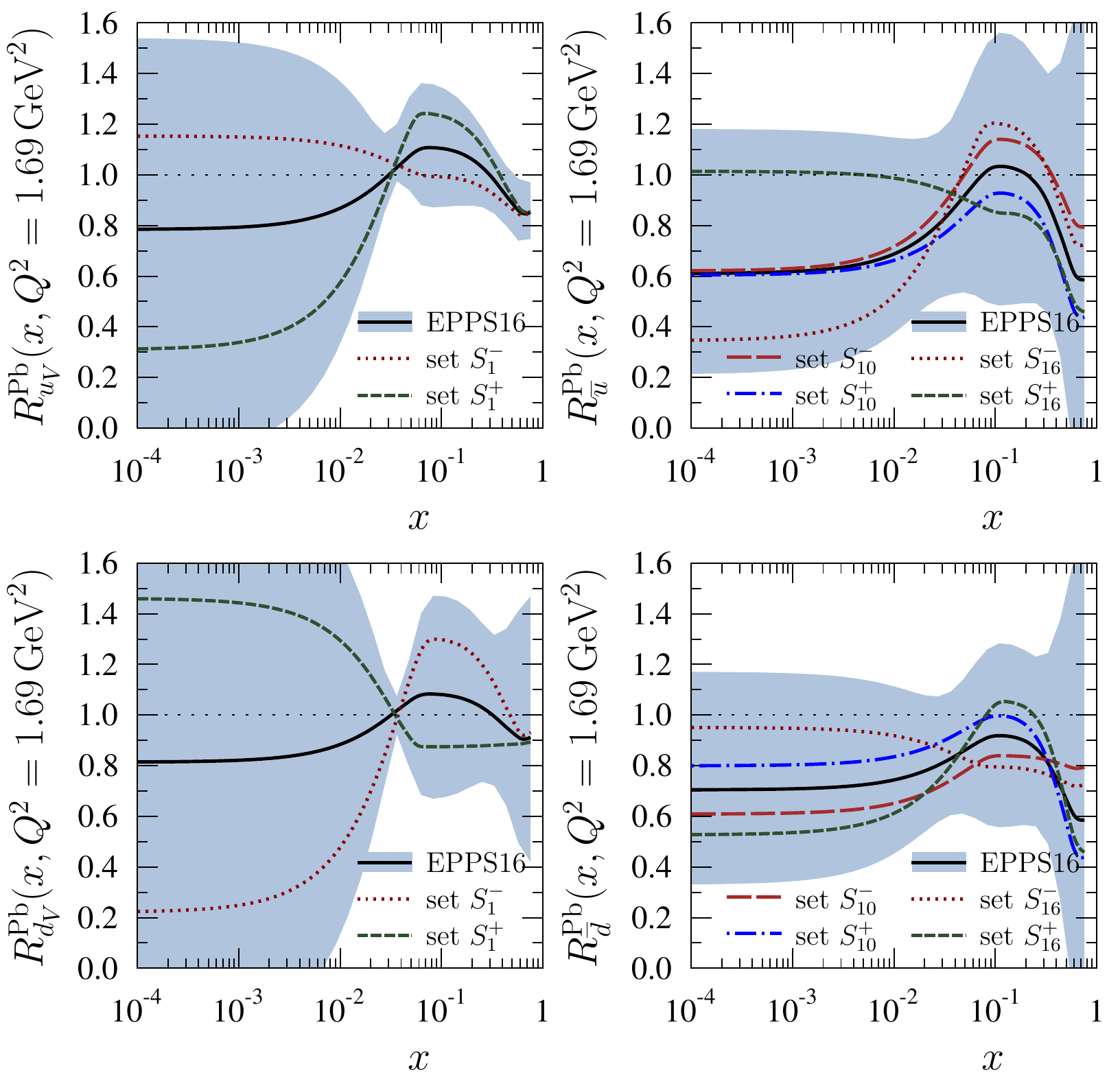}
\caption{
The EPPS16 nuclear modifications for valence and sea $u$ \& $d$ quarks for Lead at the parametrization scale $Q^2=1.69\,{\rm GeV}^2$. The solid black curves correspond to the central result and the dotted/dashed curves to the specific error sets as indicated. The total uncertainties are shown as blue bands.}
\label{fig:Pb_valencecorr}
\end{figure}

The $R_i^A$ functions themselves with error sets of Eq. \eqref{eq:2nderrset} and uncertainty bands of Eq.~\eqref{eq:asymerr}
are plotted in Fig.~\ref{fig:allsets_lowQ2} for Carbon and Lead nuclei at $Q^2=Q_0^2$ and $Q^2=10 \, {\rm GeV}^2$. Regarding these results, we make the following observations:

First, the obtained valence modifications $R_{u_{\rm V}}^A$ and $R_{d_{\rm V}}^A$ are very similar in the central set $S_0$, and strongly anticorrelated: as the average valence modification is fairly well constrained (see Fig.~\ref{fig:Pb10_with_EPS09} ahead) an error set whose, say, $R_{u_{\rm V}}^A$ is clearly below the central value has to have an $R_{d_{\rm V}^A}$ which is correspondingly above the central value, and vice versa. This is further demonstrated in Fig.~\ref{fig:Pb_valencecorr} where only the errors sets $S^\pm_1$ are shown for valence. The large error bands for $R_{u_{\rm V}}^A$ and $R_{d_{\rm V}}^A$ at small $x$ in Fig.~\ref{fig:allsets_lowQ2} reflect the fact that the flavour separation is not stringently constrained in the antishadowing region: the finite uncertainties there induce (via the sum rules) larger uncertainties in the shadowing region, see Fig.~\ref{fig:Pb_valencecorr}. 

\begin{figure*}[htb!]
\centering
\includegraphics[width=1.00\linewidth]{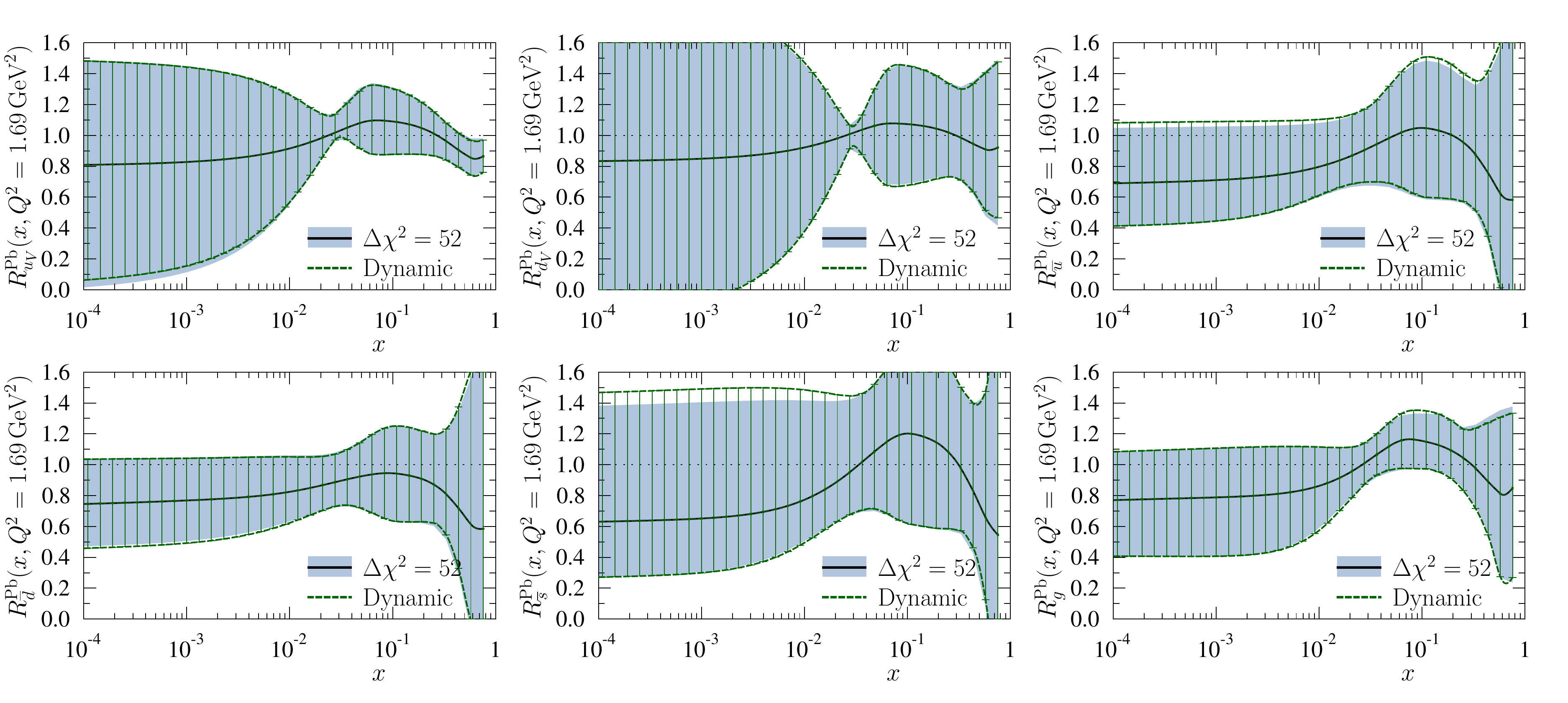}
\caption{The error bands of nuclear modifications at $Q^2=10 \, {\rm GeV}^2$ from the 
global tolerance $\Delta\chi^2=52$ used in the final
EPPS16 fit (black central line and light-blue bands) 
compared to the error bands from the dynamical tolerance determination (hatching) explained in Section \ref{Uncertaintyanalysis}.}
\label{fig:global_vs_dynamic}
\end{figure*} 

Second, interestingly also the $u$ and $d$ sea quark modifications are very similar in the central set $S_0$, and anticorrelated (except in the large-$x$ region where they were assumed to be the same at $Q^2_0$), though not as strongly as the valence quarks because also the strange-quark distribution plays some role. An example is shown in Fig.~\ref{fig:Pb_valencecorr} where the errors sets $S^\pm_{10}$ and $S^\pm_{16}$ have been plotted. In contrast to the valence quarks, individual sets are not always anticorrelated throughout all the $x$ values, but sets that are anticorrelated e.g. near $x_a$ can be very similar towards $x \rightarrow 0$. 

Third, the central value of the strange-quark nuclear modification indicates stronger nuclear effects than for the other light sea quarks. On the other hand, the uncertainty is also significant and even a large enhancement at small $x$ appears possible. While such an effect is theoretically unlikely (we would expect shadowing), it is consistent with the utilized data whose uncertainties our uncertainty bands represent. It should also be borne in mind that the determination of the strange quark in CT14 (our baseline PDF) may suffer from uncertainties (e.g. related to treatment of dimuon process in neutrino-nucleus DIS) and can, to some extent, affect the nuclear modifications we obtain. Thus, building a ``hard wall'' e.g. prohibiting an enhancement at small $x$ is not justified either. Nevertheless, the found central values of the strange-quark nuclear modifications are clearly in a sensible ballpark. 

Fourth, for gluon distributions the uncertainties are large at small $x$ at $Q_0^2$ but quickly diminish as the scale is increased. The gluon distributions in some error sets also go negative at small $x$ at low $Q^2$ but since $F_{\rm L}$ remains positive, this is allowed. 

Fifth, on average, the nuclear effects of Lead tend to be stronger than those of Carbon and also the uncertainties on Lead are larger than those on Carbon. Given that most of the data are for heavier nuclei than Carbon, especially the smaller errors for Carbon may appear a bit puzzling. The reason is in the new way of parametrizing the $A$ dependence of the nuclear effects, see Eq.~(\ref{eq:Adep}), that favours larger nuclei to exhibit larger nuclear effects. 

Sixth, the parametrization bias that our fit function entails is particularly well visible in the valence-quark panels where a narrow ``throat'' at $x\approx 0.02$ can be seen. This is an artefact of not allowing for more freedom at small $x$ while requiring the sum rules in Eq.~(\ref{eq:sum1}) and Eq.~(\ref{eq:sum2}): to satisfy the sum rule, an enhancement around $x=0.1$ must be accompanied by a depletion at small $x$ (or vice versa), and since $x_a$ for valence is fairly well determined the fit function always crosses unity near $x\approx 0.02$.

In Section \ref{Uncertaintyanalysis} we mentioned that the two error-determination options, the dynamical tolerance and fix\-ed global tolerance, lead to similar uncertainty estimates. To demonstrate this, we plot in Fig.~\ref{fig:global_vs_dynamic}
the error bands of the nuclear effects $R_i^{\rm Pb}$ at $Q^2=10$~GeV$^2$ obtained correspondingly from the error sets $S_{i}^\pm\left[{\rm dyn}\right]$ and $S_i^\pm\left[ \Delta\chi^2\right]$. Indeed, we find no significant differences between the two options.

\subsection{Comparison with data}
\label{Comparison_with_the_data}

\begin{figure*}[htb!]
\centering
\includegraphics[width=0.49\linewidth]{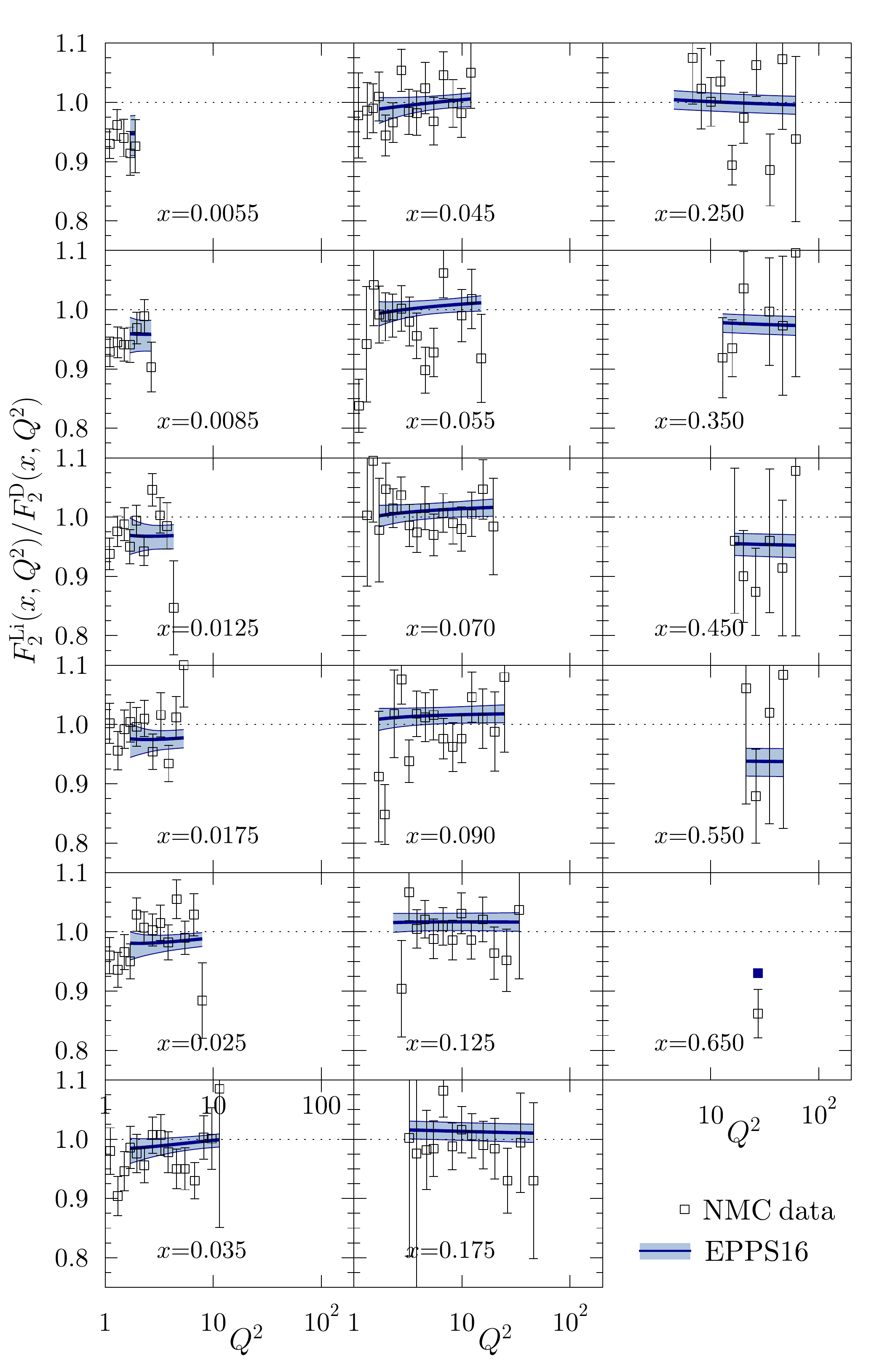} \hspace{-0.0cm}
\includegraphics[width=0.49\linewidth]{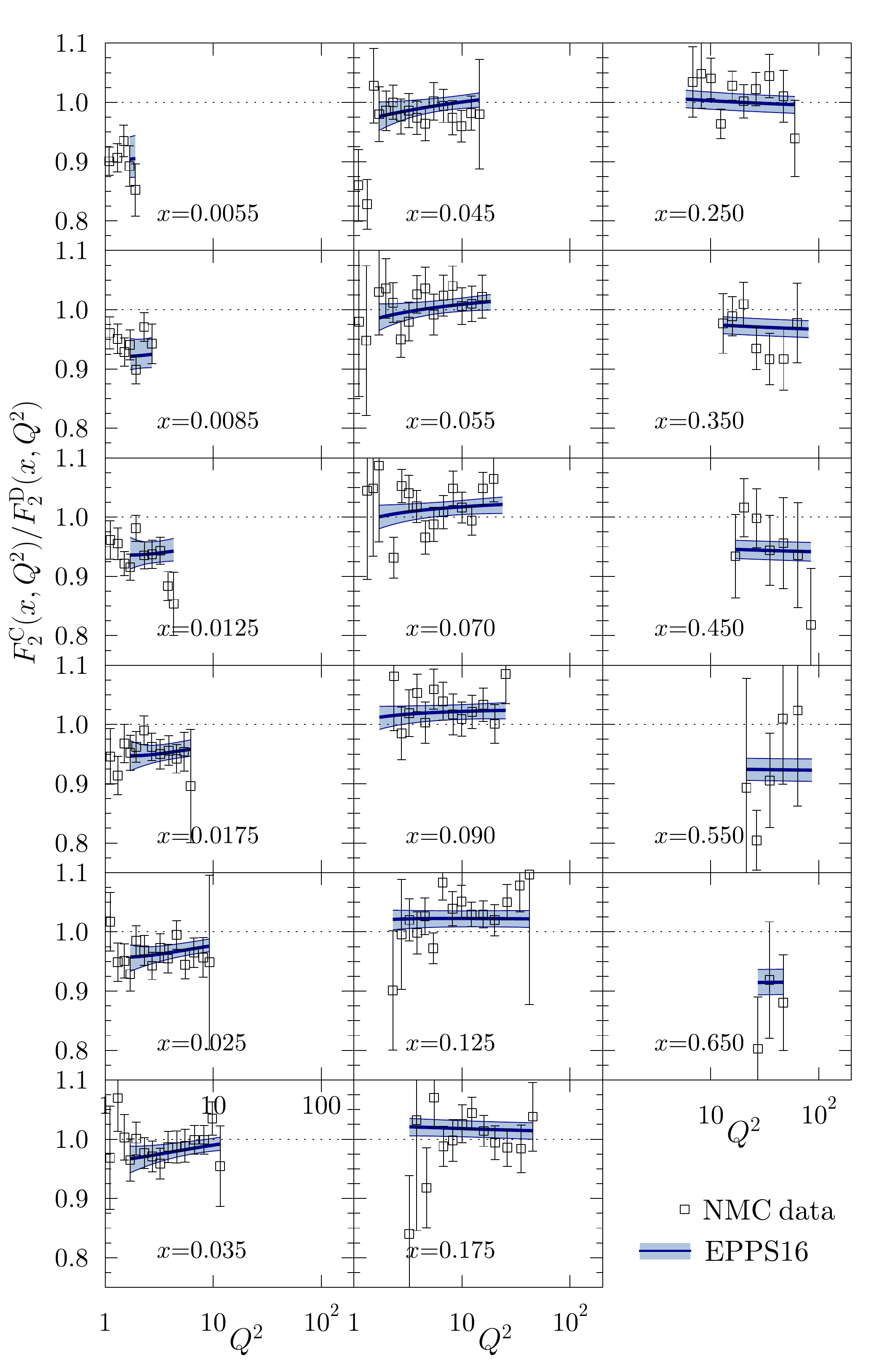}
\caption{The $Q^2$ dependence of structure function ratios as measured by the NMC collaboration \cite{Arneodo:1995cs}, compared with the EPPS16 fit.
Solid lines show our central set results, and error bands are computed from Eq.~\eqref{eq:asymerr}.}
\label{fig:CLiQ2}
\end{figure*} 

\begin{figure*}[htb!]
\centering
\includegraphics[width=1.0\linewidth]{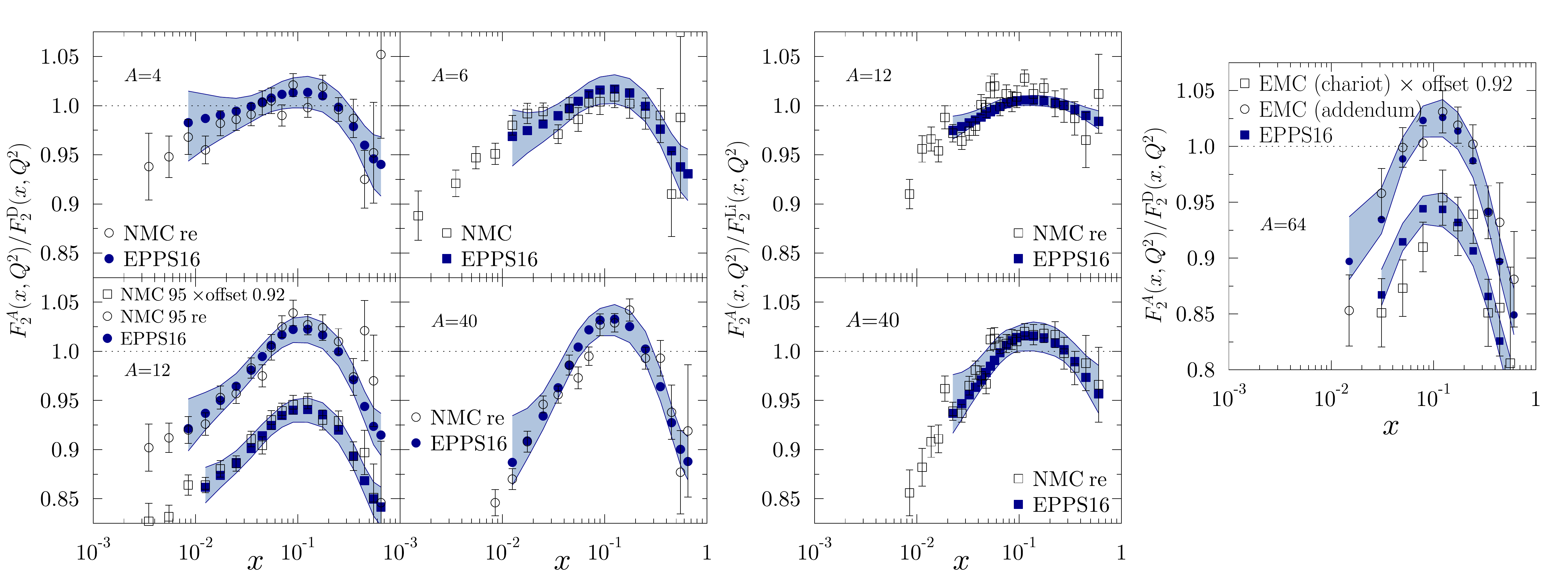}
\caption{Ratios of structure functions for various nuclei as measured by the NMC \cite{Amaudruz:1995tq,Arneodo:1995cs} and EMC \cite{Ashman:1992kv} collaborations, compared with the EPPS16 fit. In the rightmost panel the labels ``addendum'' and ``chariot'' refer to the two different experimental setups in Ref.~\cite{Ashman:1992kv}. For a better visibility, some data sets have been offset by a factor of 0.92 as indicated.
}
\label{fig:ALiDQ2}
\end{figure*} 

\begin{figure*}[htb!]
\centering
\includegraphics[width=\linewidth]{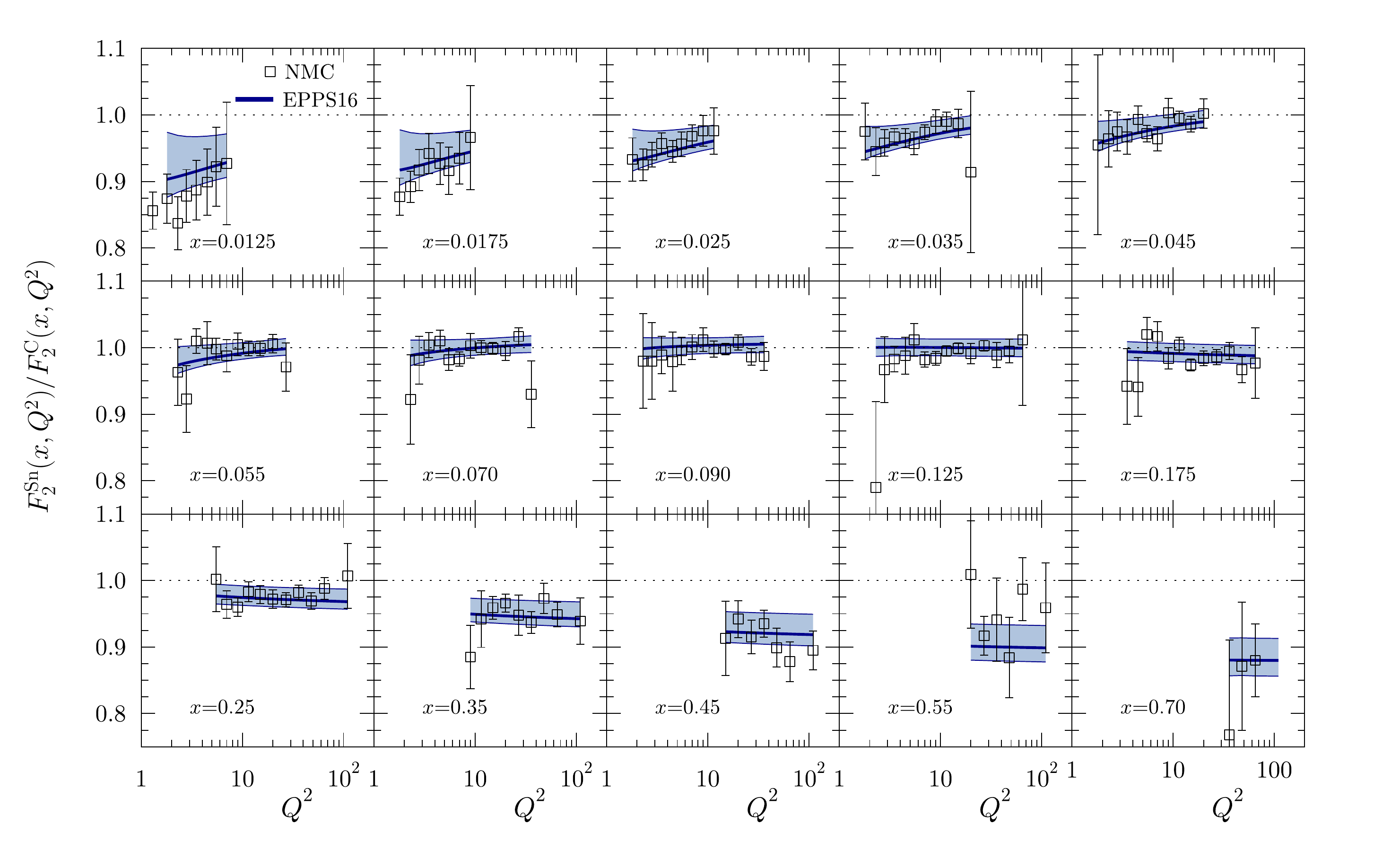}
\caption{The $Q^2$ dependence of the ratio $F_2^{\rm Sn}/F_2^{\rm C}$ for various values of $x$ as measured by NMC \cite{Arneodo:1996ru}, compared with EPPS16. 
}
\label{fig:RF2SnC}
\end{figure*} 

\begin{figure*}[htb!]
\centering
\includegraphics[width=0.464\linewidth]{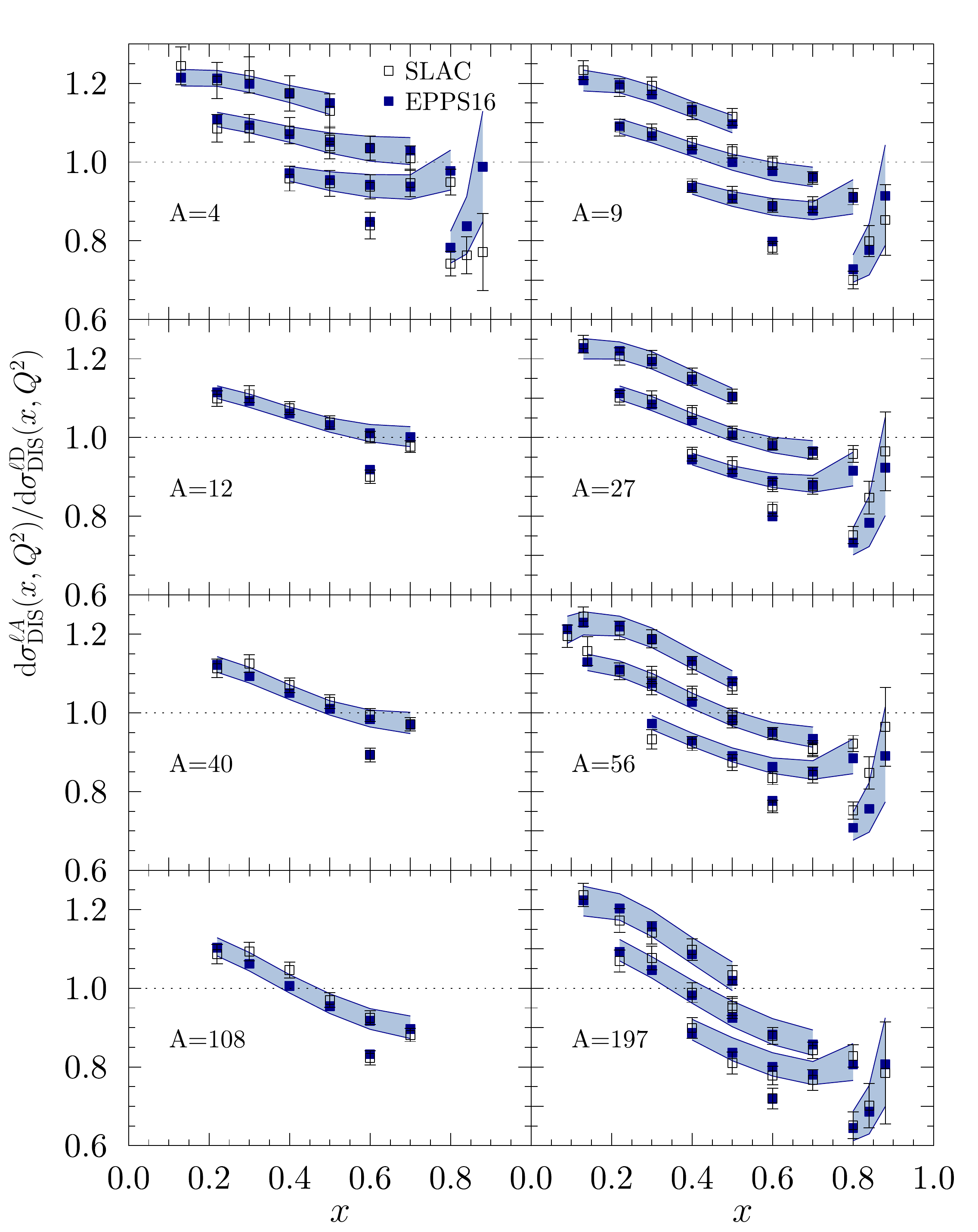} \hspace{-0.0cm}
\includegraphics[width=0.434\linewidth]{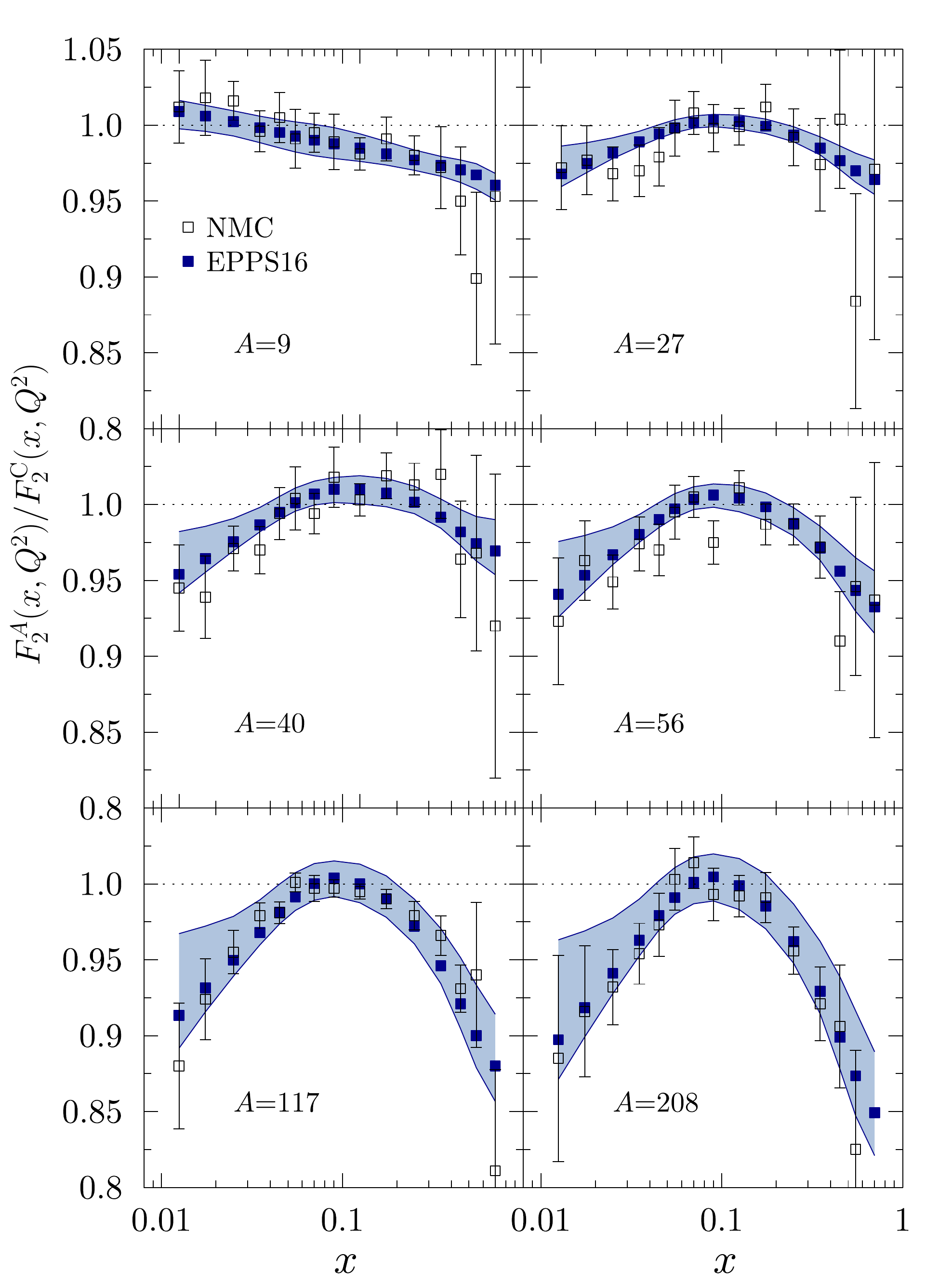}
\caption{The SLAC \cite{Gomez:1993ri} and NMC \cite{Arneodo:1996rv} data for DIS cross-section and structure-function ratios compared with the EPPS16 fit. 
For a better visibility, the SLAC data have been multiplied by 1.2, 1.1, 1.0, 0.9 for $Q^2=2\,{\rm GeV}^2$, $Q^2=5\,{\rm GeV}^2$, $Q^2=10\,{\rm GeV}^2$, $Q^2=15\,{\rm GeV}^2$, 
and the largest-$x$ set by 0.8.
}
\label{fig:SLAC_NMC}
\end{figure*}

\begin{figure*}[htb!]
\centering
\includegraphics[width=1.0\linewidth]{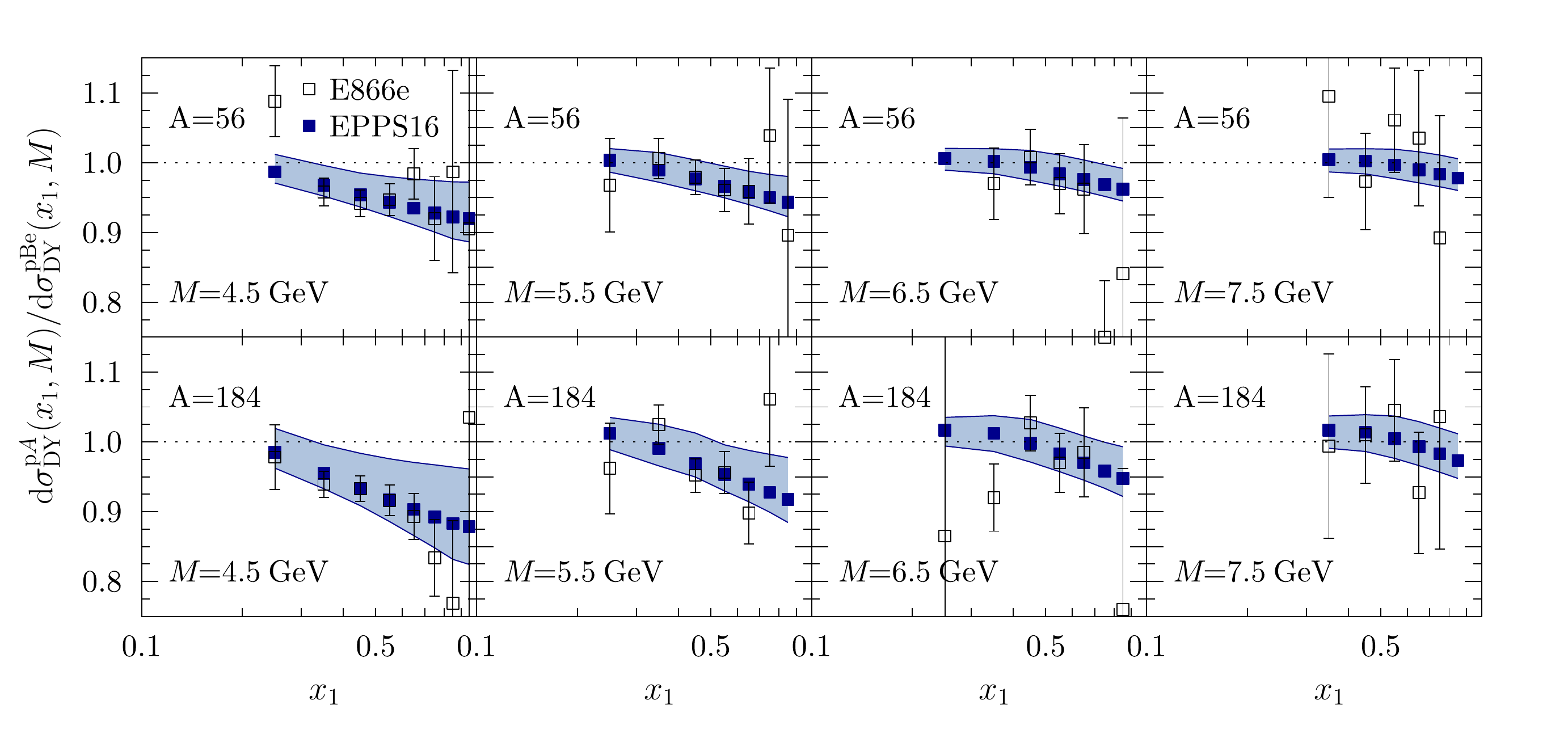}
\caption{Ratios of Drell-Yan dilepton cross sections $d\sigma^{{\rm p}A} / d\sigma^{\rm pBe}$ as a function of $x_1$ at various values of fixed $M$ 
as measured by E866 \cite{Vasilev:1999fa}, compared with EPPS16.} 
\label{fig:DYdatat}
\end{figure*} 

\begin{figure*}[htb!]
\centering
\includegraphics[width=1.0\linewidth]{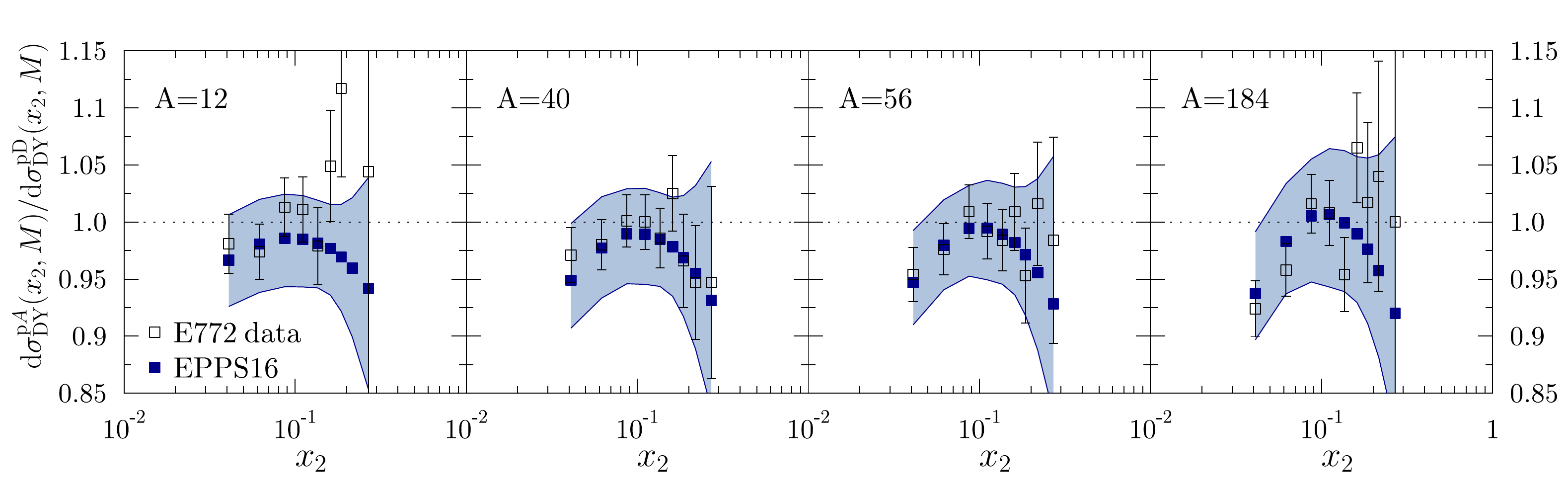}
\caption{Ratios of Drell-Yan cross sections measured by E772 
as a function of $x_2$ at fixed values of $M$, 
compared with the EPPS16 fit.}
\label{fig:RDY}
\end{figure*} 

\begin{figure*}[htb!]
\centering
\hspace{-0.5cm}
\includegraphics[width=0.40\linewidth]{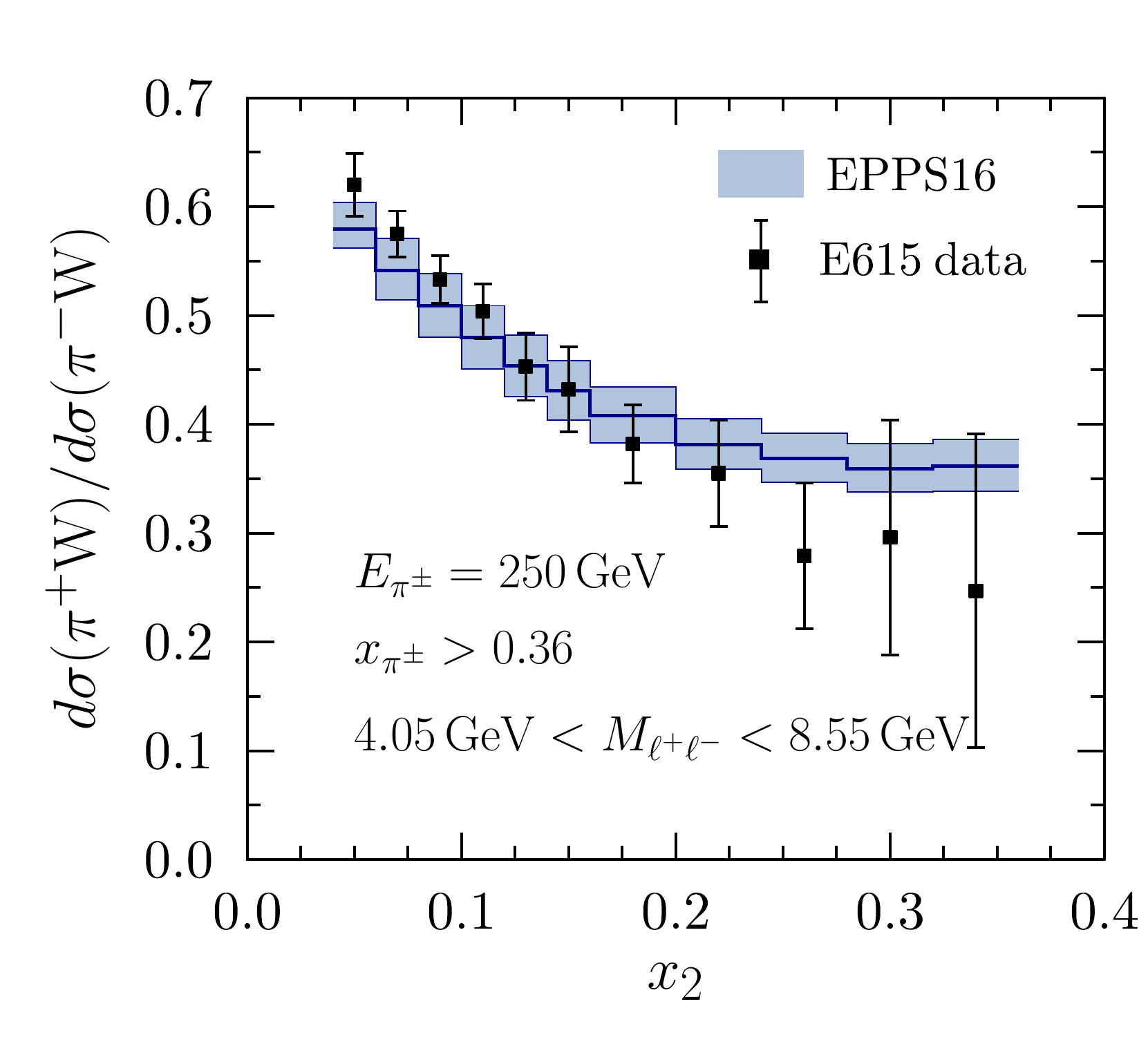} 
\includegraphics[width=0.40\linewidth]{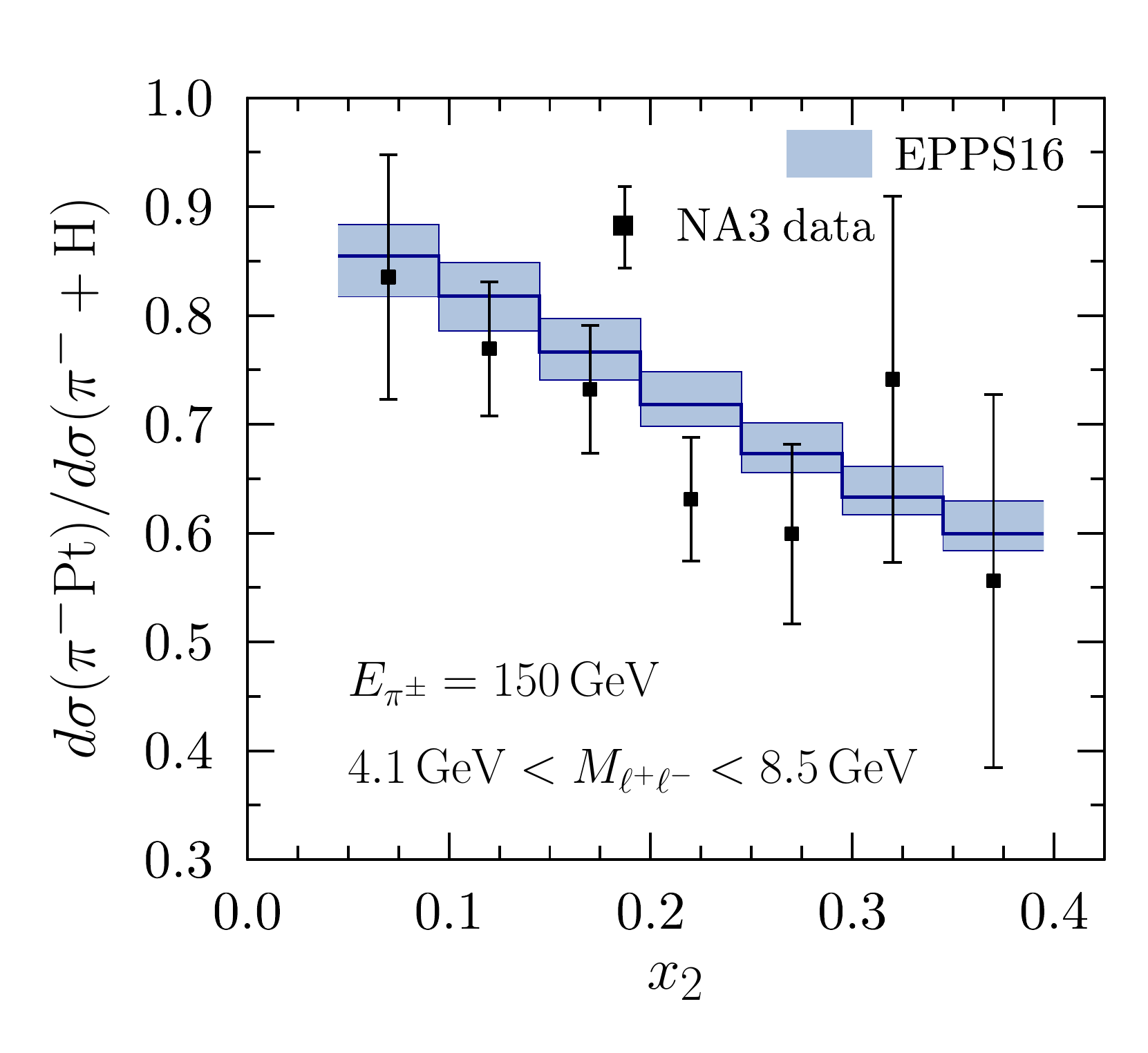}  

\vspace{-0.6cm}
\hspace{-0.8cm}\includegraphics[width=0.8\linewidth]{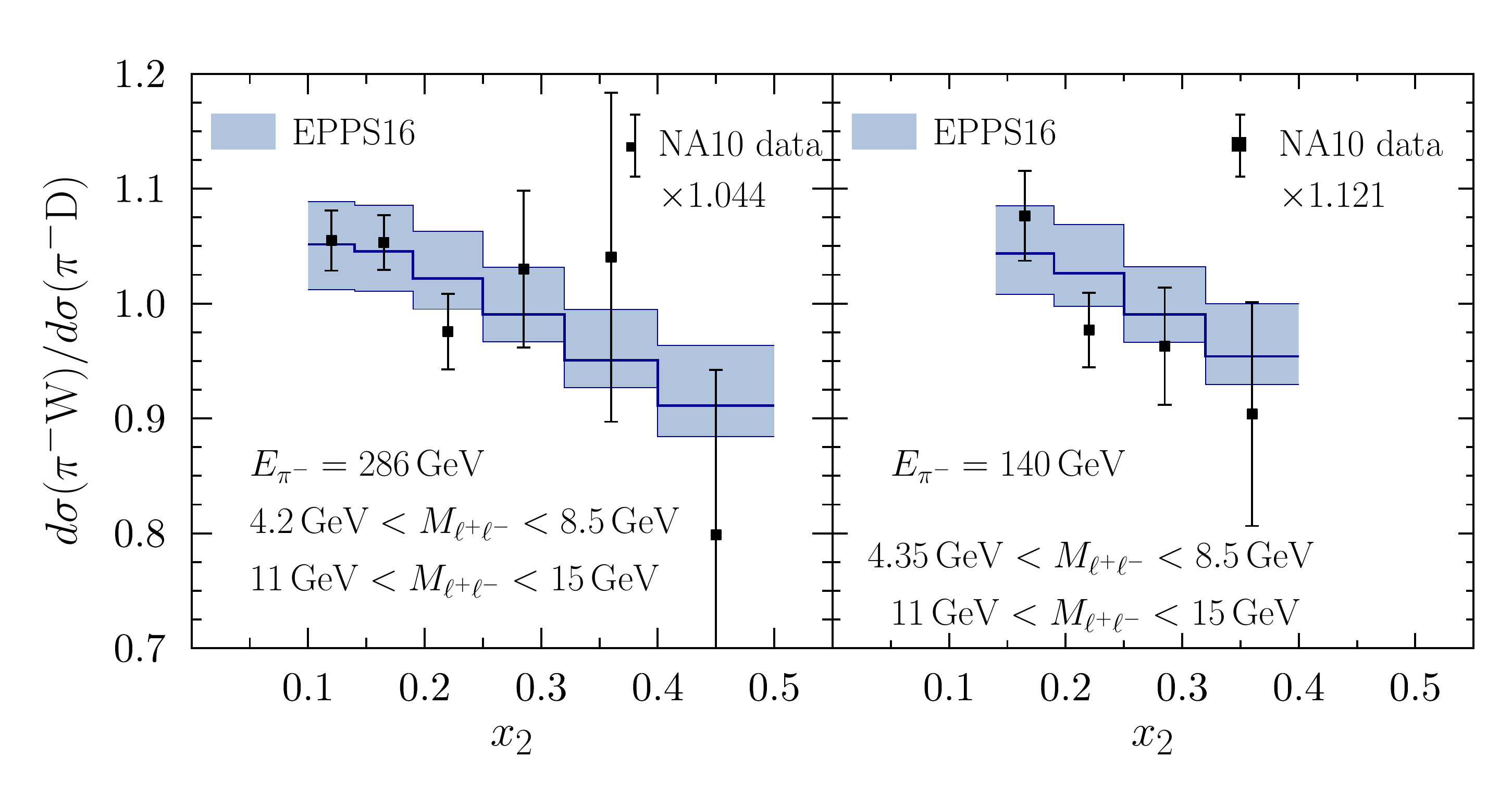} 
\caption{The $\pi^\pm$-$A$ Drell-Yan data from E615 \cite{Heinrich:1989cp}, NA3 \cite{Badier:1981ci}  and NA10 \cite{Bordalo:1987cs},  compared with the EPPS16 fit. The NA10 data have been multiplied by the optimized normalization factor $f_N$ from Eq.~(\ref{eq:chi2onlynorm}).}
\label{fig:piDYdatat}
\end{figure*} 

\begin{figure*}[htbp!]
\centering
\includegraphics[width=0.45\linewidth]{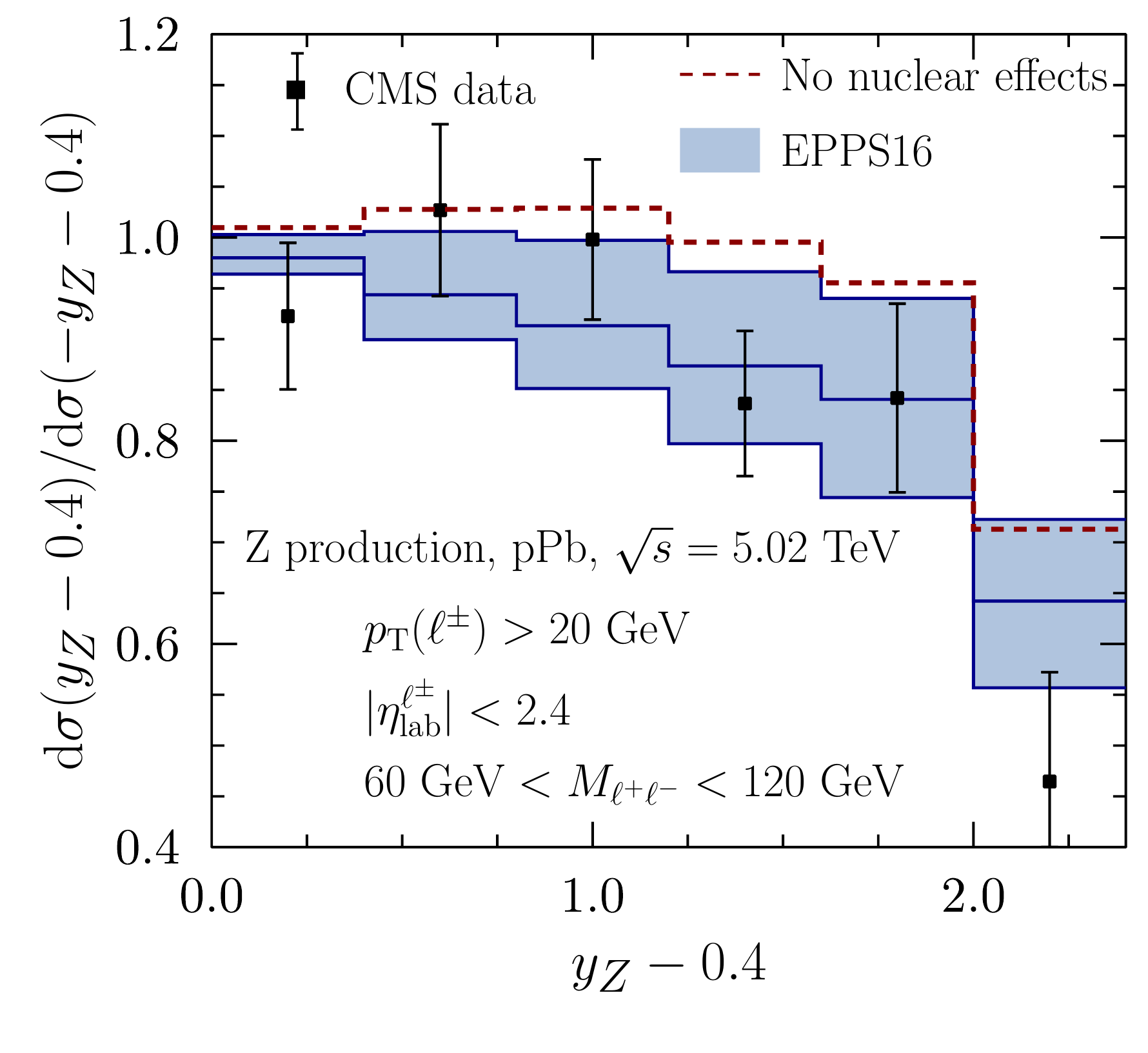}
\includegraphics[width=0.45\linewidth]{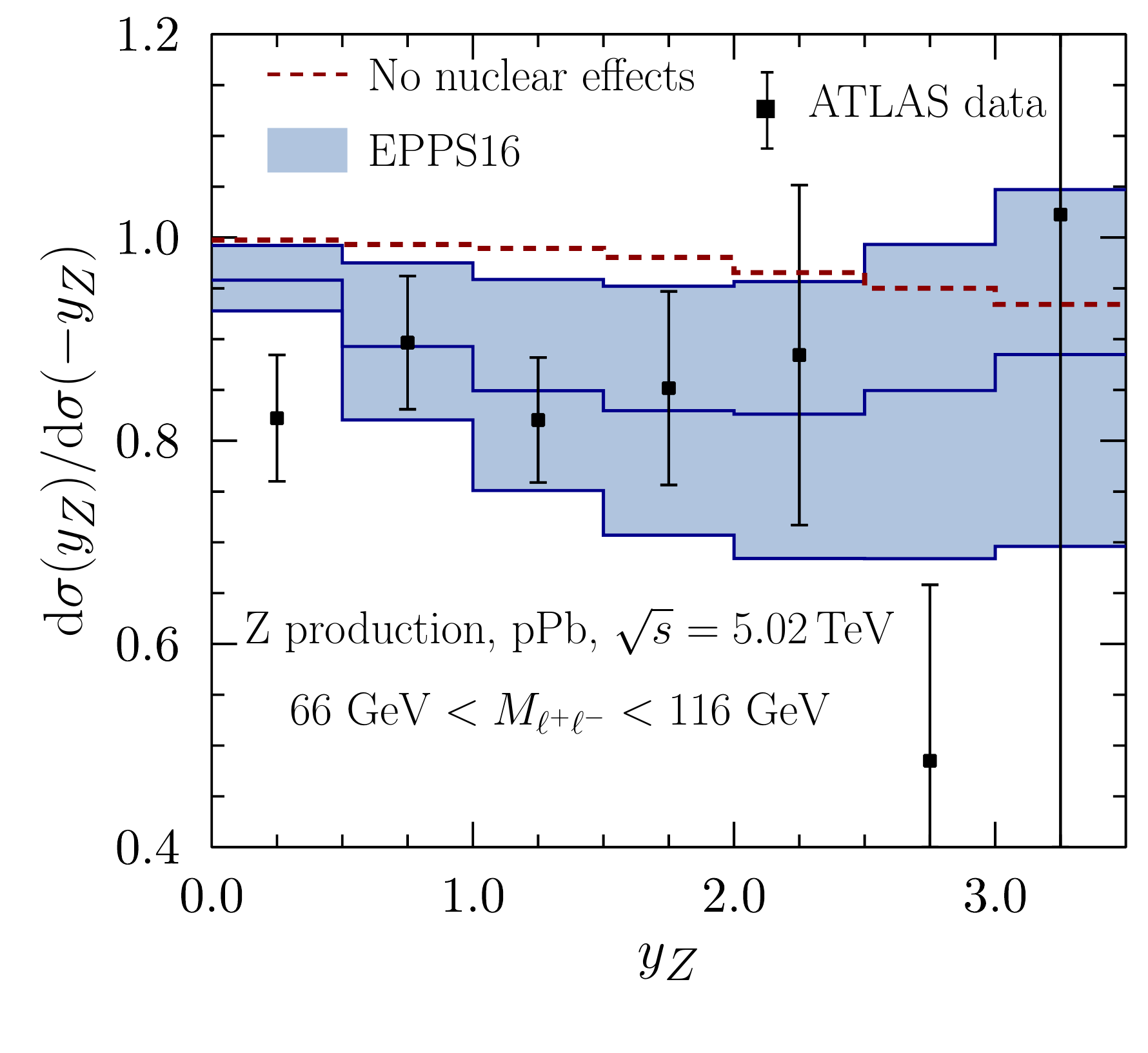}

\includegraphics[width=0.45\linewidth]{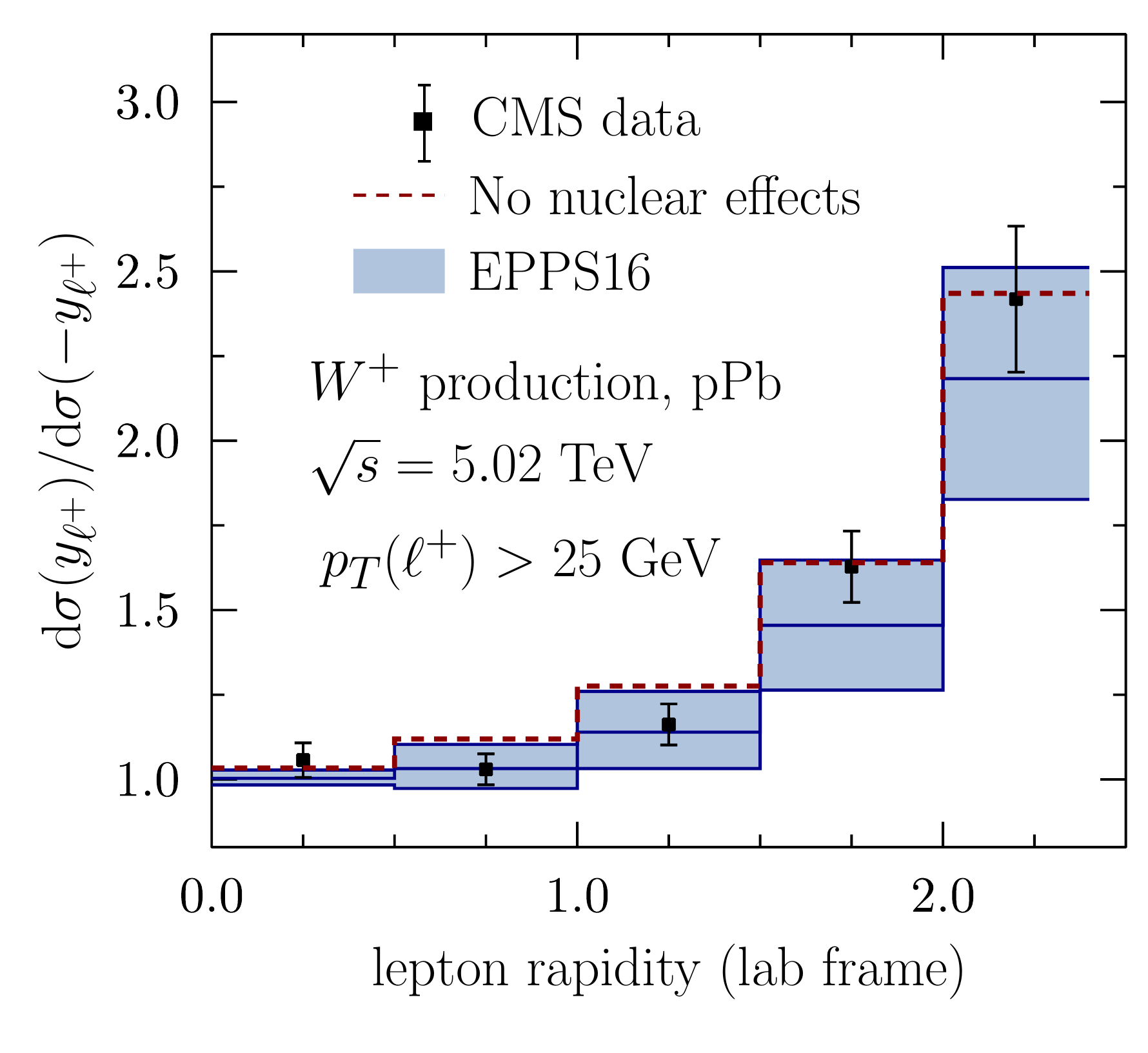}
\includegraphics[width=0.45\linewidth]{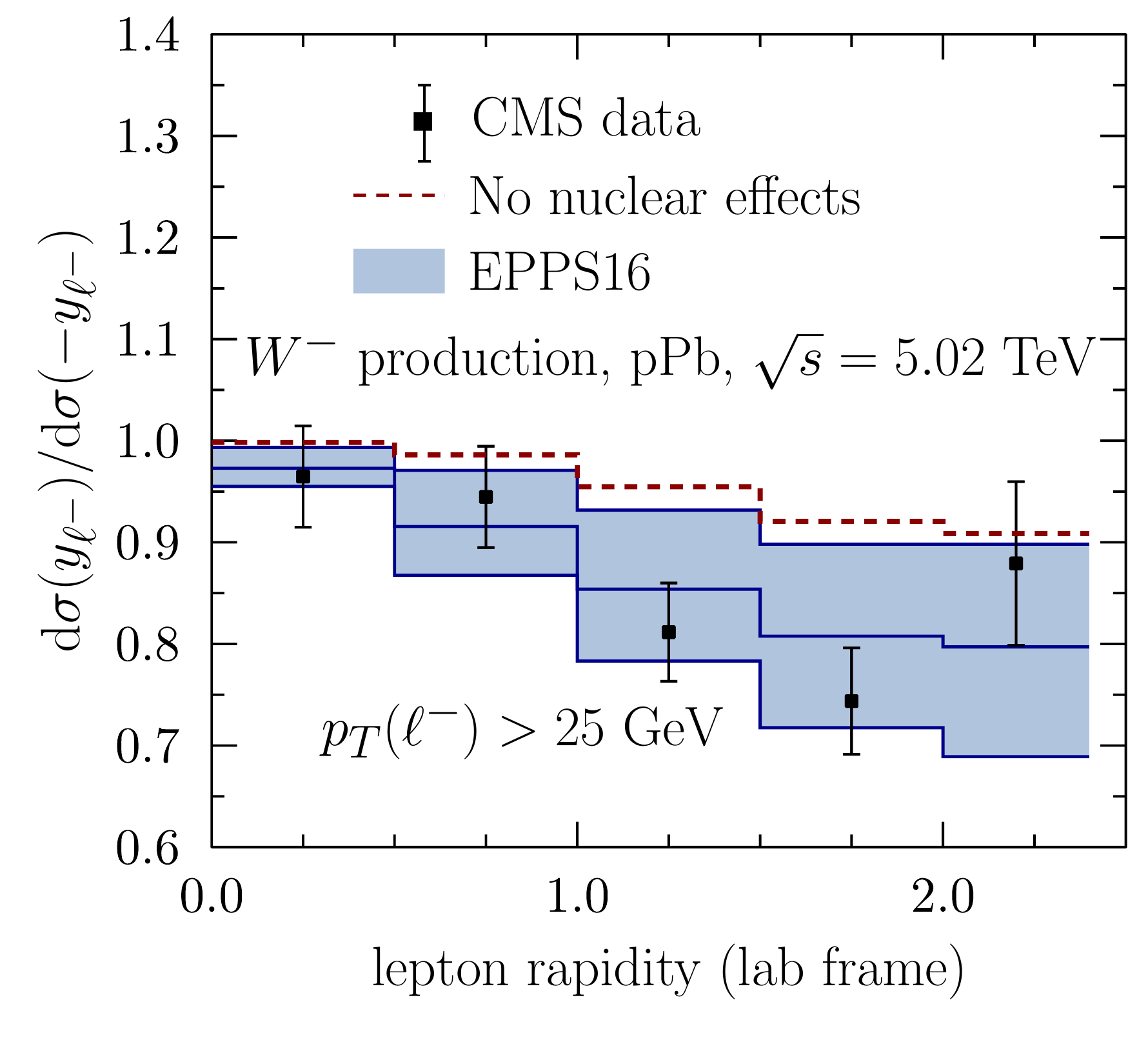}

\includegraphics[width=0.45\linewidth]{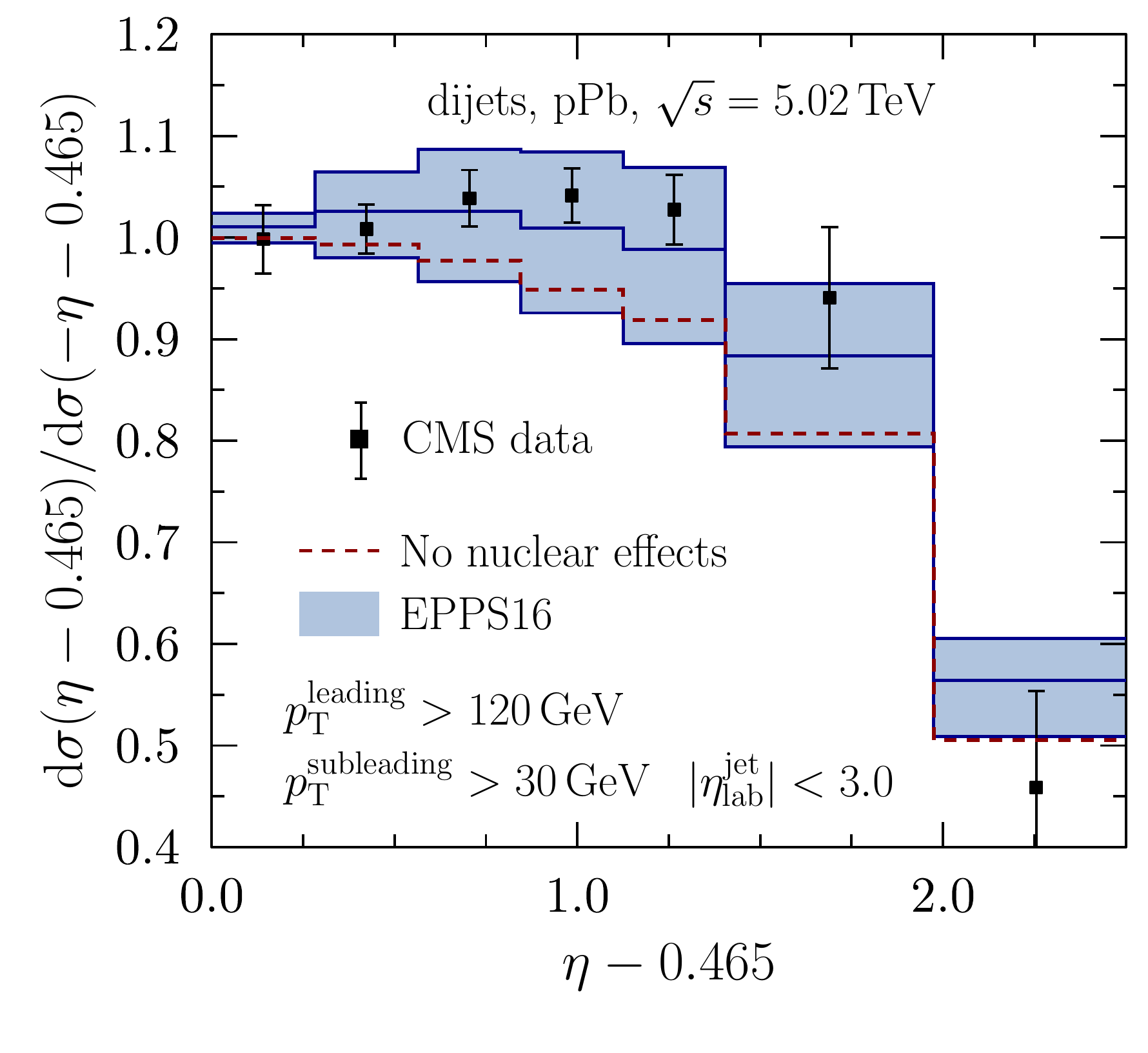}
\includegraphics[width=0.45\linewidth]{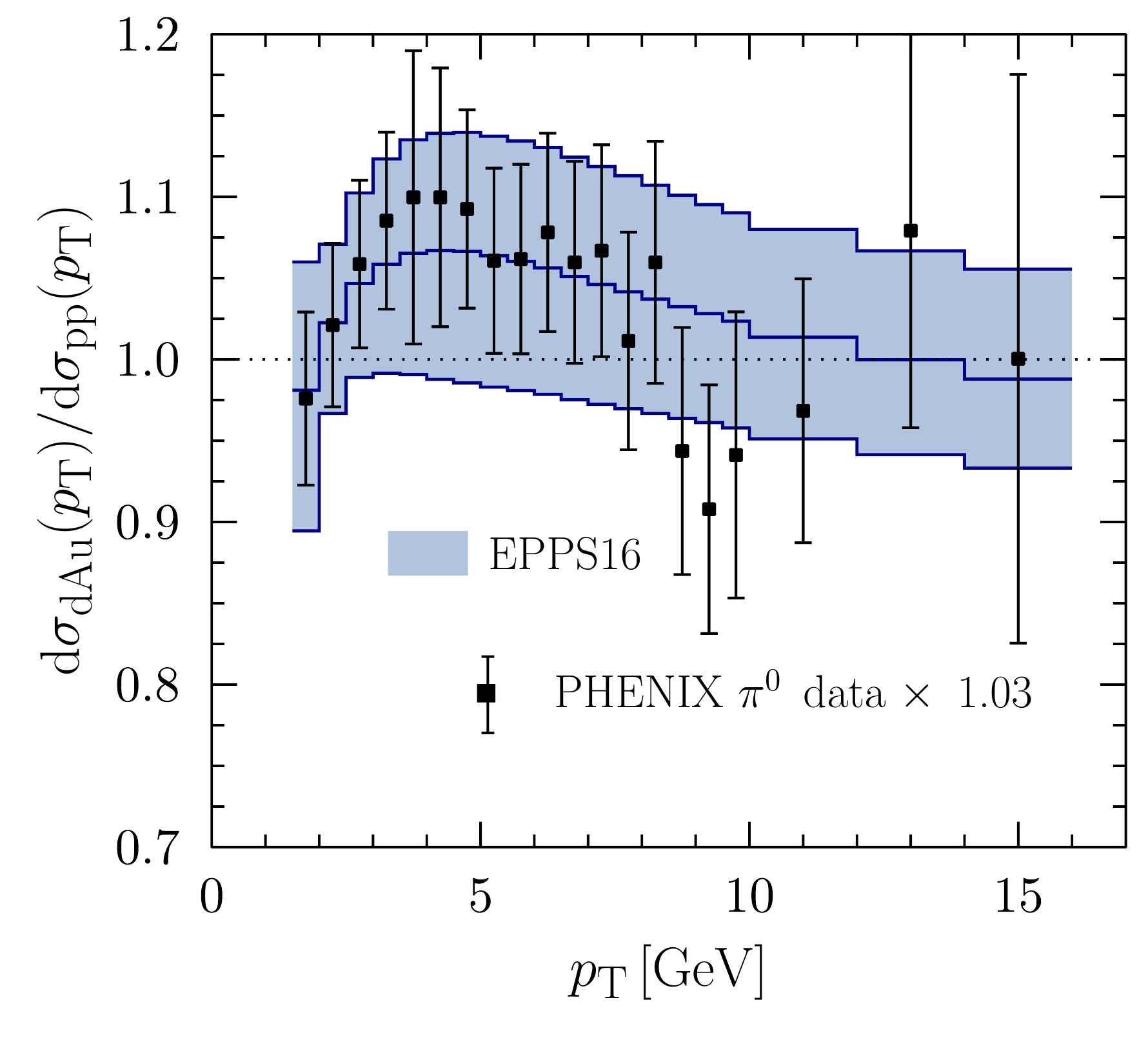}
\caption{The LHC pPb data from CMS \cite{Khachatryan:2015hha,Khachatryan:2015pzs,Chatrchyan:2014hqa} and ATLAS \cite{Aad:2015gta} for Z (upper panels) ${\rm W}^\pm$ (middle panels), and dijet production (lower left panel) compared with the EPPS16 fit. The dashed lines indicate the results with no nuclear modifications in the PDFs. The PHENIX DAu data \cite{Adler:2006wg} for inclusive pion production (lower right panel) are shown as well and have been multiplied by the optimal normalization factor $f_N=1.03$ computed by Eq.~(\ref{eq:chi2onlynorm}).}
\label{fig:LHCdatat}
\end{figure*}

\begin{figure*}[htb!]
\centering
\vspace{-0.5cm} 
\includegraphics[width=1.0\linewidth]{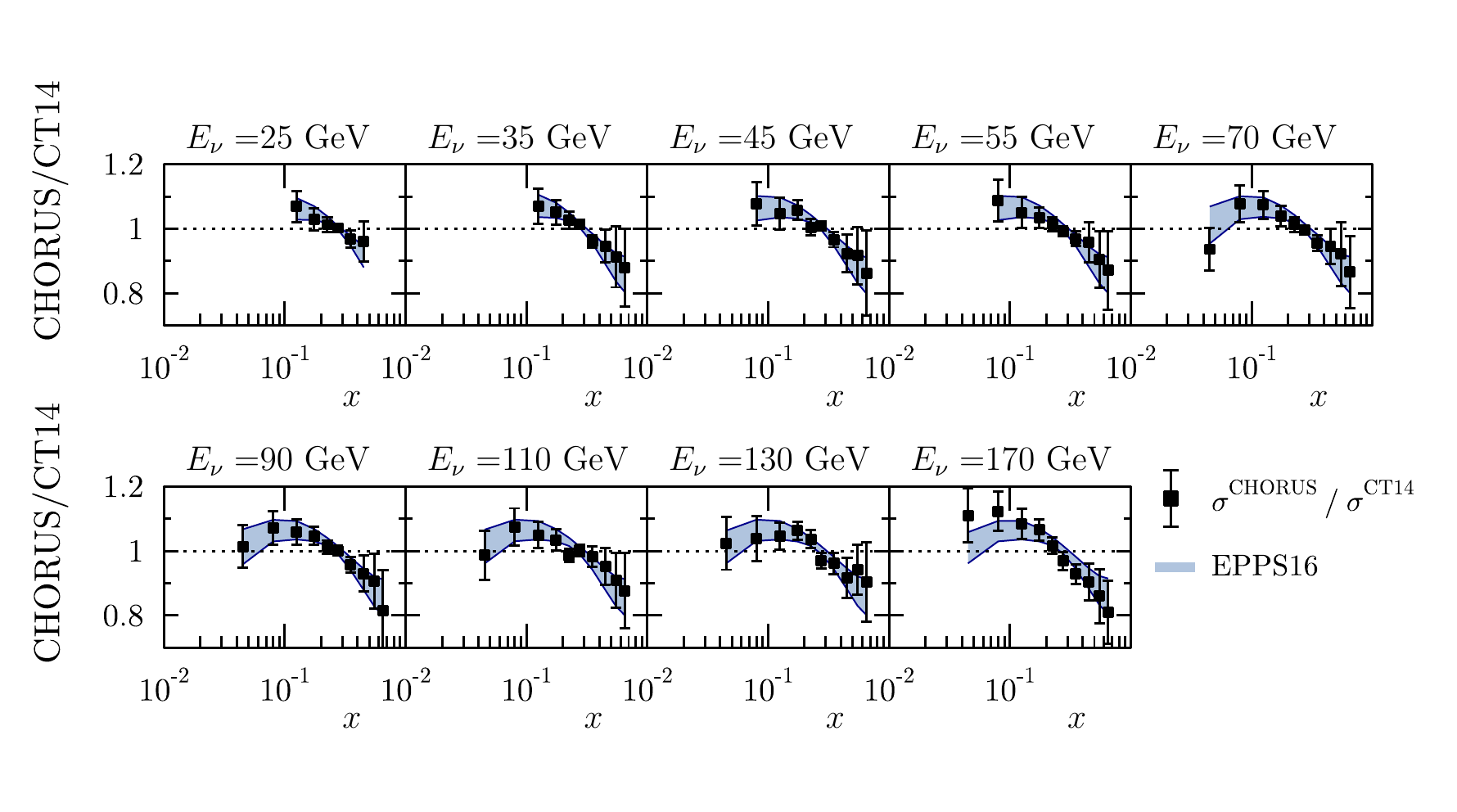} \\
\vspace{-0.5cm} 
\caption{The neutrino-nucleus DIS data based on CHORUS \cite{Onengut:2005kv} measurements, compared with the EPPS16 fit. The data as well as the theory curves have been obtained as described in Section~\ref{Treatmentoftheneutrinodata}.}
\label{fig:nudatat}
\end{figure*}

\begin{figure*}[htb!]
\centering
\vspace{-0.5cm} 
\includegraphics[width=1.0\linewidth]{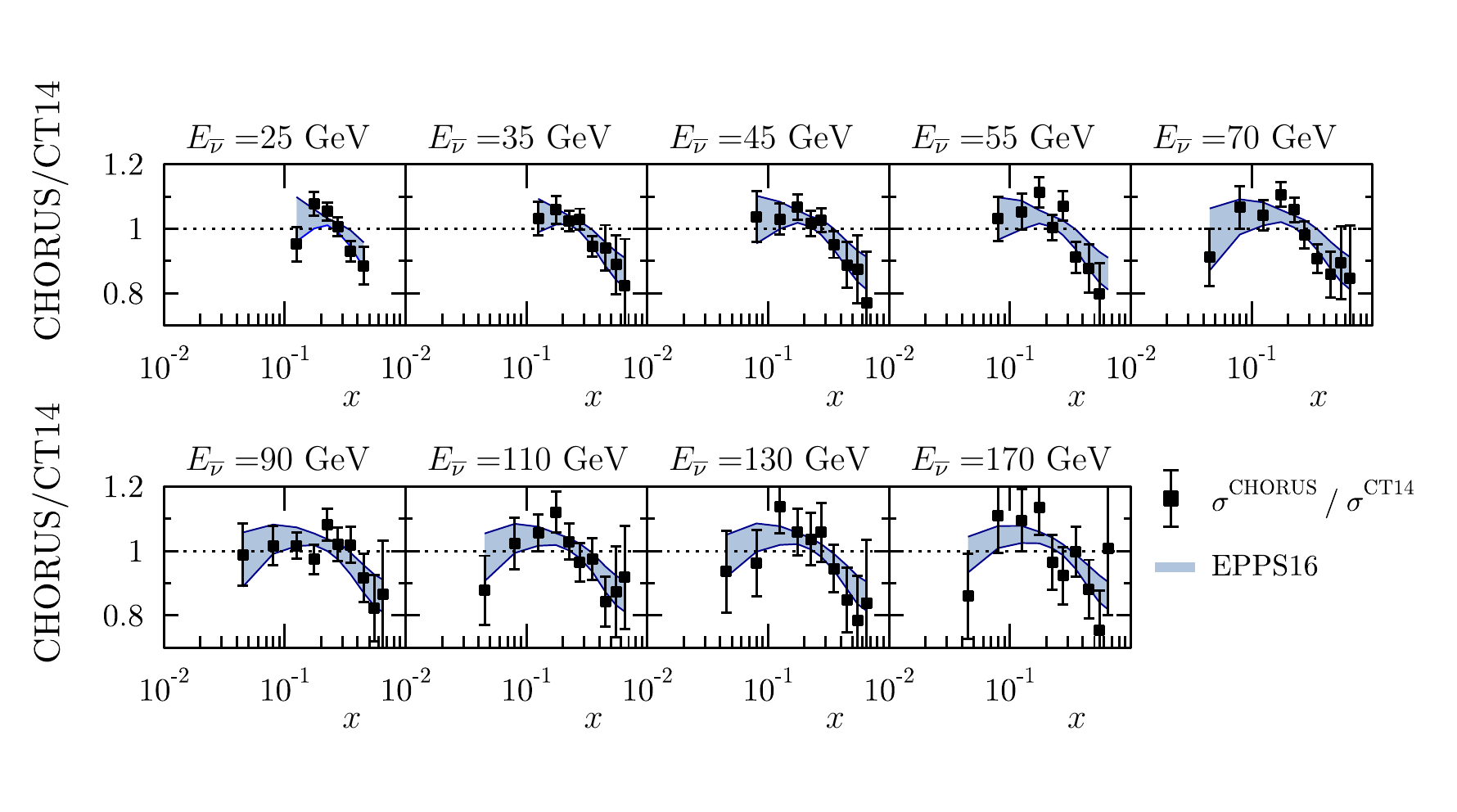}
\vspace{-1cm} 
\caption{As Fig.~\ref{fig:nudatat} but for antineutrino beam.}
\label{fig:anudatat}
\end{figure*} 

The following Figs.~\ref{fig:CLiQ2}--\ref{fig:anudatat} present a comparison of the EPPS16 fit with the experimental data
of Table~\ref{Table:Data}, computing the PDF error propagation according to Eq.~\eqref{eq:asymerr}. The error bars shown on the experimental data correspond to the statistical and systematic errors added in quadrature.
The charged-lepton DIS data are shown in Figs.~\ref{fig:CLiQ2}, \ref{fig:ALiDQ2}, \ref{fig:RF2SnC} and \ref{fig:SLAC_NMC}. We note that for undoing the isoscalar corrections as explained in Section~\ref{IsoscalarcorectionforDISdata}, the data appear somewhat different than e.g. in the EPS09 paper. On average, the data are well reproduced by the fit. In some cases the uncertainty bands are rather asymmetric (see e.g. the NMC data panel in Fig.~\ref{fig:SLAC_NMC}) which was the case in the EPS09 fit as well. This is likely to come from the fact that the $A$ dependence is parametrized only at few values of $x$ (small-$x$ limit, $x_a$, $x_e$) and in between these points the $A$ dependence appears to be somewhat lopsided in some cases. The $Q^2$ dependence of the data visible in Figs.~\ref{fig:CLiQ2} and \ref{fig:RF2SnC} is also nicely consistent with EPPS16.

The p$A$ vs. pD Drell-Yan data are shown in Figs.~\ref{fig:DYdatat} and \ref{fig:RDY}.
In the calculation of the corresponding differential NLO cross sections $d\sigma^{\rm DY}/dxdM$ we define $x_{1,2}\equiv (M/\sqrt s)e^{\pm y}$ where $M$ is the invariant mass and $y$ the rapidity of the dilepton. The scale choice in the PDFs is $Q=M$.
 While these data are well reproduced, the scatter of the data from one nucleus to another is the main reason we are unable to pin down any systematic $A$ dependence for the sea quarks at $x_a$ (some $A$ dependece develops via DGLAP evolution, however). For example, as is well visible in Fig.~\ref{fig:RDY}, it is not clear from the data whether there is a suppression or an enhancement for $x\gtrsim 0.1$. 

The pion-$A$ DY data are presented in Fig.~\ref{fig:piDYdatat}. 
As is evident from the figure, these data set into the EPPS16 fit without causing a significant tension. Overall, however, the statistical weight of these data is not enough to set stringent additional constraints to nuclear PDFs. Similarly to the findings of Ref.~\cite{Paakkinen:2016wxk}, the optimal data normalization of the lower-energy NA10 data (the lower right panel) is rather large ($f_N=1.121$), but the $x_2$ dependence of the data is well in line with the fit. 

The collider data, i.e. new LHC pPb data as well as the PHENIX DAu data, are shown in Fig.~\ref{fig:LHCdatat}. To ease the interpretation of the LHC data (forward-to-backward ratios), the baseline with no nuclear effects in PDFs is always indicated as well. The baseline deviates from unity for isospin effects (unequal amount of protons and neutrons in Pb) as well as for experimental acceptances. For the electroweak observables, the nuclear effects cause suppression in the computed forward-to-backward ratios (with respect to the baseline with no nuclear effects) as one is predominantly probing the region below $x\sim 0.1$ where the net nuclear effect of sea quarks has a downward slope towards small $x$. Very roughly, the probed nuclear $x$-regions can be estimated by $x \approx (M_{\rm W,Z}/\sqrt{s})e^{-y}$ and thus, towards more forward rapidities ($y>0$) one probes smaller $x$ than in the backward direction ($y<0$). The suppression comes about as smaller-$x$ quark distributions are divided by larger-$x$ (less-shadowed or antishadowed) quarks. In the case of dijets, the nuclear PDFs are sampled at higher $x$ and, in contrast to the electroweak bosons, an enhancement is observed. In our calculations, this follows essentially from antishadowed gluons becoming divided by EMC-suppressed gluon distributions, see Ref.~\cite{Eskola:2013aya} for more detailed discussions. The PHENIX pion data \cite{Adler:2006wg} is also well consistent with EPPS16 though, for the more precise CMS dijet data, its role is no longer as essential as it was in the EPS09 analysis.

Finally, comparisons with the CHORUS neutrino and antineutrino data are shown in Figs.~\ref{fig:nudatat} and \ref{fig:anudatat}. The data exhibit a rather typical pattern of antishadowing followed by an EMC effect at large $x$. The incident beam energies are not high enough to reach the small-$x$ region where a shadowing effect would be expected. Towards small $x$, however, the data do appear to show a slight downward bend, a possible onset of shadowing.

\subsection{Comparison with Baseline}

\begin{figure*}[htb!]
\centering
\includegraphics[width=1.00\linewidth]{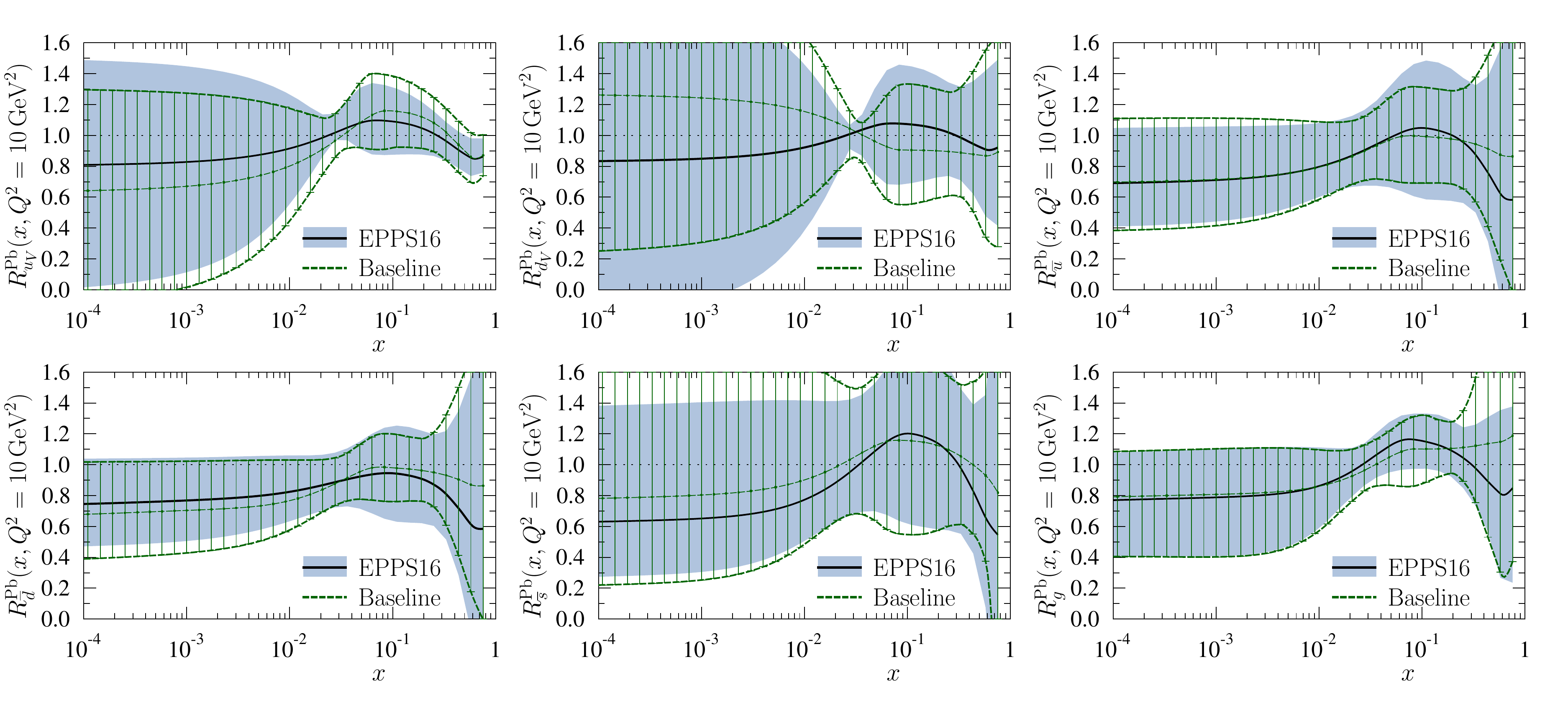}
\caption{The nuclear modifications at $Q^2=10 \, {\rm GeV}^2$ from the EPPS16 fit (black central line and light-blue bands) compared  with the Baseline fit (green curves with hatching) which uses only the data included in the EPS09 fit.}
\label{fig:compbase}
\end{figure*} 

\begin{figure*}[htb!]
\centering
\includegraphics[width=0.49\linewidth]{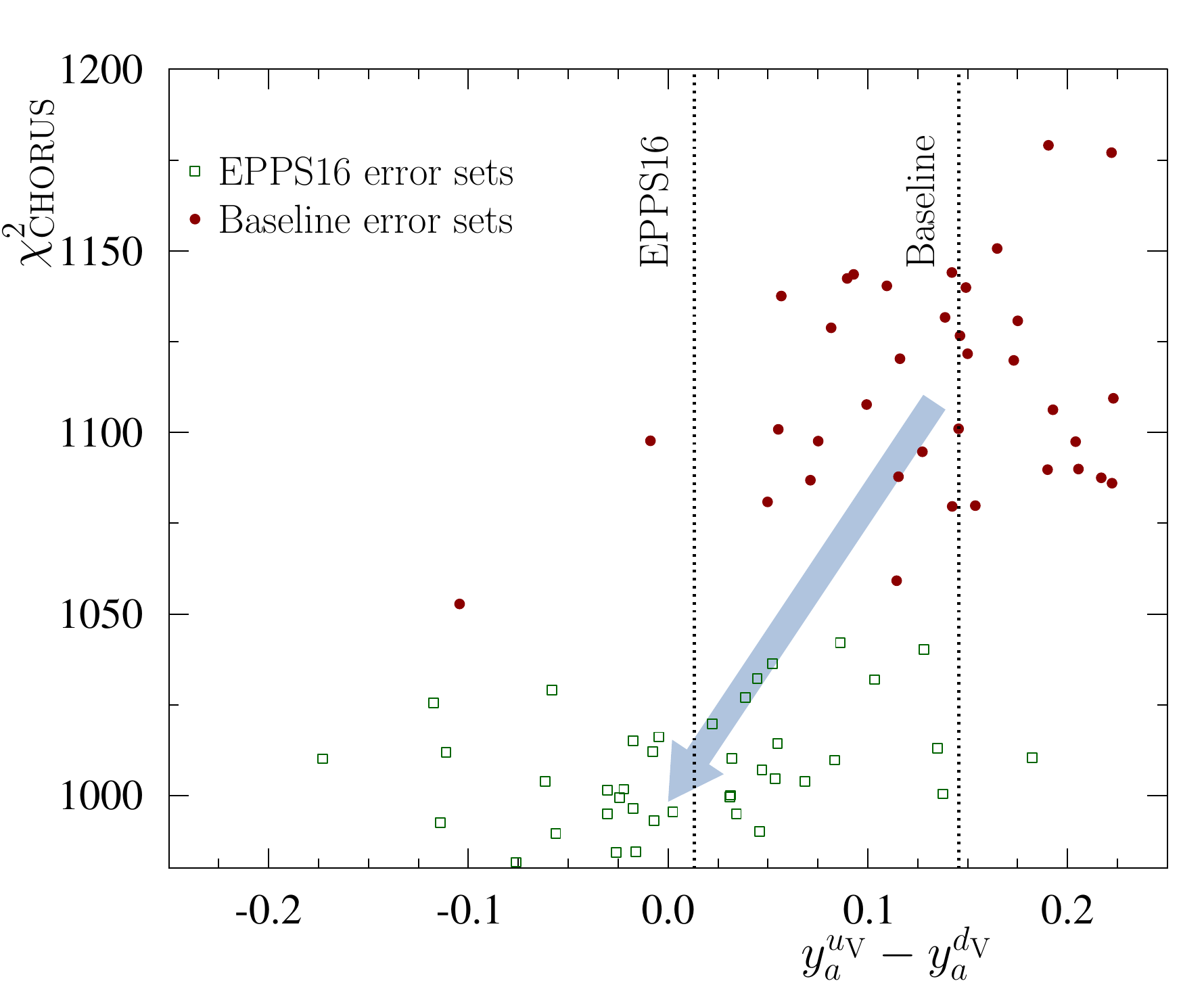}
\includegraphics[width=0.49\linewidth]{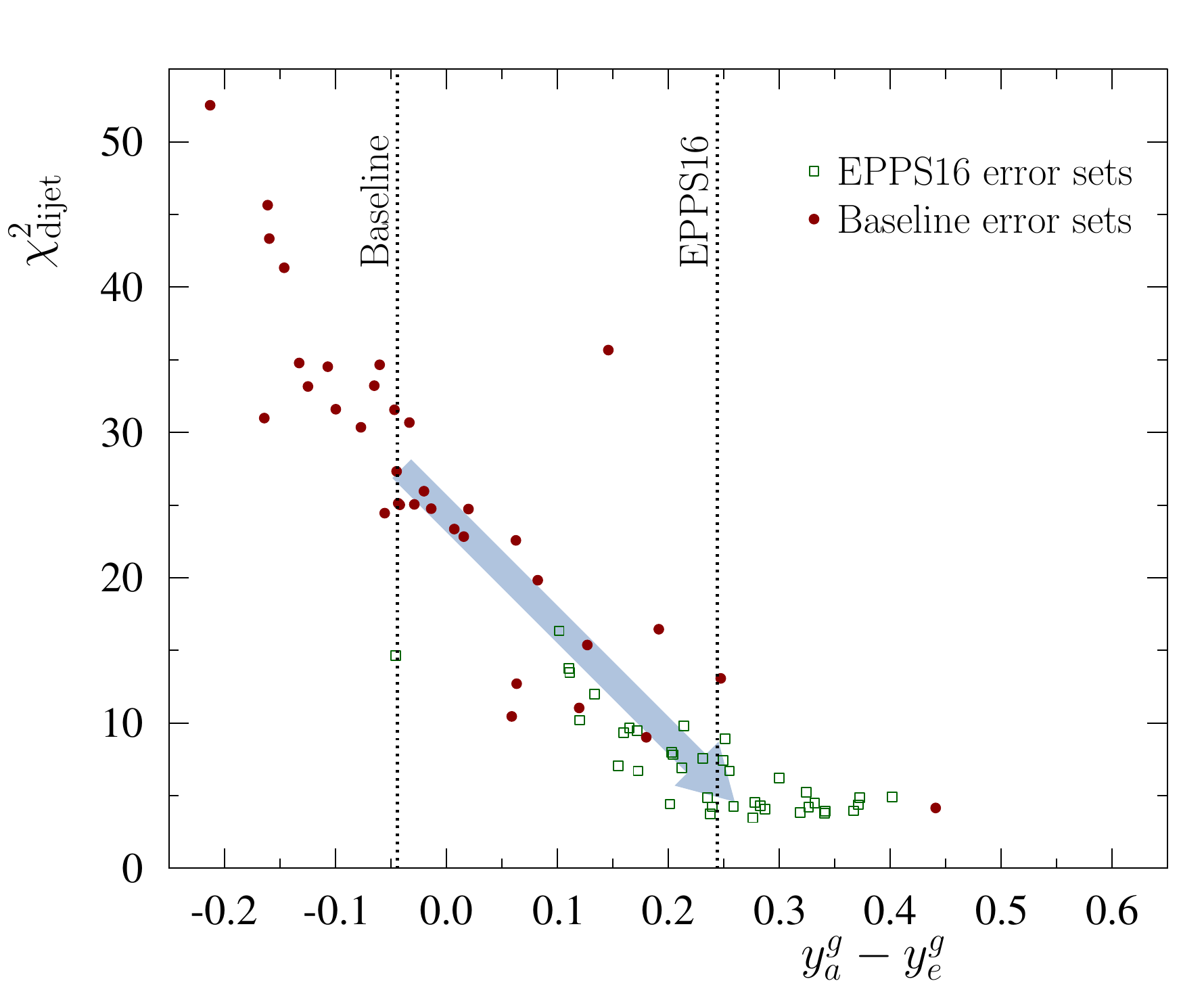}
\caption{The contribution of the CHORUS data \cite{Onengut:2005kv}  to the total $\chi^2$ as a function of $y_a^{u_{\rm V}}-y_a^{d_{\rm V}}$ (left) and the contribution of the CMS dijet data \cite{Chatrchyan:2014hqa} to the total $\chi^2$ as a function of $y_a^{g}-y_e^{g}$ (right). The green squares correspond to the EPPS16 error sets and the red circles to the error sets from the Baseline fit. The arrow indicates the direction of change induced by inclusion of these data into the analysis.}
\label{fig:CHORUS_dijet}
\end{figure*} 

\begin{figure*}[htb!]
\centering
\includegraphics[width=1.0\linewidth]{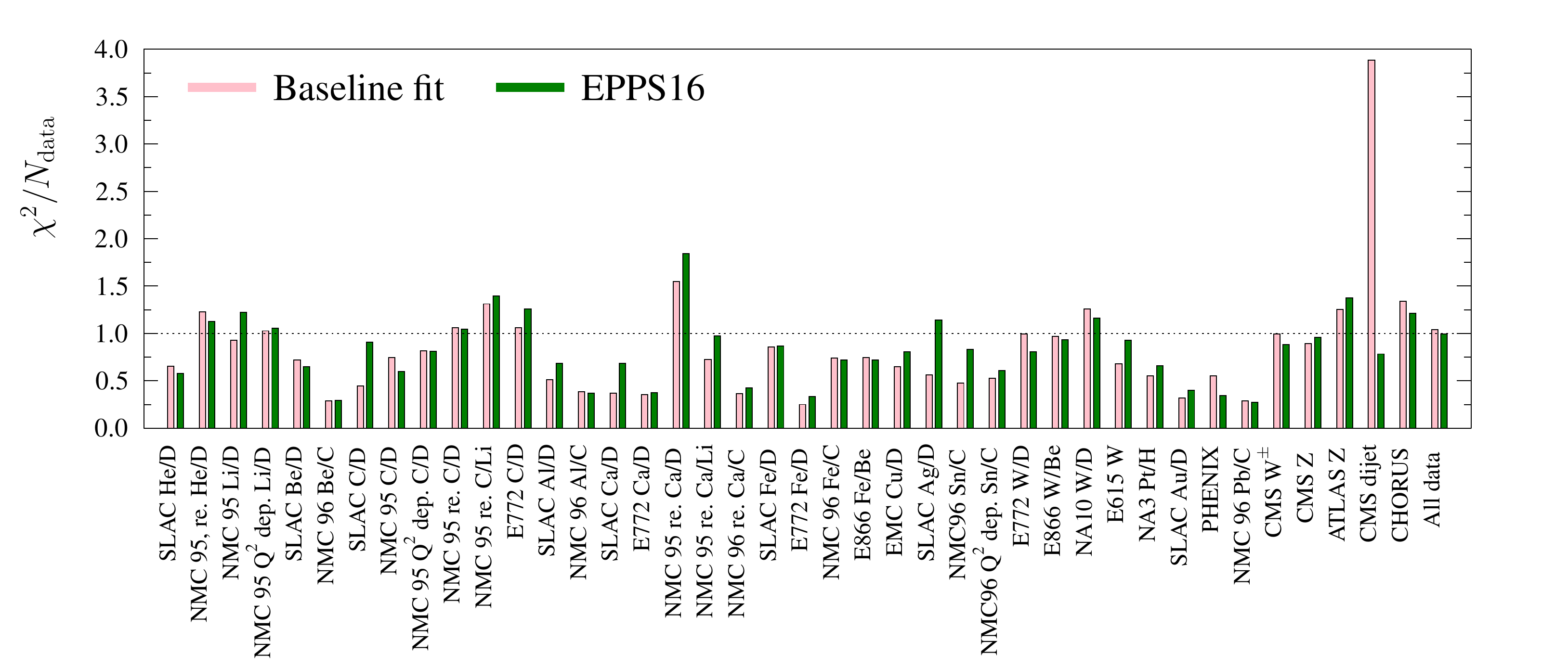}
\caption{The values of $\chi^2/N_{\rm data}$ from the Baseline fit (red bars) and EPPS16 (green bars) for data in Table~\ref{Table:Data}.}
\label{fig:old_vs_new_chi2}
\end{figure*} 

\begin{figure*}[htb!]
\centering
\includegraphics[width=1.00\linewidth]{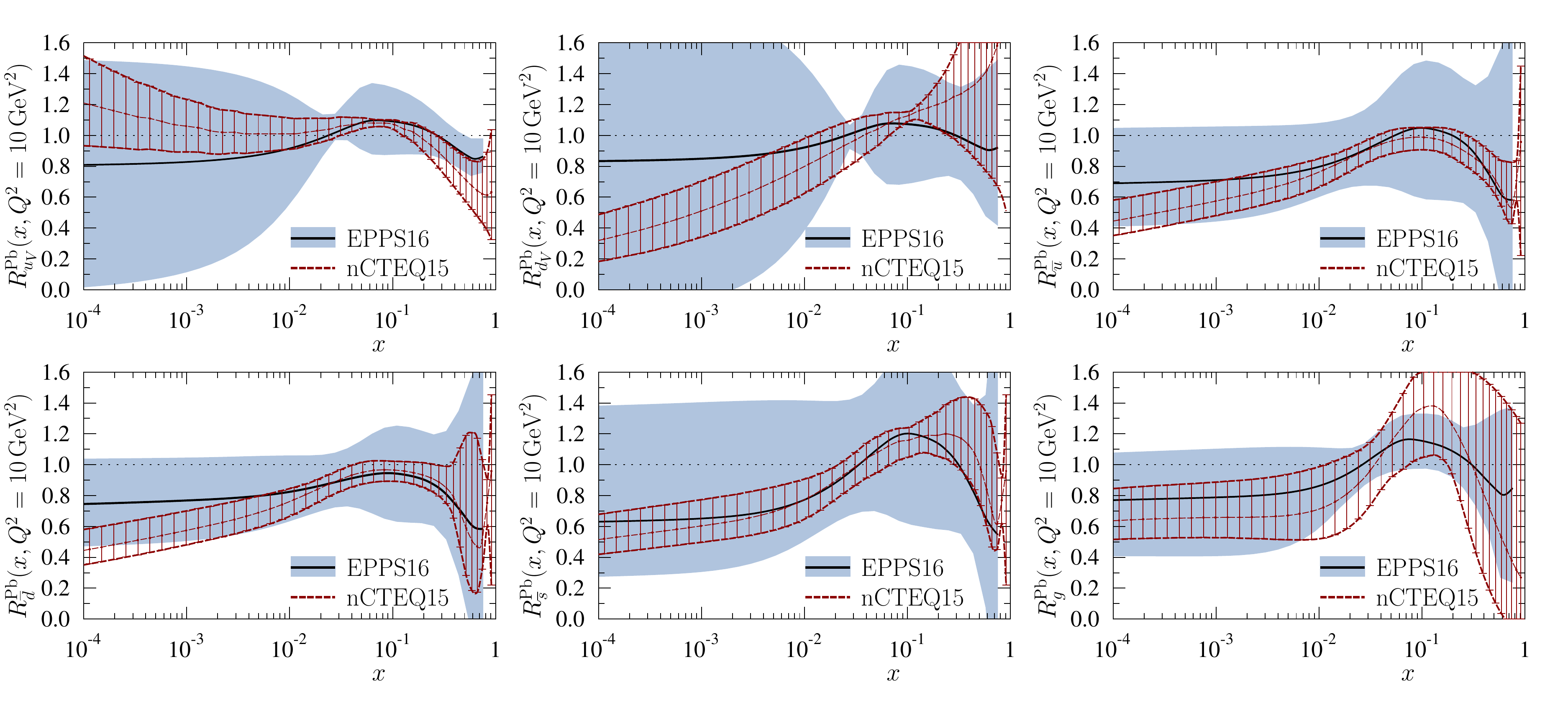}
\caption{Comparison of the EPPS16 nuclear modifications (black central curve with shaded uncertainty bands) with those from the nCTEC15 analysis \cite{Kovarik:2015cma} (red curves with hatching) at $Q^2=10\,{\rm GeV}^2$.}
\label{fig:Pb10_with_nCTEQ15}
\end{figure*}

To appreciate the effects induced by the new data (pion-$A$ DY, neutrino DIS and LHC data) in the EPPS16 fit, we have performed another fit excluding these data sets but still correcting the DIS data for the isospin effects. This fit is referred to as ``Baseline'' in what follows. 
The resulting nuclear modifications for Pb at $Q^2= 10 \, {\rm GeV}^2$ with a comparison to the EPPS16 results are shown in Fig.~\ref{fig:compbase}.
For the Baseline fit here, the global tolerance is $\Delta \chi^2_{\rm Baseline}=35$. As seen in the figure, it is not always the case that the uncertainties of EPPS16 would be smaller than those of the Baseline. This originates from the mutually different global tolerances of the two fits and from the differences of the $\chi^2$ behaviour around the minima. In any case, the uncertainty bands always overlap and both of these enclose the central values both from the Baseline fit and the full analysis. Thus, the two are consistent. Qualitatively, the most notable changes are that, in comparison to the Baseline, the EPPS16 central values of both valence-quark flavours as well as that of gluons exhibit a very similar antishadowing effect followed by an EMC pit. We have observed that this difference is mostly caused by the addition of neutrino DIS data (valence quarks) and the CMS dijet data (gluons). This is also illustrated in Fig.~\ref{fig:CHORUS_dijet} where the left-hand panel shows the $\chi^2$ contribution of the CHORUS data as a function of $y_a^{u_{\rm V}}-y_a^{d_{\rm V}}$ 
(the antishadowing peak heights for $A_{\rm ref}$ as in Table \ref{Table:Params}) and the right-hand panel the $\chi^2$ contribution of the CMS dijet data as a function of $y_a^{g}-y_e^{g}$. The individual points correspond to the EPPS16 and Baseline-fit error sets. From these panels we learn that in order to optimally reproduce the CHORUS data we need $y_a^{u_{\rm V}} \sim y_a^{d_{\rm V}}$, and an agreement with the CMS dijet data requires $y_a^{\rm g} > y_e^{\rm g}$ (EMC effect). The other new data (pion-$A$ DY, LHC electroweak data) do not generate such a strong pull away from the central set of the Baseline fit. Also the PHENIX data prefers a solution with a gluon EMC effect, but the contribution of these data in the total $\chi^2$ budget is so small that such a tendency is practically lost in the noise (in the EPS09 analysis this was compensated by giving these data an additional weight). The inclusion of the dijet data has also decreased the gluon uncertainties at large $x$, excluding the solutions with no antishadowing. In the case of ${u}$ and ${d}$ sea quarks there are no significant differences between the Baseline fit and EPPS16. It appears that the $s$-quark uncertainty at small $x$ has somewhat reduced by the inclusion of the new data, but the uncertainty is in any case large.

\begin{figure}[htb!]
\centering
\includegraphics[width=1.0\linewidth]{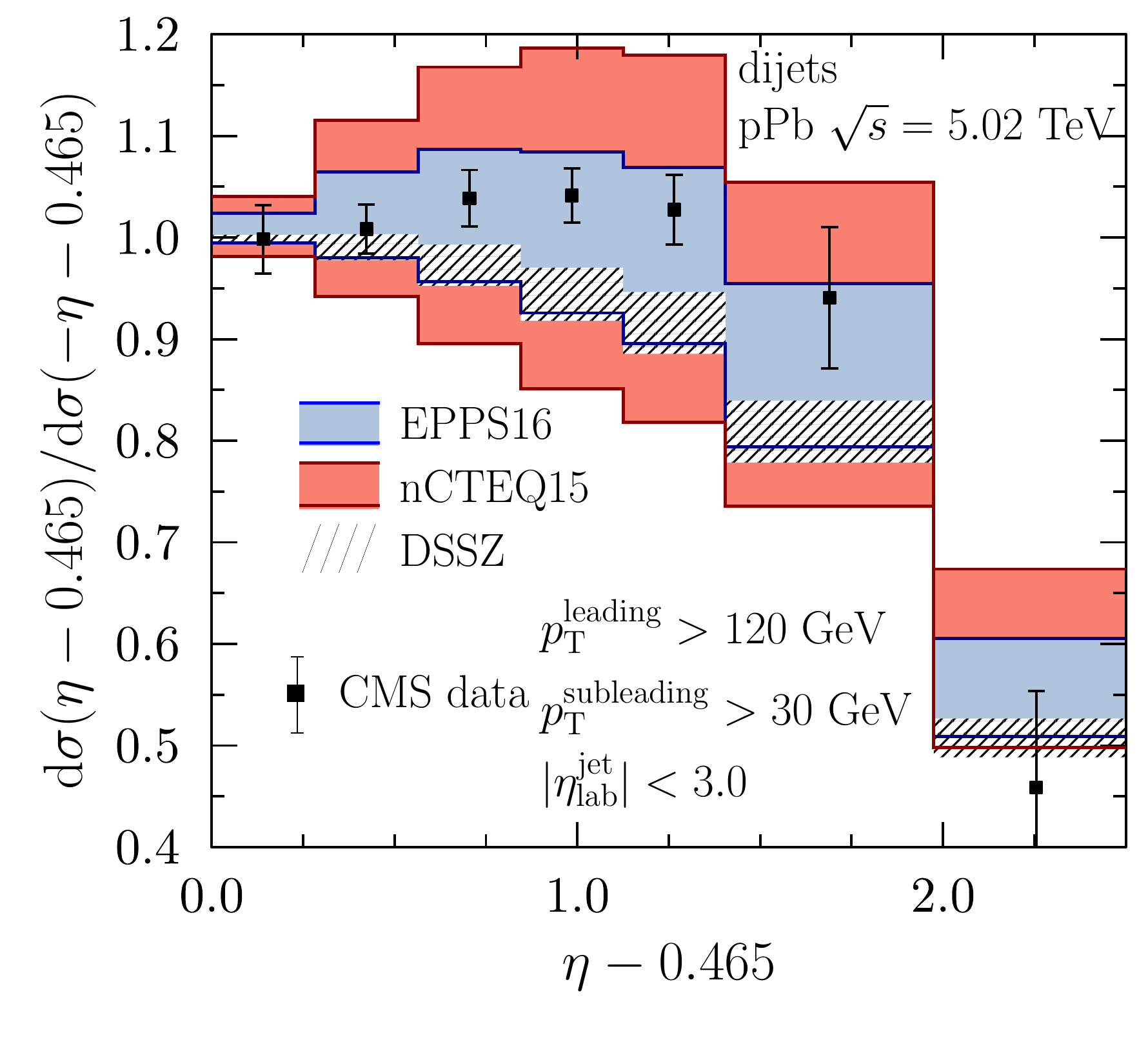}
\caption{The CMS dijet data \cite{Chatrchyan:2014hqa} compared with the results obtained with the EPPS16 (blue bands), nCTEQ15 \cite{Kovarik:2015cma} (red bands) and DSSZ \cite{deFlorian:2011fp} (hatched bands) nuclear PDFs.}
\label{fig:dijet_CMS_with_nCTEQ}
\end{figure} 

The values of $\chi^2/N_{\rm data}$ for individual data sets are shown in Fig.~\ref{fig:old_vs_new_chi2}. For the CMS dijet data the Baseline fit gives a very large value but this disagreement disappears when these data are included in the fit. However, upon including the new data no obvious conflicts with the other data sets show up and thus the new data appear consistent with the old. While it is true that on average $\chi^2/N_{\rm data}$ for the old data grows when including the new data (and this is mathematically inevitable) no disagreements ($\chi^2/N_{\rm data} \gg 1$) occur. For the NMC Ca/D data $\chi^2/N_{\rm data}$ is somewhat large but, as can be clearly seen from Fig.~\ref{fig:ALiDQ2}, there appears to be large fluctuations in the data (see the two data points below the EPPS16 error band). While the improvement in $\chi^2/N_{\rm data}$ for the CHORUS data looks smallish in Fig.~\ref{fig:old_vs_new_chi2}, for the large amount of data points (824) the absolute decrease in $\chi^2$ amounts to 106 units and is therefore significant.

\subsection{Comparison with other nuclear PDFs}

\begin{figure}[htb!]
\centering
\includegraphics[width=1.0\linewidth]{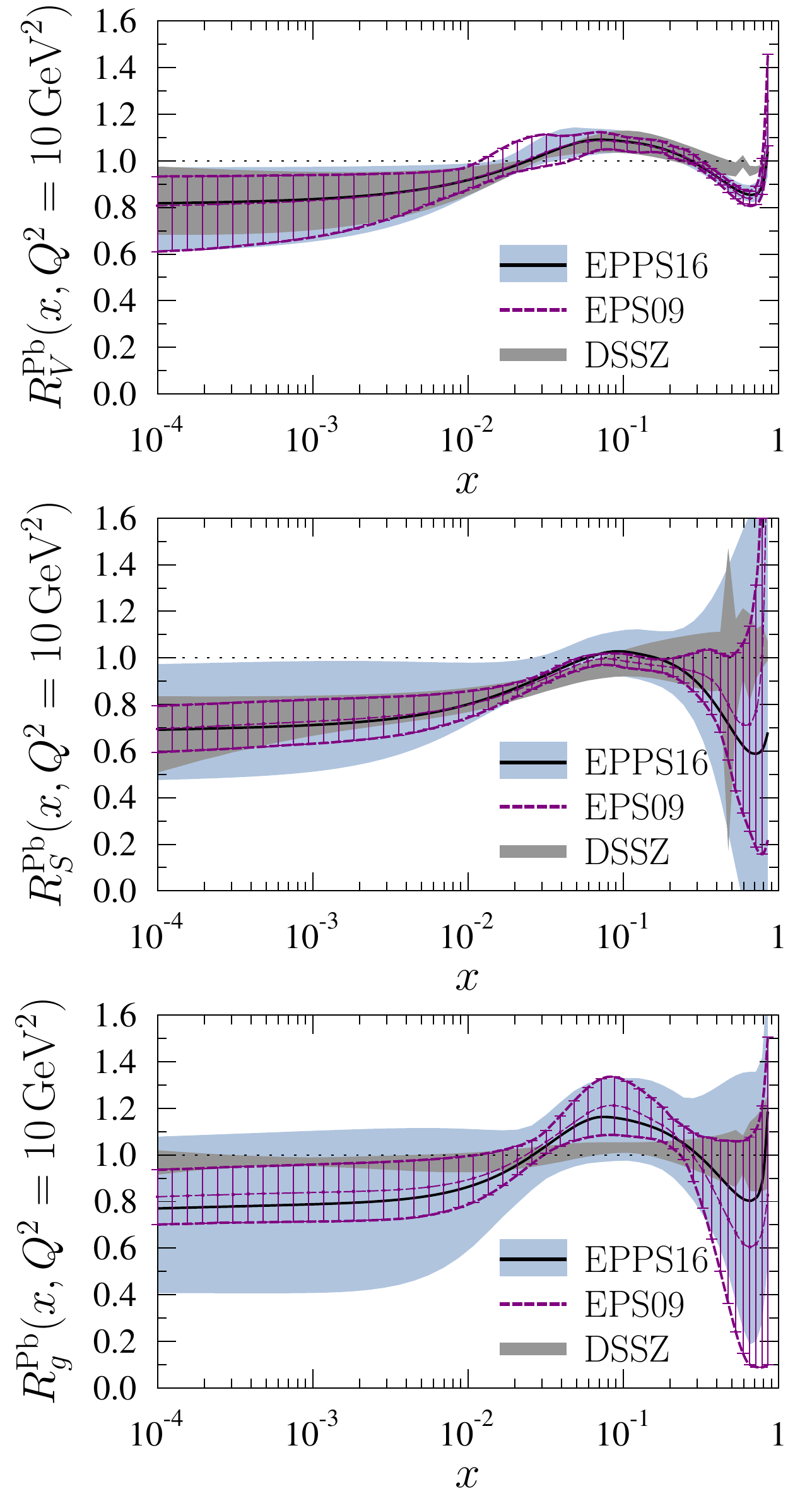}
\caption{Comparison of the EPPS16 nuclear modifications (black central curve with light-blue uncertainty bands) to those from the EPS09 analysis (purple curves with hatching) and DSSZ \cite{deFlorian:2011fp} (gray bands) at $Q^2=10\,{\rm GeV}^2$. The upper panels correspond to the average valence and sea-quark modifications of Eqs.~\eqref{eq:avval} and \eqref{eq:avsea}, the bottom panel is for gluons.
}
\label{fig:Pb10_with_EPS09}
\end{figure} 

In Fig.~\ref{fig:Pb10_with_nCTEQ15} we compare our EPPS16 results at the scale $Q^2=10\,{\rm GeV}^2$ with those of the nCTEQ15 analysis \cite{Kovarik:2015cma}. The nCTEQ15 uncertainties are defined by a fixed tolerance $\Delta \chi^2 = 35$, which is similar to our average value $\Delta \chi^2 = 52$ and in this sense one would expect uncertainty bands of comparable size. The quark PDFs were allowed to be partly flavour dependent in the nCTEQ15 analysis (although to a much lesser extent than in EPPS16), hence we show the comparison for all parametrized parton species. The two fits (as well as nCTEQ15 and our Baseline fit in Fig.~\ref{fig:compbase}) can be considered compatible since the uncertainty bands always overlap. For all the sea quarks the nCTEQ15 uncertainties appear clearly smaller than those of EPPS16 though less data was used in nCTEQ15. This follows from the more restrictive assumptions made in the nCTEQ15 analysis regarding the sea-quark fit functions: nCTEQ15 has only 2 free parameters for all sea quarks together, while EPSS16 has 9. Specifically, the nCTEQ15 analysis constrains only the sum of nuclear $\bar u + \bar d$ with an assumption that the nuclear $s$ quarks are obtained from  $\bar u + \bar d$ in a fixed way. In contrast, EPPS16 has freedom for all sea quark flavours separately, and hence also larger, but less biased, error bars. For the valence quarks, the nCTEQ15 uncertainties are somewhat larger than the EPPS16 errors around the $x$-region of the EMC effect which is most likely related to the extra constraints the EPPS16 analysis has obtained from the neutrino DIS data. Especially the central value for $d_{\rm V}$ is rather different than that of EPPS16. The very small nCTEQ15 uncertainty at $x \sim 0.1$ is presumably a similar fit-function artefact as what we have for EPPS16 at slightly smaller $x$. Such a small uncertainty is supposedly also the reason why nCTEQ15 arrives at smaller uncertainties in the shadowing region than EPPS16. For the gluons the nCTEQ15 uncertainties are clearly larger than those of EPPS16, except in the small-$x$ region. While, in part, the larger uncertainties are related to the LHC dijet data that are included in EPPS16 but not in nCTEQ15, this is not the complete explanation as around $x \sim 0.1$ the nCTEQ15 uncertainties also largely exceed the uncertainties from our Baseline fit (see Fig.~\ref{fig:compbase}). Since the data constraints for gluons in both analyses are essentially the same, the reason must lie in the more stringent $Q^2$ cut ($Q^2 > 4 \, {\rm GeV}^2$) used in the nCTEQ15 analysis, which cuts out low-$Q^2$ data points where the indirect effects of gluon distributions via parton evolution are the strongest. The inclusion of the dijet data into the nCTEQ15 analysis would clearly have a dramatic impact. This can be understood from Fig.~\ref{fig:dijet_CMS_with_nCTEQ} where we compare the CMS dijet data also with the nCTEQ15 prediction (here, we have formed the nCTEQ15 nuclear modifications from their absolute distributions and used the same dijet grid as in the EPPS16 analysis).

A comparison of EPPS16 with EPS09 \cite{Eskola:2009uj} and DSSZ \cite{deFlorian:2011fp} is presented in Fig.~\ref{fig:Pb10_with_EPS09}. In the EPS09 and DSSZ analyses the nuclear modifications of valence and sea quarks were flavour independent at the parametrization scale and, to make a fair comparison we plot, in addition to the gluons, the average nuclear modifications for the valence quarks
and light sea quarks,
\begin{align}
 R_{\rm V}^{\rm Pb} & \equiv  \frac{u^{\rm p/Pb}_{\rm V}+d^{\rm p/Pb}_{\rm V}}{u^{\rm p}_{\rm V}+d^{\rm p}_{\rm V}}, \label{eq:avval} \\
 R_{\rm S}^{\rm Pb} & \equiv  \frac{\overline{u}^{\rm p/Pb}+\overline{d}^{\rm p/Pb}+\overline{s}^{\rm p/Pb}}{\overline{u}^{\rm p}+\overline{d}^{\rm p}+\overline{s}^{\rm p}}, \label{eq:avsea}
\end{align}
instead of individual flavours. For the valence sector, all parametrizations give very similar results except DSSZ in the EMC-effect region.
As noted earlier in Sec.~\ref{IsoscalarcorectionforDISdata} and in Ref.~\cite{Paukkunen:2014nqa} this is likely to originate from ignoring the isospin corrections in the DSSZ fit. The sea-quark modifications look also mutually rather alike, the EPPS16 uncertainties being somewhat larger than the others as, being flavour-dependent, the sea quarks in EPPS16 have more degrees of freedom. 
As has been understood already some while ago \cite{Eskola:2012rg,Paukkunen:2014nqa}, the DSSZ parametrization has almost no nuclear effects in gluons as nuclear effects  were included in the FFs \cite{Sassot:2009sh} when computing inclusive pion production at RHIC. As a result, DSSZ does not reproduce the new CMS dijet measurements as shown here in Fig.~\ref{fig:dijet_CMS_with_nCTEQ}. Between EPS09 and EPPS16, the gluon uncertainties are larger in EPPS16. While EPPS16 includes more constraints for the gluons (especially the CMS dijet data), in EPS09 the PHENIX data was assigned an additional weight factor of 20. This in effect increased the importance of these data, making the uncertainties smaller than what they would have been without such a weight (the Baseline-fit gluons in Fig.~\ref{fig:compbase} serve as a representative of an unweighted case). In addition, in EPPS16 one more gluon parameter is left free ($x_a$) which also increases the uncertainties in comparison to EPS09. 

\begin{figure}[htb!]
\centering
\includegraphics[width=1.00\linewidth]{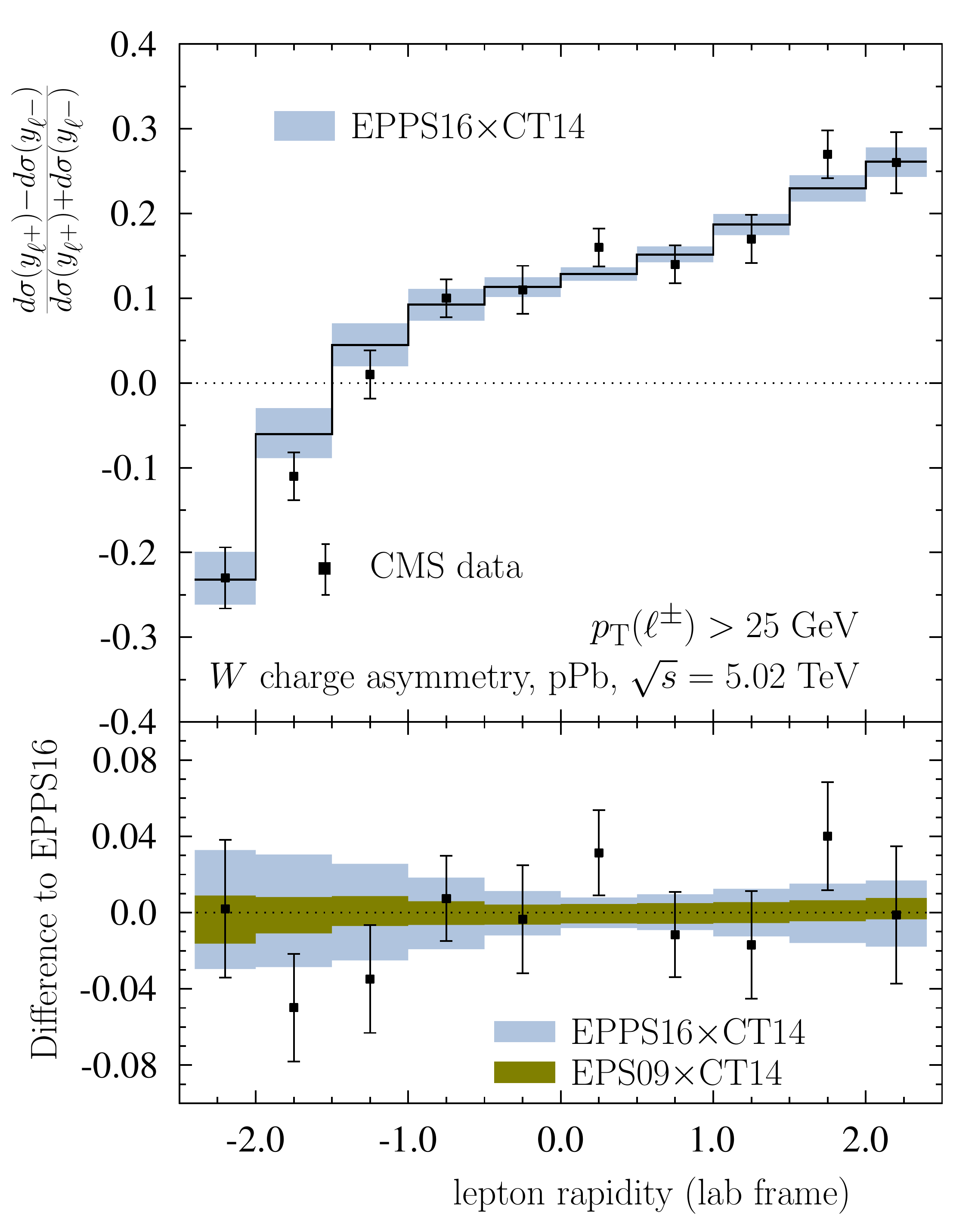}
\caption{The CMS W charge asymmetry measurement \cite{Khachatryan:2015hha} compared with the predictions using EPPS16 nuclear modifications and CT14NLO proton PDFs. In both panels the blue bands correspond to the combined EPPS16+CT14 uncertainty and in the lower panel the green band to the combined EPS09+CT14 uncertainty.}
\label{fig:Wasymmetry}
\end{figure} 

\section{Application: W charge asymmetry}

The W charge-asymmetry measurement by CMS in pPb collisions \cite{Khachatryan:2015hha} revealed some deviations from the NLO calculations in the backward direction and it was suggested that this difference could be due to flavour-dep\-endent PDF nuclear modifications. While it was shown in Ref.~\cite{Arleo:2015dba} that such a difference does not appear in the ATLAS PbPb data \cite{Aad:2014bha} at the same probed values of $x$, the situation still remains unclear. To see how large variations the new EPPS16 can accommodate, we compare in Fig.~\ref{fig:Wasymmetry} the CMS data with the EPPS16 and EPS09 predictions using the CT14NLO proton PDFs. As discussed in the original EPS09 paper \cite{Eskola:2009uj}, the total uncertainty should be computed by adding in quadrature the uncertainties stemming separately from EPPS16 and from the free-proton baseline PDFs,
\begin{equation}
 (\delta \mathcal{O}_{\rm total})^2 = (\delta \mathcal{O}_{\rm EPPS16})^2 + (\delta \mathcal{O}_{\rm baseline})^2,
\end{equation}
where $\delta \mathcal{O}_{\rm EPPS16}$ is evaluated by Eq.~(\ref{eq:asymerr}) using the uncertainty sets of EPPS16 with the central set of free-proton PDFs, and $\delta \mathcal{O}_{\rm baseline}$ by the same equation but using the free-proton error sets with the central set of EPPS16. The same has been done in the case of EPS09 results. While the differences between the central predictions of EPPS16 and EPS09 are tiny, it can be seen that the uncertainty bands of EPPS16 are clearly wider and, within the uncertainties, the data and EPPS16 are in a fair agreement. As this observable is mostly sensitive to the free-proton baseline (to first approximation the nuclear effects in PDFs cancel) we do not use these asymmetry data as a constraint in the actual fit in which we aim to expose the nuclear effects in PDFs.

\section{Summary and outlook}

We have introduced a significantly updated global analysis of NLO nuclear PDFs --- EPPS16 --- with less biased, flavour-dependent fit functions and a larger variety of data constraints than in other concurrent analyses. In particular, new LHC data from the 2013 pPb run are for the first time directly included. Another important addition here is the neutrino-nucleus DIS data. Also the older pion-nucleus DY data are now for the first time part of the analysis. From the new data, the most significant role is played by the neutrino DIS data and the LHC dijet measurements whose addition leads to a consistent picture of qualitatively similar nuclear modifications for all partonic species. Remarkably, the addition of new data types into the global fit does not generate notable tensions with the previously considered data sets. This lends support to the validity of collinear factorization and process-independent nuclear PDFs in the kinematical $x,Q^2$ region we have considered.

However, the uncertainties are still significant for all components and, clearly, more data is therefore required. In this respect, the prospects for rapid developments of nuclear PDFs are very good: It can be expected that new data from the LHC will be available soon. For example, from the 2013 pPb data taking, a more differential dijet analysis by the CMS collaboration \cite{CMS:2016kjd} as well as W data by ATLAS \cite{ATLASW} are still being prepared. In November-December 2016, the LHC has recorded pPb collisions at the highest energy ever, $\sqrt{s}=8.16$ TeV, with more than 6 times more statistics than that from the 2013 pPb run at $\sqrt{s}=5.02$ TeV.\footnote{\url{https://lpc.web.cern.ch/lumiplots_2016_PbPb.htm}} The new data from this run will provide further constraints to the nuclear PDFs in the near future. As in the case of free-proton PDFs \cite{Zenaiev:2015rfa,Gauld:2016kpd}  heavy-flavour production at forward direction \cite{LHCb:2016huj} may offer novel small-$x$ input. An interesting opportunity is also the possibility of the LHCb experiment to operate in a fixed-target mode and measure e.g. pNe (and other noble gases) collisions \cite{Zhang:2016hmo}. From other experiments, new fixed-target proton-induced Drell-Yan data from the Fermilab E-906/SeaQuest experiment \cite{Dannowitz:2016qkz} should also provide better constraints e.g. for the $A$ dependence of the sea-quark nuclear modifications. 

Further in the future, the planned Electron-Ion Collider \cite{Accardi:2012qut} (and LHeC \cite{AbelleiraFernandez:2012cc} if materialized) will provide high-precision DIS constraints for all nuclear parton flavours. In addition, the possible realization of a new forward calorimeter (FOCAL) at the ALICE experiment \cite{Peitzmann:2016gkt} would, in turn, give a possibility to measure isolated photons in the region sensitive to low $x$ gluons \cite{Helenius:2014qla}.

On the theoretical side, there is ample room for improvements as well. For example, similarly to the free-proton fits, an upgrade to next-to-NLO or inclusion of photon distributions and mixed QCD-QED parton evolution are obvious further developments. In a longer run, to avoid biases due to specific baseline proton PDFs, especially regarding the $s$ quark sector, fitting the proton PDFs and nuclear PDFs in one single analysis is ultimately needed.

\section*{Acknowledgments}

This research was supported by the Academy of Finland, Project 297058 of K.J.E.; by the European Research Council grant HotLHC ERC-2011-StG-279579; by Ministerio de Ciencia e Innovaci\'on of Spain under project FPA2014-58293-C2-1-P; and by Xunta de Galicia (Conselleria de Educacion) --- H.P. and C.A.S. are part of the Strategic Unit AGRUP\-2015/11. P.P. acknowledges the financial support from the Magnus Ehrnrooth Foundation. Part of the computing has been done in T.~Lappi's project at CSC, the IT Center for Science in Espoo, Finland.

\end{document}